\documentclass{aa}

\usepackage{graphicx}
\usepackage{txfonts}
\usepackage{hyperref}
\usepackage{amsmath}
\usepackage[dvipsnames]{xcolor}

\newcommand{\mass}{$\rm \log\,M_{\star}~[M_{\odot}]$}
\newcommand{\kms}{km~s$^{-1}$}
\newcommand{\must}{$\rm \log\,\mu_{\star}$}
\newcommand{\xgass}{xGASS}
\newcommand{\xcoldgass}{xCOLD\,GASS}
\newcommand{\hi}{\textsc{H\,i}}
\newcommand{\co}{CO}
\newcommand{\msun}{${\rm M_{\odot}}$}
\newcommand{\mstar}{${\rm M_{\star}}$}
\newcommand{\fhi}{$\rm \log\,f_{HI}$}
\newcommand{\fhtwo}{$\rm \log\,f_{H2}$}
\newcommand{\htwo}{H$_{2}$}

\newcommand{\nuvr}{$NUV$-$r$}

\defcitealias{Moran2012}{M12}

\begin{document}

    \title{xCOLD\,GASS \& xGASS: Radial metallicity gradients and global properties on the star-forming main sequence\thanks{Based on observations made with the EFOSC2 instrument on the ESO NTT telescope under the programme 091.B-0593(B) at Cerro La Silla (Chile). } \thanks{Galaxy measurements and calibrated 2D spectra of low mass galaxies are only available in electronic form
at the CDS via anonymous ftp to cdsarc.u-strasbg.fr (130.79.128.5) or via http://cdsweb.u-strasbg.fr/cgi-bin/qcat?J/A+A/} }
    \titlerunning{xCOLD\,GASS \& xGASS: Metallicity and cold gas}
    \author{K. A. Lutz \inst{1, 2, 3} \and
            A. Saintonge \inst{4} \and
            B. Catinella \inst{5, 6} \and
            L. Cortese \inst{5, 6}  \and
            F. Eisenhauer \inst{2} \and
            C. Kramer \inst{7} \and
            S. M. Moran \inst{8} \and
            L. J. Tacconi \inst{2} \and
            B. Vollmer \inst{1} \and
            J. Wang \inst{9}
            }

    \institute{Observatoire Astronomique de Strasbourg, Universit\'{e} de Strasbourg, CNRS,  UMR 7550, 67000 Strasbourg, France \\ \email{research@katha-lutz.de} \and
    Max-Plack-Institute for extraterrestrial Physics, 85741 Garching, Germany \and
    Physik Department, Technische Universit\"at M\"unchen, 85741 Garching, Germany \and
    Department of Physics and Astronomy, University College London, Gower Street, London, WC1E 6BT, UK \and
    International Centre for Radio Astronomy Research (ICRAR), M468, The University of Western Australia, 35 Stirling Highway, Crawley, WA 6009, Australia \and
    Australian Research Council, Centre of Excellence for All Sky Astrophysics in 3 Dimensions (ASTRO 3D), Australia \and
    Institut de Radioastronomie Millim\'{e}trique (IRAM), 300 rue de la Piscine, 38406 Saint Martin d'H\`{e}res, France \and
    Smithsonian Astrophysical Observatory, 60 Garden Street, Cambridge, MA 02138, USA \and
    Kavli Institute for Astronomy and Astrophysics, Peking University, Beijing 100871, China}

   \date{Received September 15, 1996; accepted March 16, 1997}

   \abstract
{The \xgass\ and \xcoldgass\ surveys have measured the atomic (\hi) and molecular gas (\htwo) content of a large and representative sample of nearby galaxies (redshift range of $0.01<z<0.05$). }
{We present optical longslit spectra for a subset of the \xgass\ and \xcoldgass\ galaxies to investigate the correlation between radial metallicity profiles and cold gas content. In addition to data from \citet{Moran2012}, this paper presents new optical spectra for 27 galaxies in the stellar mass range of $9.0 \leq$ \mass\ $\leq10.0$. }
{The longslit spectra were taken along the major axis of the galaxies, allowing us to obtain radial profiles of the gas-phase oxygen abundance (12 + log(O/H)). The slope of a linear fit to these radial profiles is defined as the metallicity gradient. We investigated correlations between these gradients and global galaxy properties, such as star formation activity and gas content. In addition, we  examined the correlation of local metallicity measurements and the global \hi\ mass fraction. }
{We obtained two main results: (i) the local metallicity is correlated with the global \hi\ mass fraction, which is in good agreement with previous results. A simple toy model suggests that this correlation points towards a 'local gas regulator model'; (ii) the primary driver of metallicity gradients appears to be stellar mass surface density (as a proxy for morphology). }
{This work comprises one of the few systematic observational studies of the influence of the cold gas on the chemical evolution of star-forming galaxies, as considered via metallicity gradients and local measurements of the gas-phase oxygen abundance. Our results suggest that local density and local \hi\ mass fraction are drivers of chemical evolution and the gas-phase metallicity. }


   \maketitle

%%%%%%%%%%%%%%%%%%%%%%%%%%%%%%%%%%%%%%%%%%%%%%%%%%

%%%%%%%%%%%%%%%%% BODY OF PAPER %%%%%%%%%%%%%%%%%%

\section{Introduction}

Theory and observations support a scenario where galaxy growth is tightly linked to the availability of cold gas. Several key galaxy scaling relations can be explained by an `equilibrium' (or `gas regulator') model, in which galaxy growth self-regulates through accretion of gas from the cosmic web, star formation, and the ejection of gas triggered by star formation and feedback from active galactic nuclei (AGN,  see e.\,g. \citealp{Lilly2013, Dave2013}). Galaxy-integrated scaling relations successfully reproduced by this model range from the mass-metallicity relation \citep{Zahid2014,Brown2018}, the baryonic mass fraction of halos \citep{Bouche2010}, and the redshift evolution of the gas contents of galaxies \citep{Saintonge2013}.

The next logical step is to explore how this gas-centric galaxy evolution model performs in explaining the resolved properties of galaxies. This is a particularly timely question as integral field spectroscopic (IFS) surveys continue to provide detailed maps of the stellar and chemical composition of large, homogeneous, representative galaxy samples. The Calar Alto Legacy Integral Field Area survey (CALIFA; \citealp{Sanchez2012}), the Sydney-AAO Multi-object Integral field galaxy survey (SAMI; \citealp{Croom2012}), and the Mapping Nearby Galaxies at Apache Point Observatory survey (MaNGA; \citealp{Bundy2015}), for instance, focus on samples of hundreds to thousands of galaxies in the nearby Universe. Similar surveys at $z>1$ are also now possible with IFS instruments operating in the near-infrared such as KMOS (e.g. the KMOS3D and KROSS surveys, \citealp{Wisnioski2015} and \citealp{Stott2016},
respectively).

Observations of colour and star formation rate (SFR) across the discs of nearby galaxies suggest a scenario in which galaxies form and quench from the inside out; overall, the outskirts of disc galaxies tend to be bluer \citep{deJong1996,Wang2011,Perez2013} and remain star-forming for longer \citep{Belfiore2017b,Medling2018}.

Alongside star formation (SF) profiles, a great deal insight can be gained by focusing on spatial variations of the chemical composition of the gas. In general, gas at the outskirts of a galaxy tends to be more metal-poor than at its centre  \citep[e.g.][]{Searle1971,Shields1974,Sanchez2014}.
Chemo-dynamical models of galaxies suggest different underlying physical processes to explain the  formation of metallicity gradients. Early on \citet{Matteucci1989} found, via Galactic models, that the inflow of metal-poor gas is vital to the formation of metallicity gradients. The chemical evolution models by \citet{Boissier1999} for the Milky Way and by \citet{Boissier2000} for disc galaxies additionally emphasise the role of radial variation of star formation rate and efficiency, as well as inside-out-growth for the formation of metallicity gradients. Furthermore, radial gas flows are found to be vital in reproducing the metallicity gradients of the Milky Way \citep{Schoenrich2009}
Within the `equilibrium' framework, such metallicity gradients would be explained by the accretion of metal-poor gas onto the outer regions of these galaxies. \citet{Pezzulli2016} showed that metallicity gradients already form in closed-box models due to the fact that denser (i.e. more central) regions of galaxies evolve faster than less dense (outer) regions. However, to arrive at realistic metallicity gradients, their analytical model requires radial gas flows and, to a lesser extent, also inside-out growth.

Metallicity gradients are common and there are many explanations of their presence. In particular, several physical processes (inside-out growth, radial flows, gradients in the star formation efficiency) have been predicted to give rise to metallicity gradients and it is difficult to assess their relative importance. It is also unclear whether or not the direction and strength of metallicity gradients depend on global galaxy properties. For example, \citet{Sanchez-Menguiano2016} and \citet{Ho2015} find no relation between metallicity gradients and the stellar mass of the galaxies. They argue that metallicity at a certain radius is only determined by local conditions and the evolutionary state of the galaxy at that radius, rather than by global properties. There are however analyses finding correlations (both positive and negative) between stellar mass and metallicity gradients. \citet{Poetrodjojo2018}, \citet{Belfiore2017a} and {}
\citet{Perez-Montero2016} find hints of flatter metallicity gradients in lower mass galaxies, while \citet{Moran2012} (hereafter M12) find that galaxies at the lower mass end of their sample have steeper gradients. We note, however, that the low-mass end of \citetalias{Moran2012} is at $10^{10}$\,M$_\odot$, where \citet{Belfiore2017a} find a turnover in the strength of the metallicity gradient, such that both more massive and less massive galaxies show flatter metallicity gradients than galaxies with masses around {}
$10^{10}$\,M$_\odot$. In addition, the sample from \citet{Poetrodjojo2018} only includes galaxies with stellar masses below $10^{10.5}$\,M$_\odot$.

As it is based on an extensive longslit spectroscopy campaign rather than IFU maps, the \citetalias{Moran2012} study may lack the full mapping of metallicity across the galaxy discs, but it does benefit from having access to direct measurements of the cold gas contents of the galaxies through the $GALEX$ Arecibo SDSS Survey (GASS) and CO Legacy Database for GASS (COLD\,GASS) surveys \citep{Catinella2010, Saintonge2011a}. Their finding is that metallicity gradients tend to be flat within the optical radius of the galaxies, but that the magnitude of any drop in metallicity in the outskirts is well-correlated with the total atomic hydrogen content of the galaxy. This provides support for a scenario where low-metallicity regions are connected to the infall of metal-poor gas, as also found by \citet{Carton2015}. Indeed, the chemical evolution models of
\citet{Ho2015}, \citet{Kudritzki2015} and \citet{Ascasibar2015} (amongst others) are able to predict metallicity gradients from radial variations in the gas-to-stellar-mass ratio. Using dust extinction maps derived from the Balmer decrement to infer local gas masses, \citet{Barrera-Ballesteros2018} find a relation between the radial profiles of gas to stellar mass ratio and metallicity that is in good agreement with the predictions from the local gas-regulator model (similar to the global model, but on local scales).

In this paper, we revisit the results of \citetalias{Moran2012} but for an increased sample of galaxies, which crucially extents the stellar mass range by an order of magnitude. This is achieved by combining new optical longslit spectra for galaxies in the stellar mass range of $9 < \log \rm{M_{\star} [M_{\odot}]} < 10$ with global cold gas measurements from the \xgass\ and \xcoldgass\ surveys \citep{Catinella2018,Saintonge2017}. The sample studied here, while lacking spatially resolved gas observations, is larger than those studied by {}
\citet{Ho2015}, \citet{Kudritzki2015} and \citet{Carton2015}, and it benefits from direct, homogeneous CO and HI observations.

This paper is organised as follows: In Sect.~\ref{sec:survey}, we present the galaxy sample and auxiliary data from the Sloan Digital Sky Survey \citep{York2000} and the \xgass/\xcoldgass\ surveys. Details on observation, data reduction and data analysis are provided in Sect.~\ref{sec:analysis}. Our results are presented in Sects.~\ref{sec:results1} and \ref{sec:results2}. We discuss these results and offer our conclusions in Sect.~\ref{sec:diss}.
Throughout the paper we assume a standard $\Lambda$CDM cosmology ($H_0= 70$\kms~Mpc$^{-1}$, $\Omega_M = 0.30$ and $\Omega_{\Lambda} = 0.70$) and a Chabrier initial mass function \citep{Chabrier2003}.

\section{Sample selection and global galaxy properties}
\label{sec:survey}

\begin{figure}
    \center
    \includegraphics[width=3.15in]{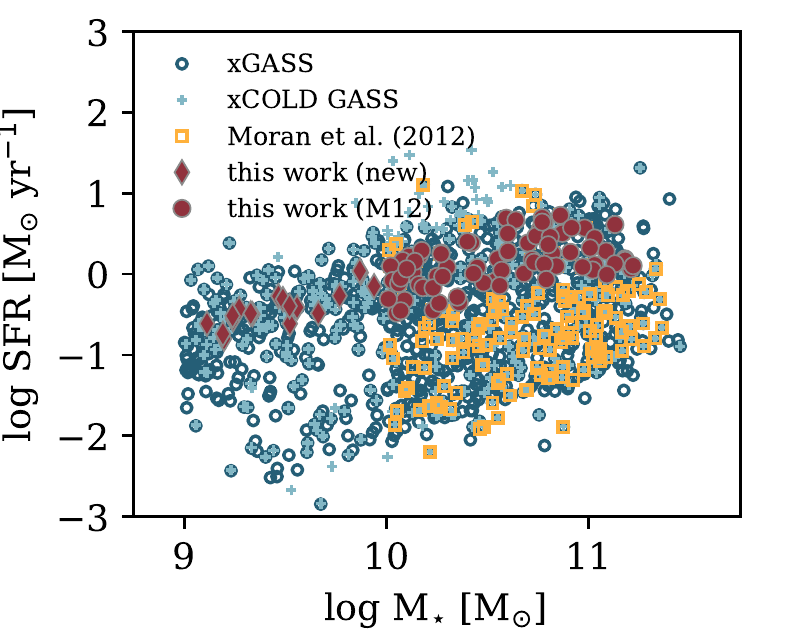}
    \caption{Distribution of the sample in the stellar mass-SFR plane. Blue open circles represent the \xgass\ sample, light blue crosses the \xcoldgass\ sample, and yellow open squares mark galaxies that were included in the \citetalias{Moran2012} analysis but not in this work. Galaxies marked with a red symbol are included in the sample used in this paper, where diamonds represent galaxies with new observations and filled circles galaxies with observations from \citetalias{Moran2012}. }
    \label{fig:sfr_vs_mstar}
\end{figure}

The extended GASS (\xgass, \citealp{Catinella2018}) and the corresponding extended COLD\,GASS (\xcoldgass, \citealp{Saintonge2017}) surveys are projects designed to provide a complete view of the cold atomic and molecular gas contents across the local galaxy population with stellar masses in excess of $10^9$\msun. The survey galaxies were randomly selected from the parent sample of objects in the SDSS DR7 spectroscopic catalogue \citep{Abazajian2009}, with
$0.01<z<0.05$ and $\log \rm{M_{\star} [M_{\odot}]} > 9.0$, and located within the footprint of the Arecibo HI ALFALFA survey \citep{Giovanelli2005a,Haynes2018}. No additional selection criteria were applied, making the sample representative of the local galaxy population. As shown in Fig. \ref{fig:sfr_vs_mstar}, it samples the entire SFR--\mstar\ plane. The \xgass\ survey provides total HI masses for 1200 galaxies and \xcoldgass\ derived total molecular gas masses from CO(1-0) observations
of a subset of 532 of these. A complete description of the sample selection, observing procedures and data products of \xgass\ and \xcoldgass\ can be found in \citet{Catinella2018} and \citet{Saintonge2017}, respectively.

In addition to the \hi\ and \co\ measurements, optical longslit spectra were obtained for a subset of the \xgass/\xcoldgass\ galaxies. For galaxies with stellar masses $\log \rm{M_{\star} [M_{\odot}]} > 10$, these data were obtained with the 6.5-m MMT telescope on Mount Hopkins, Arizona (182 galaxies) and the 3.5-m telescope at Apache Point Observatory (APO), New Mexico (51 galaxies, \citetalias{Moran2012}). For 27 galaxies in the stellar mass range of $9.0 < \log \rm{M_{\star} [M_{\odot}]} < 10.0$, optical longslit spectra have been obtained with the EFOSC2 spectrograph at the ESO New Technology Telescope (NTT) in La Silla, Chile; these are new observations, presented for the first time. These 27 galaxies were randomly selected from the \xgass\ parent sample. The only selection criterion was observability with the NTT.

In this work we combine the new NTT observations with data from \citetalias{Moran2012}. As can be seen in Fig.~\ref{fig:sfr_vs_mstar}, all low-stellar-mass galaxies with optical spectra from the new NTT observations (red diamonds) are star-forming galaxies, meaning that they are located on or nearby the star formation main sequence (SFMS). To obtain a uniform sample, only star-forming galaxies from the \citetalias{Moran2012} sample are included in this work. This is achieved by selecting only those galaxies that are within $\pm1.5 \sigma$ of the SFMS, as defined by \citet{Catinella2018} and described in more detail by \citet{Janowiecki2020}.
This selection criterion is applied at all stellar masses and a compromise between including as many low-mass galaxies as possible (20), as well as including only galaxies near a well defined SFMS. Combining the high stellar mass star-forming sample (86 galaxies) with the 20 new low-mass galaxies results in a sample of 106 galaxies. We note that for three high-mass galaxies, optical longslit data are available from both the MMT and APO. For these galaxies, we chose to use only the MMT spectra, as more reliable metallicity measurements are available at similar or larger galactocentric radii from the MMT data than  from the APO data.

\begin{figure*}
    \center
    \includegraphics[width=6.3in]{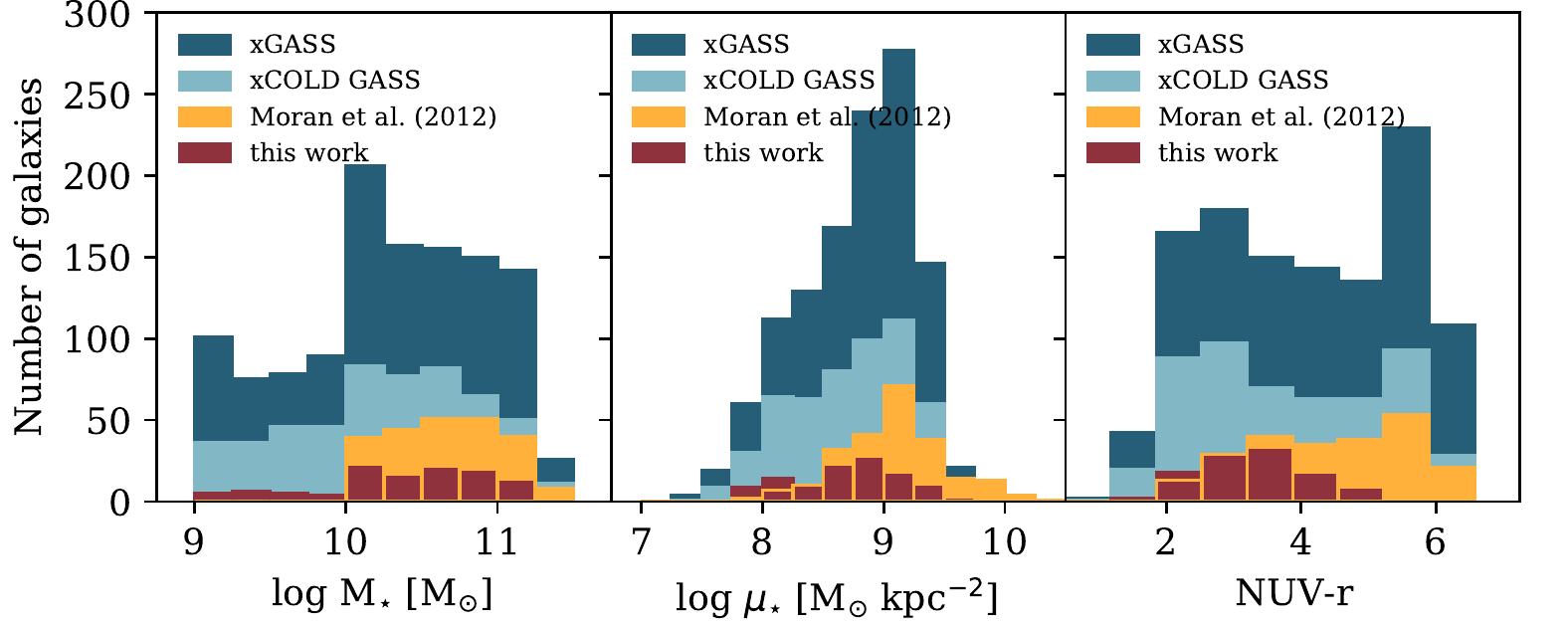}
    \caption{Distribution of stellar mass (\textbf{left panel}), stellar mass surface density (\textbf{middle panel}) and \nuvr\ colour (\textbf{right panel}) for the galaxy sample of \xgass\ (dark blue), \xcoldgass\ (light blue), \citetalias{Moran2012} (yellow) and this work (red). See Sect.~\ref{sec:grad_fit} for a description of the derivation of these quantities. }
    \label{fig:sample2}
\end{figure*}

Figure~\ref{fig:sample2} shows the distribution of stellar mass (left panel), stellar mass surface density (middle panel), and \nuvr\ colour (right panel). As can be seen here, this work expands the work by \citetalias{Moran2012} to lower stellar masses and includes more galaxies with low $\mu_\star$. As expected, due to the selection of SFMS galaxies, we include fewer high $\mu_\star$galaxies than in the \citetalias{Moran2012} analysis and no quiescent galaxies.

\begin{figure*}
    \center
    \includegraphics[width=6.3in]{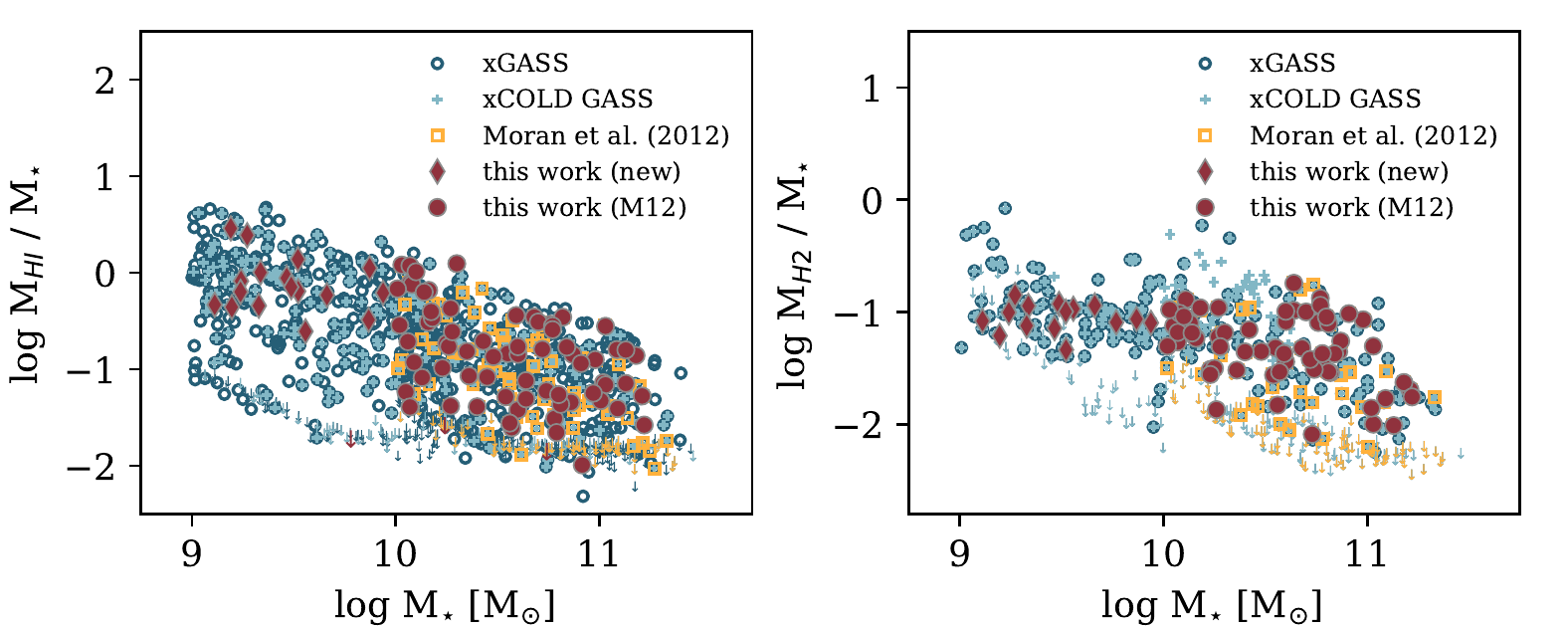}
    \caption{The \hi\ (\textbf{left panel}) and \htwo\ (\textbf{right panel}) gas-to-stellar mass ratio as a function of stellar mass. Symbols are as in Fig.~\ref{fig:sfr_vs_mstar}. If a galaxy has not been detected either in \hi\ or \co, its upper limit is shown as an arrow in the  respective panel.}
    \label{fig:fhi_vs_mstar}
\end{figure*}

The overlap between the samples with \hi, \co, and optical spectroscopic measurements is not perfect because of the timing of the various observing campaigns. Of the 106 galaxies in our sample of main-sequence galaxies with longslit optical spectra, 99 have \hi\ measurements from Arecibo and 76 have \co\ observations from the IRAM-30m telescope. When correlating measurements from optical longslit spectra with \hi\ or \co\ observations, those galaxies lacking information are excluded. In Fig.~\ref{fig:fhi_vs_mstar}, the \hi\ and \htwo\ gas mass-to-stellar mass ratios are shown as a function of stellar mass. The galaxies selected for this study have the typical gas fractions of main-sequence galaxies.

\section{NTT observations, data reduction and analysis}
\label{sec:analysis}

\subsection{Observations}

The optical longslit spectra of the 27 low-mass galaxies were obtained with the EFOSC2 spectrograph at the ESO New Technology Telescope (NTT) in La Silla, Chile in September~2012 and April~2013. The slit size was 1.5\,arcsec by 4\,arcmin and aligned along the major axis of each galaxy. In order to measure all the strong emission lines required for metallicity measurements, ranging in wavelength from [OII]372.7\,nm to H$\alpha$ at 656.3\,nm, two observations of every galaxy were needed, one for the bluer half of the spectrum (368.0\,nm to 550.0\,nm) and one for the redder half (535.0\,nm to 720.0\,nm). The overlap was used to check for consistency in flux calibration across the entire wavelength range. Individual science exposures were observed for 900\,s. The total exposure time varied according to the surface brightness of the galaxy, but amounted on average to 3600\,s per spectrum half. After including a binning factor of 2, the image size is 1024\,pixels by 1024\,pixels. The spectral resolution is 0.123\,nm and 0.113\,nm in the red and blue halves of the spectrum, respectively, which is approximately equivalent to a velocity resolution of 59 and 74\,km\,s$^{-1}$.

The raw spectra were first reduced with standard \textsc{iraf} procedures. Bias images, dome, and sky flats were taken at the beginning of each night. Bias images were subtracted from the science images, as is the dark current, which was estimated from the overscan regions of the science exposures. To obtain the overall flat field correction, both dome and sky flats were observed. First the spatial flattening was calculated from dome flats and the spectral flattening from sky flats. Then these two were multiplied to get a master flat field, which was applied to all the science images from a given night of observing. Since all exposures for one of the two spectral setups of each galaxy were obtained in one night, it was possible to stack individual frames at this point. During the stacking process cosmic rays were removed by an outlier rejection algorithm, and any remaining ones then removed manually.

Wavelength and flux calibration as well as straightening the image along the slit, were then performed on the stacked spectra. For wavelength calibration observations of a HeAr lamp in addition to the sky lines were used. The flux calibration was based on observation of multiple standard stars per night. The standard stars for the September 2012 run were EG\,21, Feige\,110 and LTT\,7987, whereas during the April 2013 run, the standard stars EG\,274, Feige\,56, LTT\,3218, LTT\,6248 were observed.

\subsection{Extraction of spatially resolved optical spectra}
\label{sec:red}

\begin{figure*}
    \center
    \includegraphics[width=160mm]{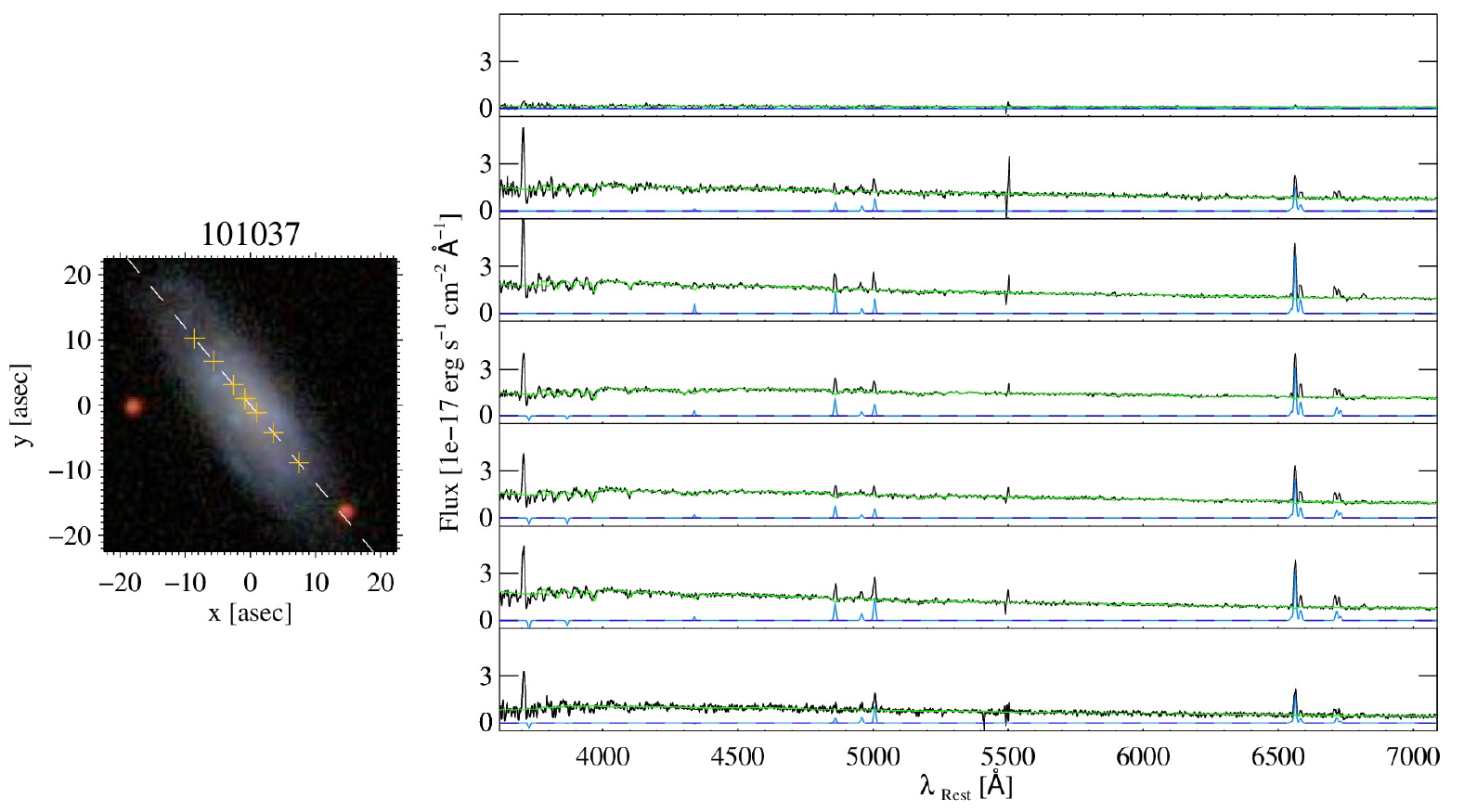}
    \caption{Example of spatially resolved spectrum. To the left a SDSS post stamp image of the galaxy (GASS 101037). North is up and east is left. The yellow crosses denote the light-weighted location of the individual spectra and the dashed line indicates the location of the slit. On the right, the individual spectra are plotted in black. Blue shows the fits to the emission lines and green the continuum fit. The spectrum plotted in the topmost panel is located farthest in the north-east of the galaxy. }
    \label{fig:example}
\end{figure*}

To extract spatially resolved, one-dimensional spectra from the reduced two-dimensional spectra, we used the same pipeline as \citetalias{Moran2012}. First, the two spectral halves were merged and possible flux mismatches were removed. Next, a rotation curve was fitted to absorption line measurements, thus it was possible to conduct all following steps in the rest frame. After that, the spectrum was spatially binned, starting in the centre and moving outwards. The size of the spatial bins was chosen such that a minimum continuum signal to noise ratio (S/N) of 5 was reached. This binning procedure resulted in one dimensional spectra covering a certain radial range at a certain radial position of the galaxy (see Fig.~\ref{fig:example}).

The stellar continuum of all the one dimensional spectra extracted in the previous step were then fitted with a superposition of simple stellar population models \citep{Bruzual2003}. The best-fitting continuum model was subtracted from the spectrum and the remaining emission lines were fitted with Gaussian functions \citep{Tremonti2004}. This process results in measurements of emission lines and stellar continuum at different galactocentric radii. An example for a typical spatially resolved spectrum with the fits to stellar continuum and emission lines is given in Fig.~\ref{fig:example}.

\subsection{Measuring radial metallicity profiles and gradients}
\label{sec:grad_fit}
With emission lines measured at different galactocentric radii, we are able to measure the gas-phase metallicity for each radial bin. After correcting the emission line fluxes for extinction \citepalias{Moran2012}, the gas-phase oxygen abundance was measured from ratios of the [OIII]$\lambda = 500.7$\,nm, H$\beta$, [NII]$\lambda =658.4$\,nm and the H$\alpha$ emission line flux following the prescription by \citet{Pettini2004}:
\begin{equation}
O3N2 = \log \left( \frac{[O III]\lambda 500.7{\rm \,nm} / H\beta }
{[N II] \lambda 658.3{\rm \,nm} / H\alpha}\right)
,\end{equation}
\begin{equation}
12 + \log (O/H) = 8.73 - 0.32 \times O3N2
.\end{equation}

There are many metallicity calibrators with different zero points, which were previously proposed and discussed in the literature (see e.\,g. \citealp{Kewley2008}). Given the available emission lines in our observations and that metallicity calibrators based on the $O3N2$ ratio are considered robust and are widely used in the literature, we focus on these types of metallicity estimators. In addition to the \citet{Pettini2004} prescription, we also use the \citet{Marino2013} $O3N2$ metallicity calibrator to make direct comparisons with\ other works (Sect.~\ref{sec:compare_manga}).
However, since the \citet{Marino2013} $O3N2$ calibrator tends to underestimate high metallicities \citep{Erroz-Ferrer2019}, we used the \citet{Pettini2004} calibrator for the majority of the {}analysis.
The same data products derived with the same pipeline are available for the \citetalias{Moran2012} galaxies. Thus, all the following analysis steps were performed for both the \citetalias{Moran2012} and the new data.

The analysis procedure described in the previous sections provided radial profiles of the gas-phase metallicity. The radial variation of these profiles was quantified by the slope of a linear fit to the metallicity as a function of radius. To account for the varying sizes of galaxies and their different distances, the galactocentric, light-weighted radius of each radial bin was normalised or converted to kpc. For normalisation, the SDSS 25\,mag\,arcsec$^{-2}$ isophotal radius ($\rm R_{25}$), Petrosian 90\,percent radius ($\rm r_{90}$) \citep{Petrosian1976} and the effective radius ($\rm r_{eff}$) were used in $r$ band. Using $u$ or $i$ band radii, that is, focusing on the young or old stellar population, does not affect the results. We therefore focus on radii measured in the $r$ band only. These radii have been published with SDSS
DR7 \citep{Abazajian2009} and are taken from the MPA-JHU catalogue\footnote{https://wwwmpa.mpa-garching.mpg.de/SDSS/DR7/}. 12 + log(O/H) was then fitted as a linear function of $r / r_{norm}$:
\begin{equation}
{\rm 12 + log (O/H)} = (\Delta {\rm 12 + log (O/H)}) \times (r / r_{norm}) + a
,\end{equation}
using the \textsc{scipy}\footnote{http://www.scipy.org/} \citep{Virtanen2020} function \texttt{curve\_fit}, which utilises the least squares-based Levenberg--Marquardt algorithm. As the first derivative of this function, $\Delta {\rm 12 + log (O/H)}$ was then defined as the radial gradient of the metallicity.

In order to improve the reliability of the results, some radial bins were excluded from the analysis. We rejected any bin where AGN emission was significantly contributing to the ionisation. Those were identified by using strong emission line ratios to place the measurement in the [NII]/H$\alpha$ versus [OIII]/H$\beta$ Baldwin -- Phillips -- Terlevich diagnostic plot (BPT, \citealp{Baldwin1981}). Any radial bin with measured line ratios falling above the empirical threshold of \citet{Kauffmann2003} was excluded from the analysis. Furthermore, we required a signal-to-noise (S/N) detection of 3 for the four emission lines [O\,III]$\lambda$\,500.7\,nm, [N\,II]$\lambda$\,658.3\,nm, H$\alpha$ and H$\beta$.

In previous studies (e.\,g. \citealp{Sanchez2014}, \citealp{Ho2015} and \citealp{Sanchez-Menguiano2016}), metallicity gradients were often calculated after discarding measurements within a certain galactocentric radius to avoid contamination by any active nucleus. While we disregard radial bins with AGN-like emission in general, we calculated metallicity gradients twice to allow for fair comparisons with these results: once using all reliable data points and once only considering measurements coming from the region outside of 0.5 times the effective $r$ band radius $\rm r_{eff,r}$. When requiring a minimum of three radial bins for gradient measurement, we measured gradients from the entire radial profile for 88 galaxies, and gradients from the radial profile between 0.5 and 2\,$\rm r_{eff,r}$ for 74 galaxies. Of these galaxies, 75 and 66 galaxies have stellar masses higher
than $10^{10}$\,M$_\odot$, respectively.

In the following sections, we investigate correlations between metallicity gradients and the stellar mass (\mass), stellar mass surface density (\must, as a proxy for morphology), the concentration index ($\rm c = r_{90} / r_{50}$, proxy for bulge to total mass ratio), specific star formation rate ($\rm sSFR = SFR / M_\star$), \nuvr\ colour, atomic and molecular gas mass fraction (gas mass fractions are defined as $\rm \log\,f_{\rm Gas} = \log\,M_{\rm Gas}/M_{\star}$), and the deficiency factor for atomic and molecular gas. Details on the derivation of these quantities are given in \citet{Saintonge2017} and {}
\citet{Catinella2018}. The deficiency factor is the difference between an estimate of the gas mass fraction from a scaling relation and the actually measured gas fraction. Here we use the best and tightest scaling relations available from the \xgass\ and \xcoldgass\ analysis. For \hi, this is the relation between \fhi\ and \nuvr\ colour, more specifically the binned medians from Table~1 in \citet{Catinella2018}. For \htwo, we used the scaling relation between \fhtwo\ and {}
log\,sSFR based on the "Binning" values for the entire \xcoldgass\ sample in Table~6 of \citet{Saintonge2017}. In both cases we extrapolated between the bins to get an expected gas mass fraction. The deficiency factor is then:
\begin{equation}
\rm def =  \log f_{expected} - \log f_{measured}
,\end{equation}
with f the gas mass to stellar mass fraction. Therefore a negative deficiency factor indicates that a galaxy is more gas-rich than the average galaxy sharing similar \nuvr\ colour or sSFR.

\section{Results: Metallicity gradients}
\label{sec:results1}

In this section, we present the results of a detailed analysis of the correlation between metallicity gradients and global galaxy properties, in particular, star formation activity and gas content.
We start by analysing and establishing which correlations between metallicity gradient and global galaxy property are of interest in our sample. In this process, we consider both gradients measured from profiles with all radial data points and gradients measured from profiles without data points inside of 0.5\,$\rm r_{eff,r}$. Then we compare our results to the literature and discuss potential differences.

\subsection{Investigating correlations between gradients and global galaxy properties}
\label{sec:correlations}
In order to test for the presence and strength of correlations between gradients and global galaxy properties, we applied multiple methods. Firstly, we calculated Spearman correlation coefficients between metallicity gradients and each global galaxy property.
    For those global galaxy properties that have correlation coefficients that significantly depart from zero, we obtained (semi-) partial Spearman correlation coefficients. These are correlation coefficients that take into account the intercorrelation between various global galaxy properties.

    Through a backward elimination based on the results of a multiple linear regression, we searched for the most important global galaxy property to determine the metallicity gradients.
    We trained a random forest model to predict metallicity gradients from those global galaxy properties that have correlation coefficients significantly different from zero.0. Then we asked the model which feature was most important in predicting the metallicity gradient.

\subsubsection{Spearman correlation coefficients}
We measured correlation coefficients between metallicity gradients and global properties as Spearman R values and calculated their errors through bootstrapping: for a sample of $n$ measurements, $0.8\times n$ measurements were randomly drawn from the sample (with replacement) and their correlation coefficient was measured. This process was repeated $0.8\times n$ times. The error of the correlation coefficient was then set to the standard deviation of the sample of $0.8\times n$ correlation coefficients. The median bootstrapping error of all measured correlation coefficients amounts to $0.1$. Therefore, for a relation to be further considered and analysed, an absolute correlation coefficient ${\rm |R|} > 0.3$ was required ($3\,\sigma$ different from 0). The absolute correlation coefficient ${\rm |R|}$ can have values between 0 and 1, where numbers closer to 1 present tighter and stronger correlations (or an anti-correlation if R is negative).

\begin{figure*}
    \center
    \includegraphics[width=6.3in]{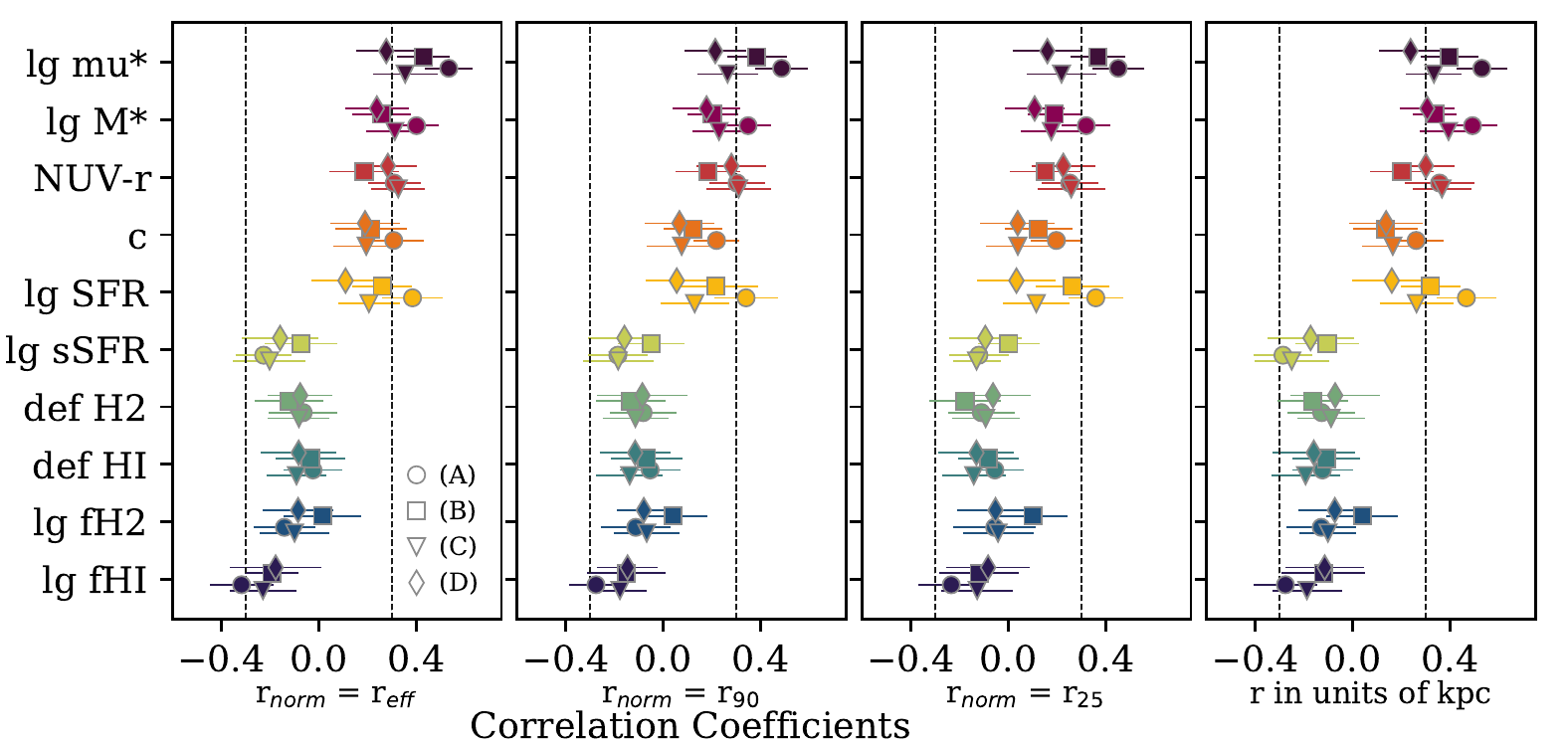}
    \caption{Correlation coefficients calculated for relations between metallicity gradients and global galaxy properties (colour code) with their error bars. The panels show, from left to right, correlation coefficients for metallicity gradients in units of dex\,$\rm r_{eff,r}^{-1}$, dex\,$\rm r_{90,r}^{-1}$, dex\,$\rm R_{25,r}^{-1}$, dex\,$\rm kpc^{-1}$. We note that all radii have been measured in the $r$ band, similar results can be obtained for radii in the $u$ and $i$ band. For each global property four different correlation coefficients are presented, and are marked by the shape of the data point. The correlation coefficient were measured with gradients based on: \textbf{(A)} the entire radial profile for all galaxies in the sample; \textbf{(B)} the entire radial profile for massive galaxies; \textbf{(C)} the radial profile outside of 0.5\,$\rm r_{eff,r}$
    {}for all galaxies in the sample; \textbf{(D)} the radial profile outside of 0.5\,$\rm r_{eff,r}$ for massive galaxies. Black dashed lines mark correlation coefficients of -0.3 and 0.3. }
 \label{fig:corr_coef}
\end{figure*}

Figure~\ref{fig:corr_coef} shows Spearman R correlation coefficients for metallicity gradients (measured with and without the central 0.5\,$\rm r_{eff,r}$) and global properties. Each data point represents the correlation coefficient between one gradient (e.\,g. the gas-phase metallicity gradient in units of dex\,$\rm r_{90, r}^{-1}$) and one global galaxy property (e.\,g. stellar mass). The shape of the data points indicates the dataset for which the correlation coefficient was measured. We note that diamonds and triangles (correlations with gradients measured from radial profiles outside of 0.5\,$\rm r_{eff,r}$) generally indicate less pronounced correlations than squares and circles (gradients measured from the full radial profile).

The median maximal radius, at which we can reliably measure metallicities, is 2.5\,$\rm r_{eff,r}$ for massive galaxies, and 1.5\,$\rm r_{eff,r}$ for low-mass galaxies. This means that when measuring metallicity gradients from the radial range between 0.5 and 2.0\,$\rm r_{eff,r}$, we mostly exclude low-mass galaxies because they do not have enough (i.\,e. three or more) radial metallicity measurements between 0.5 and 2.0\,$\rm r_{eff,r}$ to fit a gradient. In order to understand the effect of looking at massive star-forming galaxies only, we also measured correlation coefficients for massive galaxies (\mass\ > 10., cases (B) and (D) in Fig.~\ref{fig:corr_coef}). These correlation coefficients for massive galaxies are usually closer to 0 than the correlation coefficients for the entire sample. This points to a scenario in which the observed trends are amplified by low-mass galaxies.

For now, we are focusing on all relations for which the correlation coefficient is larger than $0.3$ or smaller than $-0.3$ and which are thus three times larger than the median error or in other words significantly different from zero. For gradients measured from the entire radial profile and when considering all galaxies for which a gradient could be measured (circles in Fig.~\ref{fig:corr_coef}), we find that the correlation coefficient is significantly different from zero for relations between metallicity gradient in units of dex\,$\rm r_{eff, r}^{-1}$ and \must, \mass, \fhi, \nuvr\ colour, concentration index c, and log~SFR.
The correlation coefficients for other metallicity gradients, namely, in units of dex\,$\rm r_{90, r}^{-1}$, dex\,$\rm R_{25, r}^{-1}$ and dex\,$\rm kpc^{-1}$ show similar trends, which are generally weaker, however.

Overall, we have a wide radial coverage all the way out to 2\,$\rm r_{eff,r}$ for massive galaxies but the more central measurements of metallicity often cannot be used for metallicity gradient measurement as their location on the BPT diagnostic plot indicates that the emission lines are excited by AGN-like emission rather than the emission of star-forming regions. Hence, restricting a study only to consider massive galaxies already goes in the direction of analysing correlations for metallicity gradients measured only from data points outside of 0.5\,$\rm r_{eff,r}$.

This analysis points towards a scenario in which the correlations we see between a metallicity gradient and global galaxy properties are affected by the radial location of the radial metallicity measurements, which are used for the metallicity gradient measurement. To further analyse this assumption, we only look at metallicity gradients, which were measured on radial profiles without the data in the inner 0.5\,$\rm r_{eff,r}$. These data are shown as triangles and diamonds in Fig.~\ref{fig:corr_coef}. In this case, we find no correlation remaining with absolute correlation coefficients ${\rm |R|} > 0.3$, except for the ones between metallicity gradients in units of dex\,$\rm r_{eff}^{-1}$ and \must for all galaxies and between metallicity gradients in units of dex\,kpc$^{-1}$ and \must, \mass and \nuvr\ colour.

In the following sections, we delve deeper into a statistical analysis of these correlations. We are especially interested in understanding which correlation is primary and which are secondary effects. We only focus on metallicity gradients in units of dex\,$\rm r_{eff}^{-1}$, because these are widely used in the literature, they yield the tightest correlations and the other metallicity gradients behave similarly.

\subsubsection{(semi-)Partial Spearman correlation coefficients}

To further investigate the correlations found in the last section, we also considered (semi-)partial Spearman correlation coefficients, which provide the same information as the correlation coefficients introduced above, but they allow us to fold in inter-correlations between the global galaxy properties. This is achieved by holding one measurement constant while looking at the correlation coefficient of two other measurements. In practice, this means computing correlation coefficients between residuals. Since \must\ and \fhi\ are correlated \citep{Catinella2018, Brown2015} and both appear to be correlated with the metallicity gradient, we must control for \must\ to evaluate the strength of the 'remaining' correlation between \fhi\ and the metallicity gradient. We performed this analysis for the ensemble of \must, \mass, \fhi, \nuvr\ colour, concentration index, log~sSFR and log~SFR and their correlation to the metallicity gradients in units of dex\,$\rm r_{eff,r}^{-1}$. The global galaxy properties are presented here as all those properties that showed any significant correlation in the previous section. Furthermore, we add the concentration index, c, as it was found to be correlated with the metallicity gradient by \citetalias{Moran2012}. When controlling for these properties using the \texttt{partial\_corr} implementation in the \textsc{pingouin} package\footnote{https://pingouin-stats.org}, we only find a strong correlation between \must\ and metallicity gradients. This result holds for both, metallicity gradients measured from the entire radial profile and metallicity gradients measured without data in the central 0.5\,$\rm r_{eff,r}$. For all other galaxy properties, the (semi-)partial correlation coefficients are significantly smaller than 0.3 and the majority of their 95\,percent confidence intervals includes 0, that is, both a correlation and an anti-correlation would be possible.

\subsubsection{Backward elimination}

A second method to find the one variable in a set of features that contributes most (or most optimally) to predicting a result is backward elimination. We used backward elimination in the following way: we fitted a general ordinary least squares multiple linear regression, such that the metallicity gradient in units of dex\,$\rm r_{eff,r}^{-1}$ was the dependent variable and was described as a linear combination of \must, \mass, \fhi, \nuvr\ colour, concentration index, log~sSFR, and log~SFR (the same selection of global galaxy properties as in the previous section) plus a constant. Then we took a look at the statistics of this model (as provided by the \texttt{OLS} module of the \textsc{statsmodel} package, \citealp{Seabold2010}). These statistics provide among other measures a p-value for the T-statistics of the fit. If this p-value is large for one of the variables, then this variable is likely not useful in the fit. In the context of our backward elimination, we used the p-value in the following, iterative way: after the first multiple linear regression, we eliminated the variable with the largest p-value, then ran the fit again without the eliminated variable. We continued to eliminate variables and run the fit until the p-values of all remaining variables were below 0.05, which is a p-value commonly judged as statistically significant.

Applying this procedure to our data returned the following results: for metallicity gradients measured on the entire radial profile, the backward elimination leaves \must\ and \fhi, with the importance (i.\,e. the coefficient) of \must\ twice the one of \fhi. For metallicity gradients measured without data points at radii smaller than 0.5\,$\rm r_{eff,r}$, only \must\ remains with a p-value smaller than 0.05.

A caveat of this method is the underlying assumption of linear relations between metallicity gradients and global galaxy properties, which is not necessarily the case. We improve on this caveat in the next section by using a random forest regression.

\subsubsection{Random Forest model}
A random forest \citep{Ho1995} is a non-parametric, supervised machine learning technique made up of a set of decision trees. The result of a random forest is the mean of all decision trees in the forest and is thus generally more robust than a single decision tree. The aim of this analysis is to train a random forest to predict the metallicity gradient in units of dex\,r$_{eff,r}^{-1}$ from \must, \mass, \fhi, \nuvr\ colour, concentration index c, log~sSFR, and log~SFR. Once the model is fully trained, we can ask what is the relative contribution of the different features to predicting the metallicity gradient.

We used the implementation provided by \textsc{scikit-learn} \citep{Pedregosa2011} and trained the model to optimise the mean squared error. We allow for a maximum of 20 leaf nodes in the decision trees, use 160 decision trees and leave the default settings for all other parameters. As mentioned above, not all galaxies in our sample are equipped with all measurements and sometimes metallicity gradients could not be measured due to too few radial bins with sufficient emission line detections. Thus the samples to work with contain 81 (67) galaxies for metallicity gradients measured on the entire radial profile (only from data points outside of 0.5\,$\rm r_{eff,r}$). Of each sample, we use 80\,percent of the galaxies for training purposes and 20\,percent to test the resulting model. Tests after the training showed that metallicity gradients can be predicted with a mean absolute error of 0.06 (0.07)\,dex\,$\rm r_{eff,r}^{-1}$ for metallicity gradients measured on the entire radial profile (only from data points outside of 0.5\,$\rm r_{eff,r}$), and the most relevant features for the prediction are \fhi\ and \must\ in both cases.

\subsection{Correlation between the metallicity gradient, stellar mass surface density, stellar mass, and f$_{HI}$}
\label{sec:detailed_corrs}
\begin{figure*}
    \center
    \includegraphics[width=6.3in]{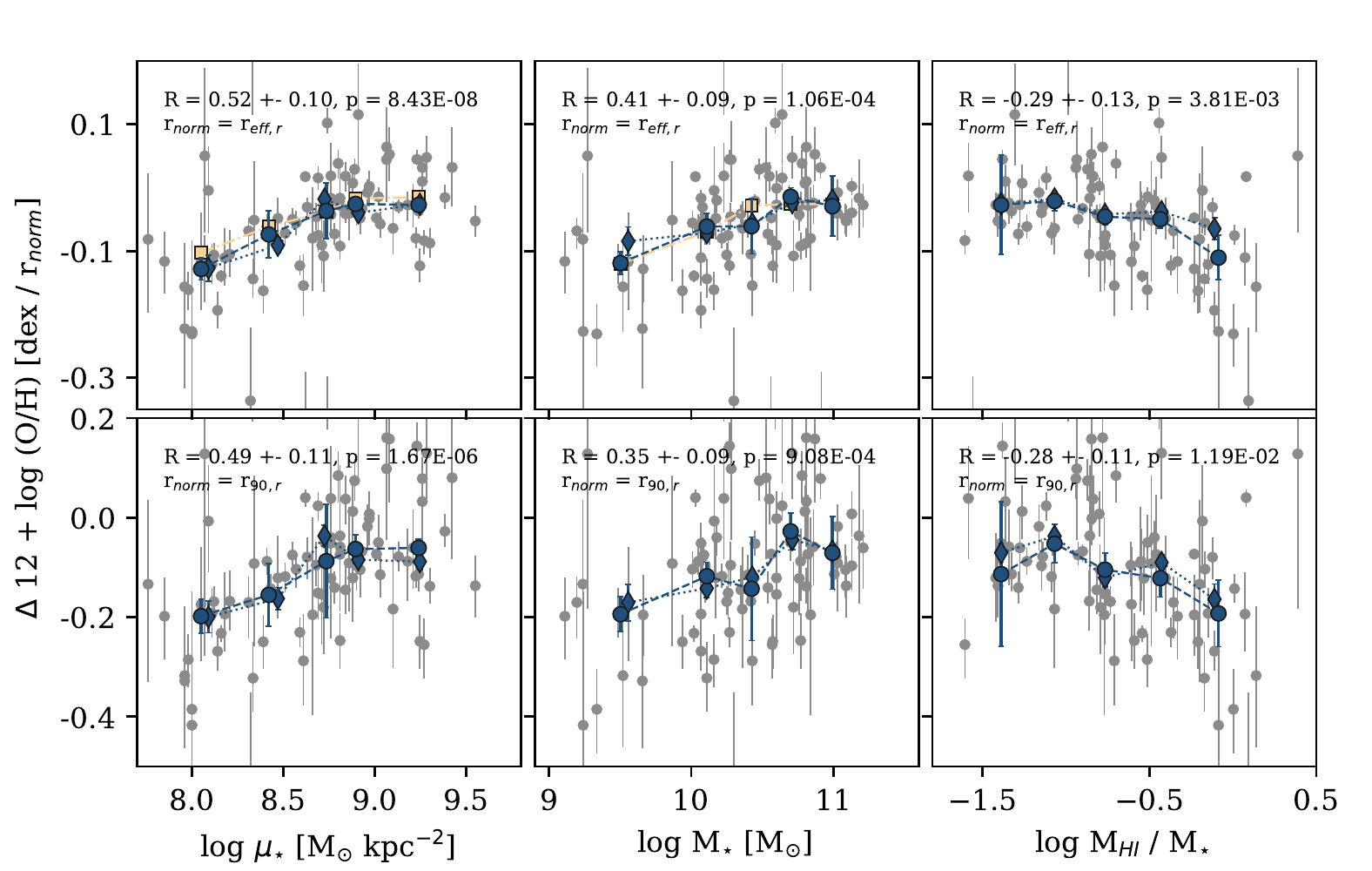}
    \caption{Metallicity gradients shown as a function of stellar mass surface density (left), stellar mass (middle), and atomic gas mass fraction (right). Coloured circles connected with dashed lines present the median gradient of individual profiles. The error bars in y-direction give the bootstrapping error. Coloured diamonds connected by dotted lines show the metallicity gradient of the stacked profiles. The grey data points in the background represent the metallicity gradients measured per galaxy. From top to bottom, the radius used when fitting the gradient is normalised by $\rm r_{eff}$ and r$_{90}$. The text in the upper part of each panel provides the Spearman R correlation coefficient for all galaxies, its bootstrapping error and the corresponding p-value. The coefficient is calculated between the data points for individual galaxies rather than the binned values. Where available, the yellow squares show an expected average metallicity gradient; see the text for more details.}
    \label{fig:grad_with}
\end{figure*}

\begin{figure*}
    \center
    \includegraphics[width=6.3in]{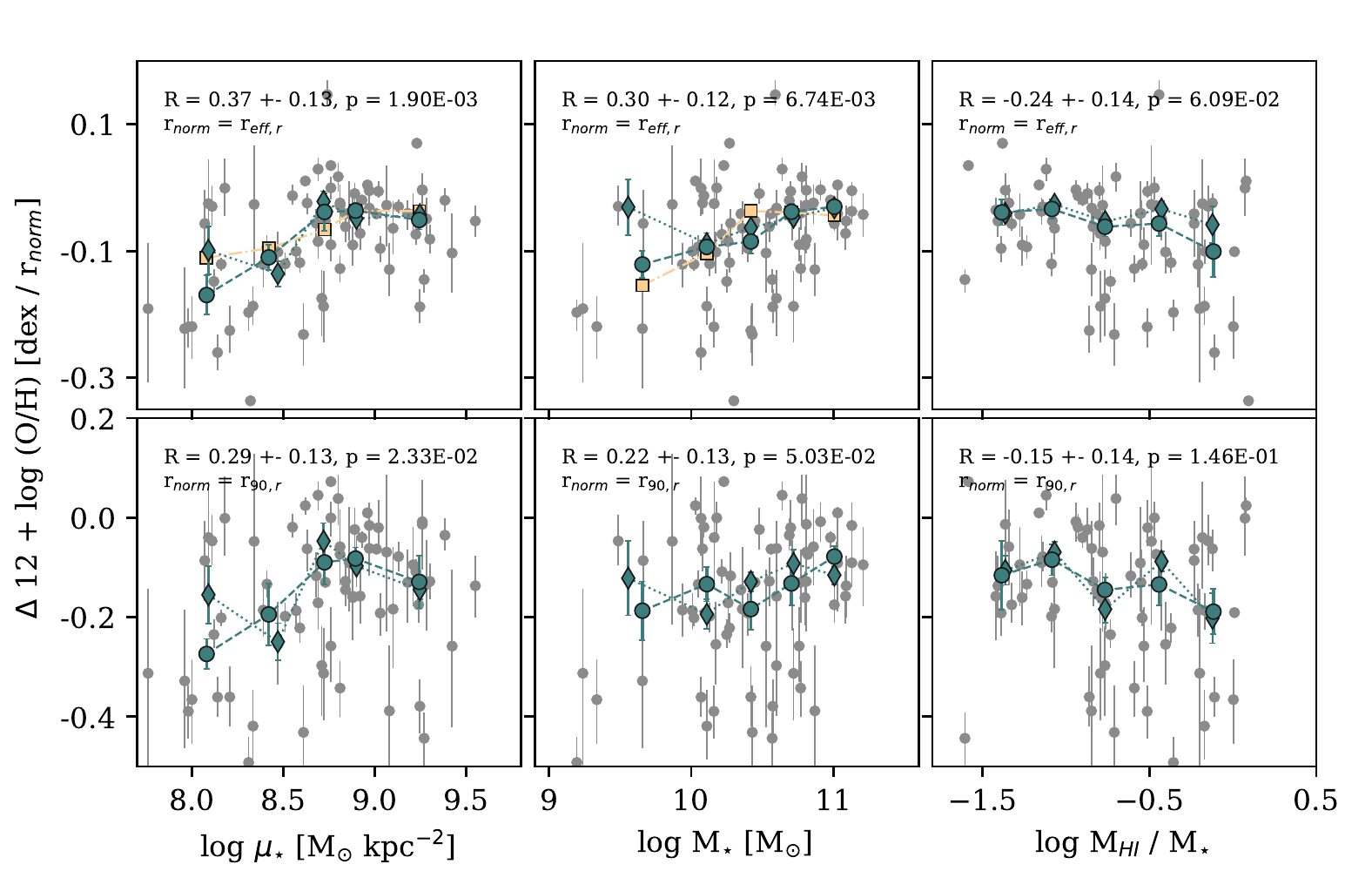}
    \caption{Metallicity gradients shown as a function of stellar mass surface
density (left), stellar mass (middle), and atomic gas mass fraction (right), as in Fig. 6, but here the gradients have been measured from profiles without the central 0.5\,$\rm r_{eff}$. We note how most correlations from Fig.~\ref{fig:grad_with} turn into scatter plots and gradients of stacked profiles deviate from average gradients of individual profiles.}
    \label{fig:grad_wo}
\end{figure*}

With the statistical tests shown in the last sections, a scenario is building in which \must\ is the main driver of metallicity gradients and \fhi\ may play a secondary rule. We present further investigations of these correlations and the correlation with stellar mass, because this one is best studied in the literature. We examined them by dividing the sample into five bins of the global property, such that each bin contained about the same number of galaxies. The average radial gradient in each bin was then estimated in two different ways: (i) a gradient measured from the average stacked metallicity profile based on the radial metallicity measurements of all galaxies in the bin, to be called a 'gradient of a stacked profile' and (ii) the median of all gradients measured from individual galaxy metallicity profiles, to be called 'the average gradient of individual profiles'. To obtain the gradient of the stacked profile, we take all radial data points of all galaxies within one bin and fit a line to all radial metallicity measurements that fulfil our quality criteria. Radial data points for all galaxies are weighted equally and radii are normalised or measured in kpc. The resulting correlations are shown in Fig.~\ref{fig:grad_with} (metallicity gradients measured on the entire radial metallicity profile) and Fig.~\ref{fig:grad_wo} (gradient measured from the radial metallicity profile outside of 0.5\,$\rm r_{eff,r}$).

The strongest correlation between global galaxy properties and metallicity gradients in our samples, in all cases, is observed with \must: generally, galaxies with lower $\mu_\star$ have steeper metallicity gradients than high $\mu_\star$ galaxies, regardless of the radius normalisation or unit. When comparing these correlations to the ones obtained when measuring the metallicity gradients from radial profiles without the inner 0.5\,$\rm r_{eff,r}$ (Fig.~\ref{fig:grad_wo}, left column), an increase in scatter and overall flattening of the trends can be seen. This is also reflected in the correlation coefficients, which generally decrease from around 0.5 to around 0.2. A notable exception is the correlation coefficient between \must\ and the metallicity gradient in units of
{}dex\,$\rm r_{eff,r}^{-1}$ (top row, left panel in Fig~\ref{fig:grad_wo}), which is the only correlation coefficient that is larger than 0.3 in the analysis of gradients measured from profiles without the central 0.5\,$\rm r_{eff,r}$.

The relation between M$_\star$ and metallicity gradients is weaker than the one between \must\ and metallicity gradients. For gradients measured on the entire radial metallicity profile, we find correlation coefficients larger than 0.3 for all normalising radii (middle column, Fig.~\ref{fig:grad_with}). The scatter is larger for gradients in units of dex\,R$_{25, r}^{-1}$ or dex\,r$_{90, r}^{-1}$. In particular, for gradients in units of dex\,$\rm r_{eff,r}^{-1}$, it can be seen that the trend between stellar mass and metallicity gradients measured from the entire radial profile is driven by low-mass galaxies. Once we remove the inner 0.5\,$\rm r_{eff,r}$ from the metallicity profile for gradient measurement
(middle column, Fig.~\ref{fig:grad_wo}), the resulting gradients do not correlate with stellar mass any longer: the scatter of gradients from individual profiles increases and both approaches to measuring binned, average gradients either produce a flat relation or scatter throughout the parameter space.

A further test whether stellar mass or $\mu_\star$ is the more defining factor in determining the metallicity gradient was inspired by Fig.~5 from \citet{Belfiore2017a} but the results are inconclusive. The yellow symbols and lines in Fig.~\ref{fig:grad_with} and \ref{fig:grad_wo} show the expected gradients. For the relation between $\mu_\star$ and metallicity gradient, we calculate the average stellar mass in each bin of $\mu_\star$ and then extrapolate between the nearest M$_\star$ bins to get the expected metallicity gradient and vice versa. As the expected average metallicity gradients match with the measured average metallicity gradients, this test does not provide additional insights.

The third global property that we are considering here is the \hi\ gas mass fraction. The strongest correlations are measured between gradients in units of dex\,$\rm r_{eff,r}^{-1}$. While correlation coefficients for gradients with normalising radii $\rm r_{eff}$ are at least $3\,\sigma$ different from zero, we find that gradients with normalising radii r$_{90}$, R$_{25}$ and in units of dex\,kpc$^{-1}$ are only 2 to 3\,$\sigma$ different from zero. Once moving from gradients measured on the entire radial metallicity profiles to gradients measured without the central 0.5\,$\rm r_{eff,r}$, we find a similar behaviour as observed for the correlations between stellar mass and metallicity gradients: the scatter of the individual gradients increases, correlations of binned values either flatten or their scatter increases as well.

The sample selection and the resulting distribution of global galaxy properties can also affect correlations between metallicity gradients and global galaxy properties. As can be seen, in Fig.~\ref{fig:sample2}, for example the stellar mass range $9.0 \le \log {\rm M_{\star} [M_{\odot}]} \le 10.0$, is more sparsely sampled. Hence, individual extreme and low-mass galaxies might significantly drive correlations. To show that this is not the case, we show the individual metallicity gradients in Figs.~\ref{fig:grad_with} and \ref{fig:grad_wo} as small grey symbols.

\subsection{Comparison to MaNGA}
\label{sec:compare_manga}
\begin{figure*}
    \center
    \includegraphics[width=6.3in]{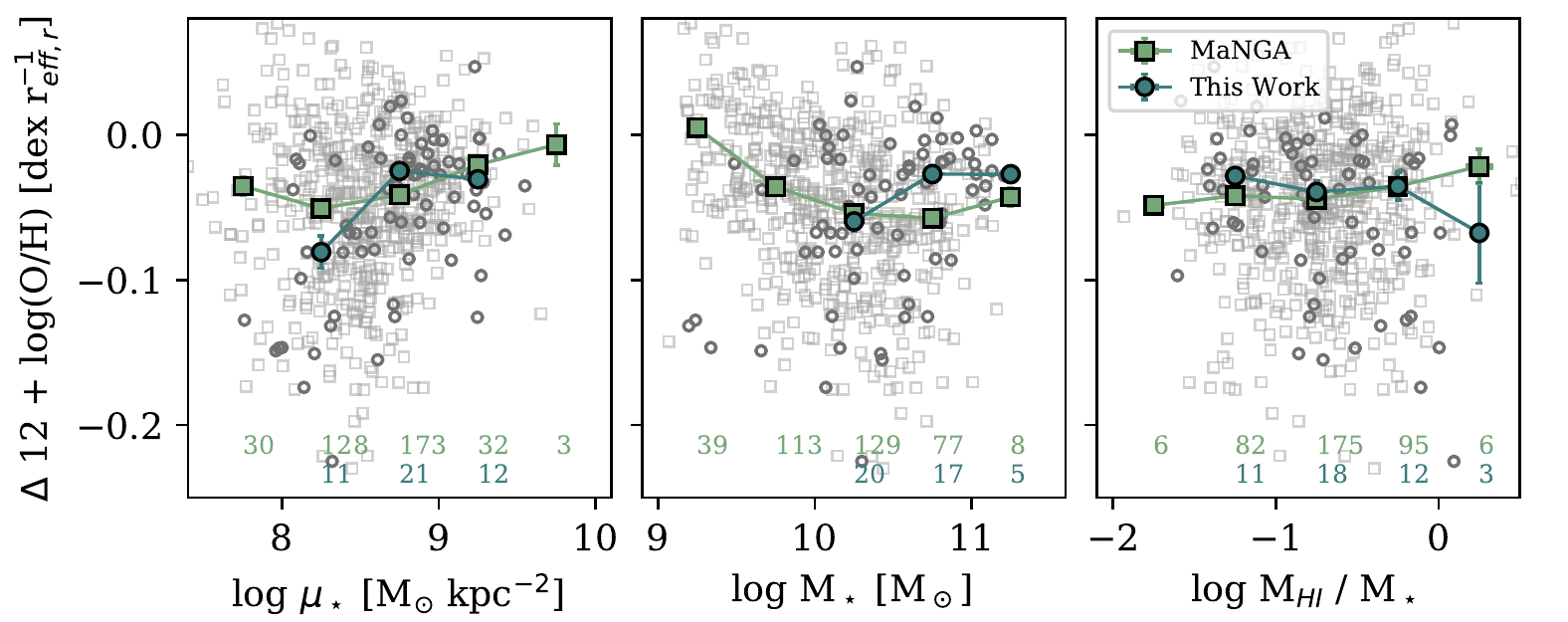}
    \caption{Comparison with MaNGA: all panels show the metallicity gradient as a function of global properties: \must, log~M$_\star,$ and \fhi\ (from left to right). Trimmed mean gradients agree with median gradients within the standard deviation (error bars), thus we only show the median gradients. Grey, open squares in the background show individual MaNGA metallicity gradients and green, filled squares show MaNGA median gradients. Grey, open circles in the background, and teal filled circles show the data from this work for the case that metallicity gradients were measured from profiles without radial data points inside of 0.5\,$\rm r_{eff}$. The bins within which median gradients were measured, were set to be equidistant in order to mitigate any effects of different distributions of the global properties. We note that we use the M13 $O3N2$ calibration in this figure.}
    \label{fig:manga}
\end{figure*}

As indicated in the introduction, the rise of large IFU surveys has provided large samples of local star-forming galaxies for which a metallicity gradient can be measured. In Fig.~\ref{fig:manga}, the results of this paper are compared to data from MaNGA.

The MaNGA data for this comparison comes from the data release 15 and includes two value added catalogues: MaNGA Pipe3D value added catalog: Spatially resolved and integrated properties of galaxies for DR15 \citep{Sanchez2018a,Sanchez2016,Sanchez2016a} and HI-MaNGA Data Release 1 \citep{Masters2019}. The Pipe3D catalogue includes gas-phase metallicity gradients measured in units of dex\,$\rm r_{eff,r}^{-1}$, the local gas-phase metallicity measured at the effective radius, the total stellar mass, and global star formation rates (from H$\alpha$ emission lines). We note that the metallicity estimator used in this MaNGA catalogue is the $O3N2$ estimator by \citet{Marino2013}. While both the
\citet{Marino2013} (M13) and the \citet{Pettini2004} (PP04, used in this paper) metallicity prescription are based on the $O3N2$ line ratio, their normalisation is slightly different. For a consistent comparison, we use the M13 $O3N2$ method in every figure that includes MaNGA data (and note in the figure caption when this is the case).

We combine the Pipe3D catalogue with the SDSS DR7 MPA-JHU catalogue\footnote{https://wwwmpa.mpa-garching.mpg.de/SDSS/DR7/} to obtain 50\,percent Petrosian radii for all galaxies. Together with the stellar mass as given by the MaNGA team, we are thus able to calculate $\mu_{\star}$ (see Sect.~\ref{sec:survey}). In addition, we use the \hi\ masses and upper mass limits as provided by \citet{Masters2019}.

If we are only selecting those galaxies that have measured metallicity gradients, as well as at least an upper limit for the \hi\ mass, a match in the MPA-JHU catalogue (for $\mu_{\star}$ measurements), and which are within $\pm1.5 \sigma$ of the \citet{Catinella2018} SFMS, we get a sample of 544\,galaxies from the MaNGA data sets. For simplicity, we treat \hi\ mass upper limits as their true value. In Fig.~\ref{fig:manga}, we show the different correlations between metallicity gradient and global galaxy properties (from left to right: stellar mass surface density, stellar mass and \hi\ mass fraction). As the distribution of our and the MaNGA galaxies in these properties are different, we fix the widths (0.5\,dex) and centres of the bins of global galaxy properties. This approach simulates a flat distribution in stellar mass surface density, stellar mass and \hi\ mass fraction for both the MaNGA and our sample. In each of these bins, we only use galaxies within a certain metallicity gradient percentile range (within the 16-84 percentile range) to remove extreme outliers, and refer to the corresponding quantities, for example, the mean gradient, as 'trimmed'. For each bin of galaxies, we thus calculate a trimmed mean gradient, standard deviation, error of the trimmed mean, and a median gradient.
In Fig.~\ref{fig:manga}, the median gradients are shown at the centre of the bin. The numbers in the bottom indicate how many galaxies contributed to the median. As trimmed means and medians agree within the standard deviation, we only show the median gradients.

We also use a second, more stringent percentile range (40-60) as a check of the initial, broader range. Although the numbers of galaxies drop significantly for the 40-60 percentile cut, the mean and median gradients trimmed in this way agree well between the 40-60 percentile cut method and the 16-84 percentile cut method. This means that the mild cut is sufficient to estimate a robust mean. The resulting trends agree with observations in Fig.\ref{fig:grad_wo}. In the MaNGA data, correlations between the metallicity gradients and these global properties can also be seen. In most cases, our metallicity gradients, which were measured without data points at radii smaller than 0.5\,$\rm r_{eff,r}$ agree with with the MaNGA data, except for low-mass, low $\mu_\star$ systems. It is interesting to note that generally the trend between metallicity gradients and \must\ (galaxies with lower $\mu_\star$ have steeper declining metallicity gradients), is also seen in the MaNGA data, except for the lowest \must\ bin. Considering the distribution and location of the individual gradient measurements (grey symbols in Fig.~\ref{fig:manga}), this shows that overall our sample covers a similar parameter space as the MaNGA measurements. There are three low-mass galaxies with steeper gradients than most MaNGA galaxies at the same stellar mass. We have placed both the MaNGA galaxies and our sample on various scaling relations to understand whether these galaxies are special with respect to star formation or gas content. However, this is not the case
here (see also App.~\ref{app:compare_manga} and Fig.~\ref{fig:manga_sample}).

\citet{Belfiore2017a} investigated the correlation between metallicity gradients and stellar mass based on MaNGA data and found that the steepest (most negative) metallicity gradients are measured for galaxies with stellar masses around $10^{10}$ to $10^{10.5}$\,M$_{\odot}$. Galaxies at lower and higher stellar masses have flatter radial metallicity profiles. This trend can also be seen in the middle column of Fig.~\ref{fig:manga}. This is particularly interesting as the \citet{Belfiore2017a} results are not based on the Sanchez et al. value-added catalogue that we use on the present work.

\subsection{Radial variation of the gas-phase metallicity}
\label{sec:radial_profs}
\begin{figure*}
    \center
    \includegraphics[width=6in]{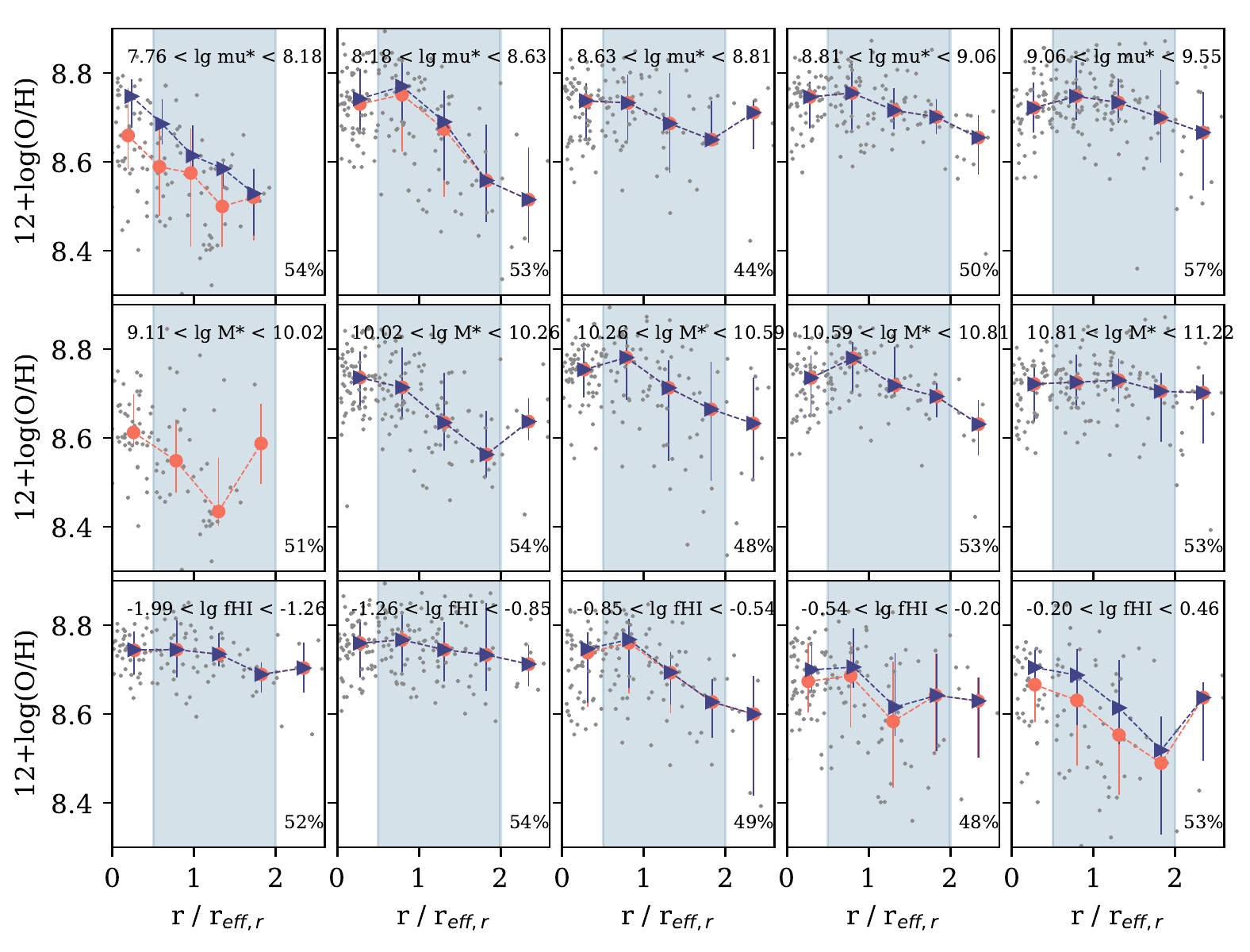}
    \caption{Median metallicity profiles in bins of different global galaxy properties. Each row of plots corresponds to one global galaxy property, from top to bottom: $\mu_\star$, M$_\star,$ and f$_{HI}$. Each panel in a row shows average radial metallicity profiles of all galaxies within the bin of the global galaxy property, with the range given at the top of the panel. The dark shaded region corresponds to $\rm 0.5 \leq r_{eff,r} \leq 2.0$, which is the radial region within which MaNGA computes metallicity gradients. Circles show the median metallicity profiles of all galaxies and triangles the median metallicity profile of massive galaxies only (M$_\star > 10^{10}$\,M$_\odot$). These profiles have been computed in radial bins, all of which have the same radial width. The small grey dots show the individual radial metallicity data points. The number in the bottom right corner of each panel indicates the percentage of radial data points located within the range $0.5 \leq \rm r_{eff,r} \leq 2.0$. }
    \label{fig:metal_profile}
\end{figure*}

One reason why the correlations between metallicity gradient and M$_\star$, f$_{HI}$ (and $\mu_\star$) change depending on how the metallicity gradient is measured and which sub-sample is considered can be seen when taking the average shape of the radial metallicity profiles into account. To do so, a median radial metallicity profile was calculated for each bin of $\mu_\star$, M$_\star$ and f$_{HI}$. One profile is the running median of all data points meeting the criteria to be included in the gradient fit of all galaxies within one bin of global galaxy property. These profiles are shown in Fig.~\ref{fig:metal_profile}. As we can see, it is not only the metallicity gradient, but also the shape and y-axis intercept of the median metallicity profile that vary with $\mu_\star$, M$_\star$ and f$_{HI}$. Overall, galaxies with lower $\mu_\star$, lower stellar masses and higher \hi\ mass fractions have lower central metallicities, which is in agreement with the mass--metallicity relation (see e.g. \citealp{Tremonti2004,Bothwell2013,Brown2018}).
In addition, the median profiles of higher $\mu_\star$ galaxies with higher stellar masses and lower \hi\ mass fractions show a plateau or even decrease of metallicity within approximately $0.5\,\rm r_{eff,r}$. When fitting a line to such a profile with a plateau in the centre, the resulting slope will be flatter. Thus, the central metallicity measurements within $0.5\,\rm r_{eff,r}$ affect the resulting metallicity gradient. This effect is enhanced by the fact that only approximately 50\,percent of all radial data points are within the radial range $0.5 \leq \rm r_{eff,r} \leq 2.0$. Another 30-40 \,percent of our radial metallicity data points are at radii smaller than
0.5\,$\rm r_{eff,r}$. When measuring metallicity gradients including the inner 0.5\,$\rm r_{eff,r}$, the results are thus significantly affected by these data. This effect has already been observed before by, for example, \citet{Rosales-Ortega2009} and \citet{Sanchez2014} and it is one reason why some studies dismiss central metallicity measurements in their gradient estimation.

We computed these profiles for different subsets of our galaxy sample (circles: all galaxies, triangles: only massive galaxies, i.e. M${_\star}>10^{10}$\,M$_\odot$). In particular, for bulgy and relatively \hi-poor galaxies, the radial profiles are dominated by massive galaxies and are consistent between the median profiles of all galaxies and massive galaxies only. For the lowest $\mu_\star$, \hi-rich galaxies, we find that the median profile of all galaxies differs from the median profile of massive galaxies (see top left and bottom right panel in Fig.~\ref{fig:metal_profile}). Thus, the largest effect of low-mass galaxies on the measurements of average gradients is seen in these bins (smallest $\mu_\star$, most \hi-rich). This effect together with low number statistics and the fact that some of our lowest mass galaxies have relatively steep gradients contribute to the discrepancy with MaNGA at low stellar mass surface densities and high \hi\ mass fractions.

At this point, we found that all correlations are to some degree dependent on the sample selection and the definition of the metallicity gradient. Our analysis suggests that there is a correlation between metallicity gradients and $\mu_\star$ for galaxies on the star formation main sequence, especially when measuring the gradient from the entire radial metallicity profile. Before we move on to discussing this finding in greater detail in Sect.~\ref{sec:diss}, we explore the relation between local metallicity and global \hi\ content.

\section{Results: Local metallicity and global H{\small I} content}
\label{sec:results2}
We now focus on the correlation between local gas-phase metallicity measurements and the global \hi\ mass fraction as found by \citetalias{Moran2012}. With the new NTT data presented here, we are able to build on the findings of previous works.

\label{sec:local_metal}
\begin{figure}
    \center
    \includegraphics[width=3.15in]{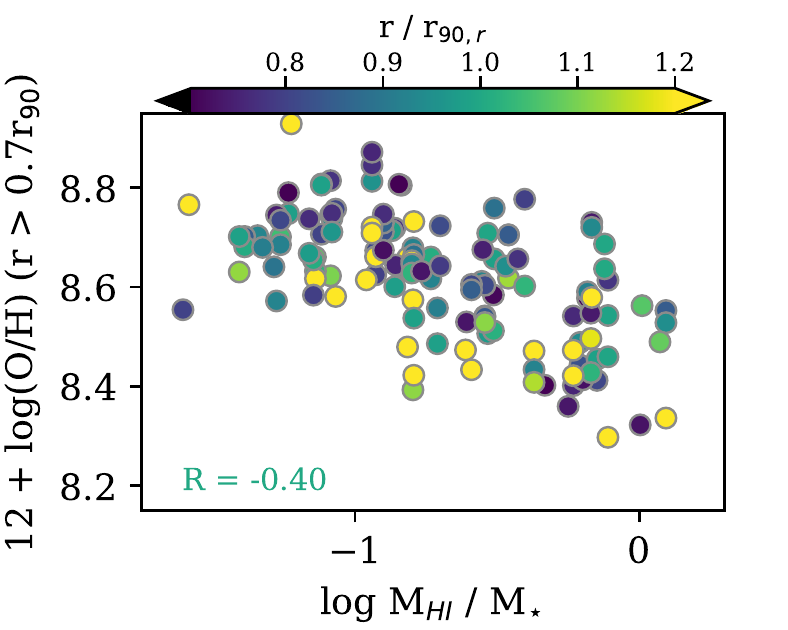}
    \caption{Local metallicity at the outskirts of galaxies as a function of the global \hi\ mass fraction. Circles show the global \hi\ mass fraction as a function of metallicity measurements outside of $0.7{\rm r_{90,r}}$. The data points are colour coded according to their galactocentric radius normalised by ${\rm r_{90,r}}$. We note that some galaxies have multiple metallicity measurements outside of $0.7{\rm r_{90,r}}$ and would thus appear multiple times. The number in the lower left corner provides the Spearman correlation coefficient. }
    \label{fig:outermetal_vs_fhi}
\end{figure}

\begin{figure}
    \center
    \includegraphics[width=3.15in]{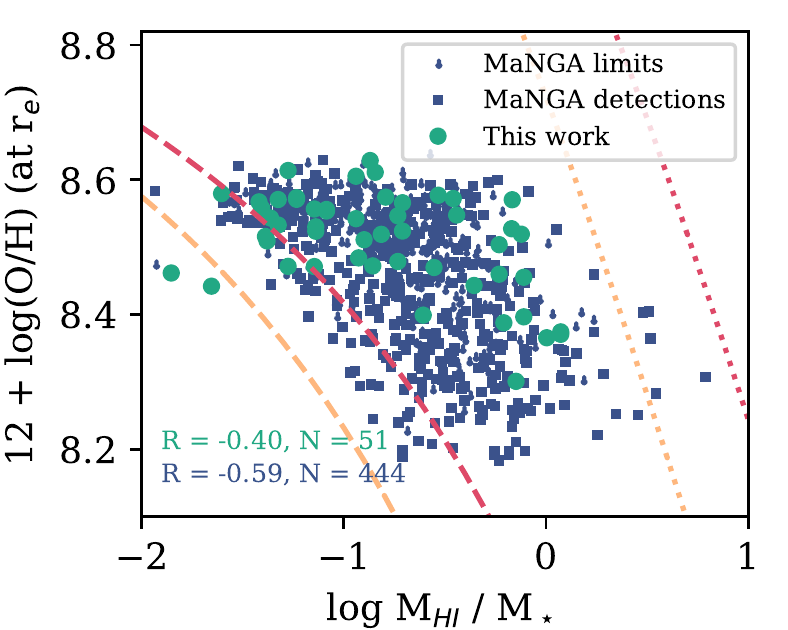}
    \caption{Local metallicity at and around $\pm10$\,percent of the effective radius as a function of the global \hi\ mass fraction for MaNGA (dark blue squares and arrows) and our sample (green circles), respectively. The numbers in the bottom left provide the Spearman correlation coefficient R and the number of galaxies used to calculated the statistic. For MaNGA we only used \hi\ detections in the computation of R. The yellow and red line show our model for different ratios of \hi\ to stellar radius, namely, 3.3 and 5.6, respectively. The underlying model of the dashed lines assumes an effective yield of 0.00268 \citep{Pilyugin2004} and the dotted lines a stellar yield of 0.037 (\citealp{Vincenzo2016} and references therein). We note that we use the M13 $O3N2$ calibration in this figure.}
    \label{fig:re_metal_vs_fhi}
\end{figure}

Previously, \citetalias{Moran2012}  reported a correlation between the local metallicity at the edge of the stellar disc and the global \hi\ mass fraction. In Fig.~\ref{fig:outermetal_vs_fhi}, we added the data of the new low-mass galaxies and find that the correlation holds. For MaNGA galaxies, only local metallicity measurements at one effective radius are provided in the value-added catalogues. Together with all those galaxies from our sample, which have a metallicity measurement within $\pm10$\,percent of the effective radius in $r$ band, the correlation between local metallicity around the effective radius and the global \hi\ mass fraction is shown in Fig.~\ref{fig:re_metal_vs_fhi}. Again, a correlation is recovered. In summary, we find that local metallicity correlates with global \hi\ mass fraction.

One way to explain these correlations between local metallicity and global \hi\ mass fraction is suggested by the following simple model. We assume an exponential stellar disc, such that the stellar mass surface density is given by:
\begin{equation}
    \label{equ:stellar_suf_dens}
\Sigma_\star = \Sigma_{0,\star} \times e^{-r / {\rm r_{0, \star}}},
\end{equation}
and the total stellar mass by:
\begin{equation}
    {\rm M_\star} = 2\pi \times \Sigma_{0,\star} \times {\rm r_{0,\star}}^2,
\end{equation}
where $\rm r_{0, \star}$ is the stellar scale length and $\Sigma_{0,\star}$ the central stellar column density. For the \hi\ disc we describe the \hi\ mass by:
\begin{equation}
    {\rm M_{HI}} = \pi \times \ \Sigma_{0,{\rm HI}} \times {\rm r_{HI}}^2
,\end{equation}
where $\rm r_{HI}$ is the \hi\ disc size and $\Sigma_{0,{\rm HI}}$ the (central/ constant) \hi\ column density. Furthermore, we assume a local closed-box model where the local metallicity $Z$ at radius $r$ can be described as (see e.\,g. \citealp{Mo2010}):
\begin{equation}
    \label{equ:metal_rad}
    Z(r) = - y_{eff} \ln \left(\frac{\Sigma_{{\rm HI}}(r) + \Sigma_{{\rm H2}}(r)}{\Sigma_{{\rm HI}}(r) + \Sigma_{{\rm H2}}(r)+\Sigma_\star(r)}\right)
,\end{equation}
with $y_{eff}$ the effective yield and the $\Sigma(r)$ the local column densities of \hi, \htwo\ and stars at radius $r$. To evaluate this equation at the effective radius $\rm r_{eff}$, we take into account that (i) the \hi\ and \htwo\ column densities are approximately equal at $\rm r_{eff}$ \citep{Bigiel2012}; (ii) the \hi\ column density at the effective radius is approximately the same as in the centre, which is suggested by the tight \hi\ mass size relation \citep{Wang2016,Broeils1997}; and (iii) $\rm r_{eff} \approx 1.7 \times r_{0, \star}$ (and use Eq.~\ref{equ:stellar_suf_dens}).
We thus obtain
\begin{align}
    Z(r=r_e) &= -y_{eff} \ln \left(\frac{2 \times \Sigma_{{\rm HI}}}{2 \times \Sigma_{{\rm HI}}+0.18 \Sigma_{0,\star}}\right),\\
    {} &= -y_{eff} \ln \left(\frac{\pi \times {\rm r_{HI}}^2 \times \Sigma_{{\rm HI}}}{\pi \times {\rm r_{HI}}^2 \times \Sigma_{{\rm HI}}+0.09\pi \times {\rm r_{HI}}^2  \Sigma_{0,\star}}\right),\\
    {} &= -y_{eff} \ln \left(\frac{{\rm M}_{{\rm HI}}}{{\rm M}_{{\rm HI}}+0.045 \times {\rm M}_\star \times ({\rm {\rm r_{HI}} / r_{0,\star}})^2 }\right) .
\end{align}
According to \citet{Broeils1997}, for instance, there is a good correlation between the radius of the \hi\ and the stellar disc for spiral galaxies. Thus, this (local closed-box) model suggests indeed a correlation between the local metallicity at the effective radius and the \hi\ mass fraction. Since the stellar scale length is also tightly correlated to r$_{90}$, a similar calculation can be carried out for Fig.~\ref{fig:outermetal_vs_fhi}.
In Fig.~\ref{fig:re_metal_vs_fhi}, we added the model prediction assuming a stellar oxygen yield of 0.037 \citet{Vincenzo2016} from the \citet{Romano2010} and \citet{Nomoto2013} stellar models assuming a
\citet{Chabrier2003} initial mass (dotted lines) and an effective oxygen yield of 0.00268 measured by \citep{Pilyugin2004} (dashed lines) in spiral galaxies. Furthermore, we follow \citet{DeVis2017} to convert between metallicity mass fraction and metallicity number density fractions. We note that we show both an example for true stellar yields and one for an effective yield. When using the effective yield small amounts of in- and outflows are included in this toy model. With the stellar yield this is a pure closed-box model. In addition, we show two different ratios of $\rm r_{HI} / r_{0,\star}$ = 3.3, and 5.6 in yellow and red, respectively.
These ratios are approximately equivalent to $\rm r_{HI} / R_{25}$ = 1.0, and 1.7, with $\rm r_{HI} / R_{25}$ = 1.7 (red line) the preferred value by observations of spiral galaxies \citep{Broeils1997}. More recent measurements of this ratio by \citet{Wang2016} suggest a range of values $\rm 0.6\lessapprox r_{HI} / R_{25} \lessapprox 5$.

\section{Discussion}
\label{sec:diss}

\subsection{Metallicity gradients}

We analysed the radial metallicity profile of a sample of star formation main sequence galaxies from the \xgass\ and \xcoldgass\ galaxy sample and investigated the correlation with global galaxy properties, such as \hi\ and \htwo\ gas mass fraction, stellar mass, morphology and star formation activity. Depending on the method for measuring the gradients and the radial region in which the gradient is measured, we get the following results. %\LEt{ The paragraphs below can remain as they are, but please remove the dashed bullets at the start of each.}

\textbf{Firstly, measuring the metallicity gradient from the entire radial profile:} We find correlations between metallicity gradients and multiple global galaxy properties, which have correlation coefficients significantly different from zero. However, the correlation coefficients get closer to 0, when considering only massive galaxies (M$_\star$ > 10$^{10}$\,M$_{\odot}$). The correlations between metallicity gradients and global galaxy properties are tightest for metallicity gradients measured in units of dex\,$\rm r_{eff,r}^{-1}$. However, also when we are normalising the galactocentric radius with Petrosian r$_{90}$, the isophotal radius R$_{25}$ or measuring the radius in kpc, we recover these correlations. The correlations are such that less massive, more \hi-rich galaxies with smaller $\mu_\star$ have steeper metallicity gradients than more massive, higher $\mu_\star$, more \hi-poor galaxies.

\textbf{Secondly, measuring the metallicity gradient from the radial profile without the central 0.5\,$\rm r_{eff,r}$}: In this case, we only recover a correlation coefficient significantly different from zero for \must\ and metallicity gradient in units of dex\,$\rm r_{eff,r}^{-1}$. All other relations between metallicity gradients and global galaxy properties are either flat or the data are too scattered across the parameter space. This implies that the stellar mass surface density not only shapes the radial metallicity profile in the centre of galaxies but also the steepness of the metallicity decline towards the outskirts.

\begin{figure*}
    \center
    \includegraphics[width=6.3in]{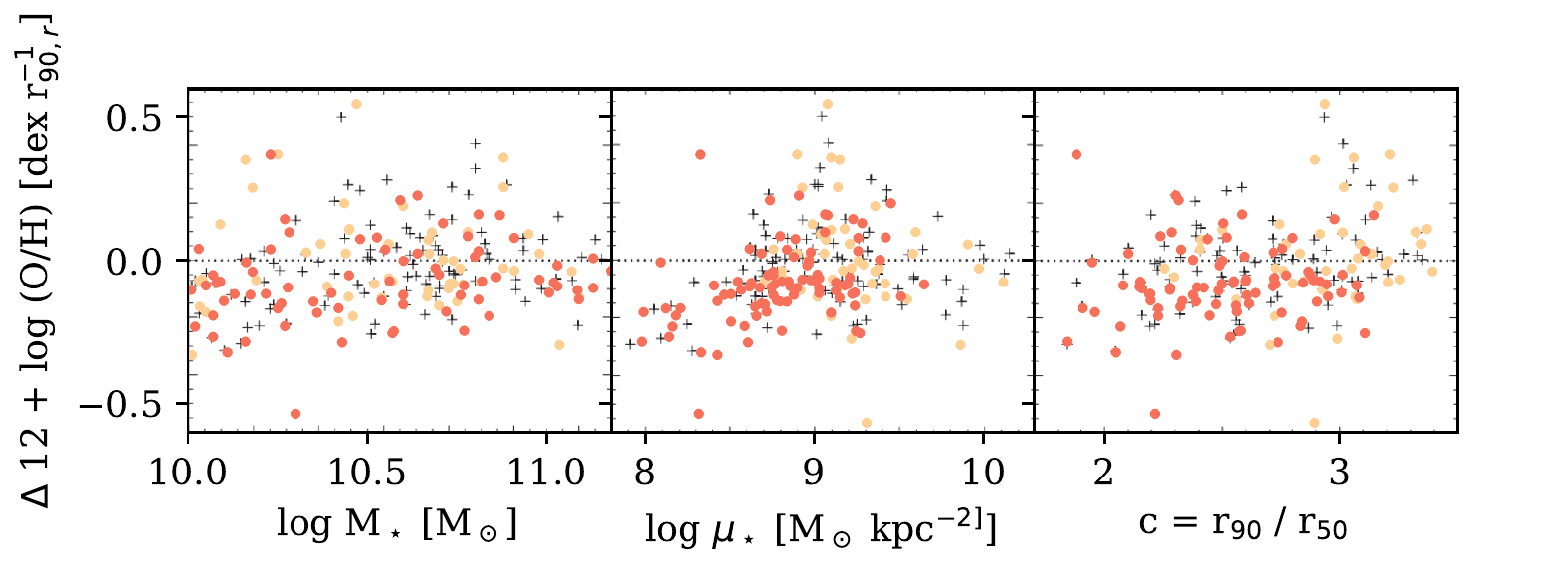}
    \caption{Direct comparison to Figure 5 of \citetalias{Moran2012}. The dark crosses in the background are data by \citetalias{Moran2012}, the yellow points our linear fits to the metallicity profiles of massive, quiescent galaxies (off the SFMS) and the orange points our linear fits to the metallicity profiles of massive, star-forming galaxies (on the SFMS).}
    \label{fig:moran2012}
\end{figure*}

In both cases, a deeper analysis of inter-correlations between the global galaxy properties revealed that only \must\ determines metallicity gradients. All other correlations appear to be driven by the relation between \must\ and other global galaxy properties. \citetalias{Moran2012} found that the concentration c (as a proxy for bulge to total mass ratio) is closer related to metallicity gradient than $\mu_\star$. To understand these differences between two studies that use the same underlying data set, we compared the distribution of c and \must\ for the \citetalias{Moran2012} and our sample (see Fig.~\ref{fig:moran2012}).

While we use the same radially binned spectra for galaxies with stellar masses greater than $10^{10}$\,M$_\odot$ as \citetalias{Moran2012}, in this paper, we use a different method to fit the gradients and we only use galaxies within $\pm1.5 \sigma$ of the star formation main sequence as defined by \citet{Catinella2018}. Generally, we recover the same trends as \citetalias{Moran2012}. As can be seen from the middle and right panel of Fig.~\ref{fig:moran2012}, the correlations between the metallicity gradients and stellar mass surface density $\mu_\star$ or the concentration index c are different for our work than for
\citetalias{Moran2012}. Where the removal of quiescent galaxies emphasises a correlation between metallicity gradients and $\mu_\star$, the same step wipes out the correlation between metallicity gradient and c. Thus, selecting only SFMS galaxies, as we did, preferentially removes galaxies with flat or positive gradients and large concentration indices and galaxies with all types of gradients and large stellar mass surface densities from the \citetalias{Moran2012} sample. Thus slightly different trends are induced. Overall, the results from \citetalias{Moran2012} and our results agree in the sense that 'more bulge-dominated' galaxies have flatter radial metallicity profiles than 'more disc-dominated' galaxies. This is in contrast to results based on CALIFA data, which did not find any correlation between the metallicity gradient and Hubble type \citep{Sanchez2014,Sanchez-Menguiano2016}.

Overall, we observe that the steepness of metallicity gradients and the shape of radial metallicity profiles are driven by the stellar mass surface density. We also see correlations with stellar mass but our statistical tests suggest that stellar mass surface density is a more important driver. To understand what this finding implies for galaxy evolution, we consider two chemo-dynamical models by \citet{Pezzulli2016} and \citet{Boissier2000}.

\citet{Pezzulli2016} consider models with growing exponential stellar disks. They find that in models, in which gas accretes from the intergalactic medium (IGM) such that the disk grows with a constant exponential scale length, galaxies form metallicity gradients that are not compatible with observations. Once adding radial flows, gradients become more realistic. When considering IGM gas accretion plus radial flows plus inside-out growth, realistic gradients are  formed  and  less IGM accretion is needed than in the previous case. Overall, the steepness of their  metallicity gradients is driven by the angular momentum misalignment of accreted gas with respect to the disc. The more miss-aligned the accreted gas, the larger radial gas flows, the steeper metallicity gradients. In the context of our observational findings, these models suggest that galaxies with smaller $\mu_\star$ would have larger radial flows, as indicated by their steeper metallicity gradients. Galaxies with larger $\mu_\star$ have smaller radial flows, which would mean that less and less gas arrives at their centres. Once these galaxies use up the gas in their centres, inside-out quenching would set in. Shortly afterwards, these galaxies would reach equilibrium in their centres. In our observations, this equilibrium state is reflected in the flattening of the radial metallicity profiles towards galaxy centres. Such a saturation effect has also been suggested in such works as \citet{Koppen1999}.

The chemo-dynamical models of \citet{Boissier2000} investigate the galaxy evolution as a function of halo spin parameter $\lambda$ and rotation velocity, which is a proxy for mass. These models rely purely on IGM accretion and inside-out growth of an exponential disk. No radial flows are implemented. The central surface brightness in their model galaxies is determined by the halo spin parameter, such that galaxies with smaller central surface brightness tend to reside in haloes with larger spins. This is also found in other simulations and models (e.\,g. \citealp{Kim2013}). Quantitatively their metallicity gradients are steeper than commonly measured. However, qualitatively their Fig.~15 shows that their model galaxies form steeper gradients the higher the halo spin parameter, and thus the lower the central surface brightness. Furthermore, galaxies with very low halo spin and thus high central surface brightness appear to form a metallicity plateau in their centres. These results agree with our observations. Once more the flattening of the radial metallicity profile in the centre can be explained with different accretion patterns in low and higher $\mu_\star$ galaxies. The IGM accretion onto more massive galaxies with higher total surface density is higher in the beginning but shuts down faster than for less massive and less dense galaxies (their Fig.~3). With the decrease in gas supply, once more metallicity converges towards an equilibrium value, as can be seen in the centres of our high $\mu_\star$ galaxies.

The comparison to two chemo-dynamical models suggests that our observational finding of steeper metallicity gradients in galaxies with lower $\mu_\star$ can be explained. Our recovered relation can either be interpreted as the impact of (i) the halo spin parameter on the inside-out growth of exponential disks or (ii) smaller radial velocities in galaxies of earlier type.

The correlation between metallicity gradient and stellar mass has often been discussed in the literature. The CALIFA team \citep{Sanchez2014,Sanchez-Menguiano2016} as well as \citet{Kudritzki2015} and \citet{Ho2015} find a universal metallicity gradient, that is, no correlation with stellar mass or morphology.  On the other hand, in the MaNGA data, \citet{Belfiore2017a} find the steepest declining metallicity profiles, that is, the steepest metallicity gradients
for galaxies around stellar masses of $10^{10}$ to $10^{10.5}$\,M$_\odot$ and flatter metallicity profiles in lower and higher mass galaxies. We recover these trends in the MaNGA data, which we use for comparison with our sample (middle column Fig.\ref{fig:manga}). All these studies measure the gradients from radial metallicity profiles within the radial range of $0.5 \leq$ $\rm r_{eff,r}$ $\leq 2.0$. When measuring metallicity gradients for our sample from the radial profiles without the central 0.5\,$\rm r_{eff,r}$, we recover a relatively flat correlation with large scatter (see in particular middle column in Fig.~\ref{fig:grad_wo}) and thus agree with previous studies. Interestingly, \citet{Bresolin2019} have studied metallicity gradients in low-mass spirals with longslit spectroscopy and also found relatively steep metallicity gradients. These are consistent with or steeper than our measurements (see e.\,g. their Fig.~8).

\citet{Poetrodjojo2018} measured metallicity gradients for a small number of SAMI galaxies using the entire radial metallicity profile and find that low-mass galaxies have flatter metallicity gradients than more massive galaxies. We note, however, that their upper stellar mass limit is 10$^{10.5}$\,M$_\odot$. They furthermore caution that the stellar mass distribution of the sample heavily impacts on the observed trends between metallicity gradients and stellar mass. In addition, diffuse ionised gas might pose a problem \citep{Poetrodjojo2019}.
% Further results by Poetrodjojo et al. (private communication), suggest that galaxies around $\approx 10^{9.5}$ to $10^{10}$\,M$_\odot$ have the steepest metallicity gradients, while metallicity gradients for more and less massive galaxies are flatter. Considering, that the galaxy sample investigated by \citetalias{Moran2012} only contained galaxies more massive than  $10^{10}$\,M$_\odot$ and that they observed more massive galaxies to have flatter metallicity gradients, the works by \citetalias{Moran2012} and Poetrodjojo et al. (private communication) agree well.
The lower stellar mass limit of the galaxy sample investigated by \citetalias{Moran2012} is at $10^{10}$\,M$_\odot$ and metallicity gradients were also measured on the entire radial metallicity profiles. They observed more massive galaxies to have flatter metallicity gradients. These two results are not mutually exclusive and given that \citet{Belfiore2017a} observe a change of trend around the stellar mass limits of the \citet{Poetrodjojo2018} and \citetalias{Moran2012} stellar mass limits, the two results might even be complementary. In this case, we would expect to observe this turnover in our results.
For our sample, however, the stellar mass range 9.0 $<$ \mass\ $<$ 10.0 is more sparsely sampled than higher stellar masses and we only measured average gradients in one stellar mass bin in this stellar mass range (see in particular middle column, second row from top in Fig.~\ref{fig:grad_with}). Thus, a turnover can not be robustly recovered from our data. Nonetheless, our analysis shows that in addition to the stellar mass distribution of the sample (as observed by \citealp{Poetrodjojo2018}), the radial location where the metallicity gradient is measured affects results regarding the correlation between gradients and global galaxy properties.

{}Until today, there have only been few studies, aside from that of \citetalias{Moran2012},
investigating the link between metallicity and \hi\ content. \citet{Brown2018}, \citet{Bothwell2013}, and \citet{Hughes2013} find that larger \hi\ content leads to lower (central) metallicities, \citet{Bothwell2016} reported that molecular gas is more relevant in determining the metallicity.
 With respect to gradients (rather than central metallicities as in \citealp{Bothwell2016}), we find that \hi\ is more tightly correlated to metallicity than \htwo. \citet{Carton2015} investigated metallicity gradients in a sample of massive galaxies and they find, in contrast to our results, that more \hi-rich galaxies have flatter gradients. Their sample, however, covers a smaller stellar mass range than our sample, doesn't reach as high \hi\ mass fractions as our sample and they use a different metallicity estimator. Hence, the comparison is difficult. Nonetheless, we only observe the same trends as \citet{Carton2015} when considering our analysis of the MaNGA sample: higher \hi\ mass fractions come with flatter metallicity gradients.

{}Overall, we find that \must\ determines metallicity gradients in our sample of SFMS galaxies, which reflects predictions from the chemo-dynamical evolution models by \citet{Pezzulli2016} and \citet{Boissier2000}. Correlations with stellar mass and \hi\ mass fraction are less robust and a more detailed analysis suggests that these trends are induced due to correlations between \must\ and stellar mass as well as \fhi.

\subsection{Local metallicity and global HI mass fraction in local closed-box models}
Based on the previous findings of \citetalias{Moran2012}, we investigated the correlation between local metallicity and global \hi\ mass fraction. Here, we consider local metallicities measured in the vicinity of either $\rm r_{eff,r}$ (for our sample and for a sample of MaNGA galaxies with \hi\ mass) or r$_{90, r}$ (only for our sample). In both cases, we find a correlation between the \textbf{local} metallicity and the \textbf{global} \hi\ to stellar mass ratio. When comparing the observed correlation to the relation expected for a local closed-box model utilising a the true stellar yield, we find that metallicities, as expected, are significantly overestimated. When using an effective yield, which accounts for in- and outflows and turns the model in a gas regulator model, we find that this model is in better agreement with the data. The detailed choices are discussed below.
Simulations \citep{Forbes2014a} have shown that these radial gas flows are vital for the evolution of galaxies but they are in equilibrium around a redshift of 0. Observations \citep{Schmidt2016} of radial flows, which bring metal-poor gas towards the centres of galaxies, in the \hi\ kinematics, show that they exist but are not detected in every galaxy, mostly likely because they are small. Also, the \citet{Pezzulli2016} model suggests that small radial flows are necessary but not the main driver of metallicity gradients.

To compare the model to the data, we have to make assumptions for the (effective) yield and the ratio of \hi\ to stellar disc size. The ratio of \hi\ to stellar disc size has not yet been studied extensively. \citet{Broeils1997} find a remarkably tight correlation between \hi\ disc size and 25\,mag\,arcsec$^{-2}$ isophotal radius $\rm R_{25}$ for spiral galaxies, with the average radius ratio being 1.7. However, galaxies with higher $\mu_\star$ contain less \hi\ and, thus, the ratio between \hi\ and stellar disc size likely decreases. An extensive analysis by \citet{Wang2016} find a range of radius ratios: $\rm 0.6\lessapprox r_{HI} / R_{25} \lessapprox 5$.
Thus, we also show the model results with an \hi\ to 25\,mag\,arcsec$^{-2}$ isophotal radius ratio of 1.0. For the yield, we chose two different values: 0.00268, an effective yield obtained by \citet{Pilyugin2004} for spiral galaxies, and 0.037, a stellar yield obtained by \citet{Vincenzo2016} from the \citet{Romano2010} and \citet{Nomoto2013} stellar models assuming a
\citet{Chabrier2003} initial mass function and the average gas phase metallicities of our galaxies. Being a measure of the true stellar yield, the prediction based on the \citet{Vincenzo2016} yield is an upper limit. Thus, indeed outflows of metal-rich gas or inflow of metal-poor gas must have taken place in our sample galaxies. The \citet{Pilyugin2004} appears at the lower end of our data, which might imply that in- and outflows in our sample galaxies is less effective or pronounced than in the spiral galaxies analysed by
\citet{Pilyugin2004}. In addition the differing metallicity estimators between our work and \citet{Pilyugin2004} might induce differences \citep{Vincenzo2016}. Overall, this model works well to explain the correlation between a \textbf{local} metallicity measurement and the \textbf{global} \hi-to-stellar-mass ratio.

Recent large surveys of the \hi\ fraction and its correlation to other global properties of galaxies suggest that the morphology (as described by the stellar mass surface density $\mu_\star$) is one defining factor (secondary to \nuvr\ colour) in setting the \hi\ mass fraction \citep{Catinella2013,Catinella2018,Brown2015}. Together with the analysis of the primary driver of metallicity gradient, this might explain why \fhi\ correlates with metallicity gradients. Another approach might be provided by our simple calculations in Sect.~\ref{sec:local_metal}, which show that the global \hi\ mass fraction sets the local metallicity at specified radii
(here $\rm r_{eff}$ and $\rm r_{90}$). Once the global \hi\ mass fraction determines the metallicity at, for example,\, $\rm r_{eff}$ and $\rm r_{90}$, f$_{HI}$ also determines the rate at which the metallicity changes from $\rm r_{eff}$ to $\rm r_{90}$ and, thus, the metallicity gradients. In this way, our simple model could also explain why the metallicity gradient seems to correlate with \hi\ mass fraction.

\citet{Barrera-Ballesteros2018} did not look at the correlation between metallicity gradients or local metallicity and global \hi\ content but the authors did offer their report that local metallicity depends on local cold gas mass fractions (estimates based on the optical extinction $A_V$). In particular, they found lower metallicities in regions where the ratio of local gas to local total mass is high. As we assume constant \hi\ column density across a exponentially declining stellar disc, our model also suggests lower metallicities where the \hi\ to stellar surface density is higher. Thus, both our simple model and our data agree with the findings by \citet{Barrera-Ballesteros2018}. We are furthermore able to specify that \hi\ is more important than \htwo\ in defining the metallicity. In light of these results, it will be interesting to follow up on these investigations once resolved \hi\ and metallicity observations are available for a large number of galaxies, in particular, through combinations of surveys such as MaNGA and Apertif (Adams et al. in prep)\footnote{https://www.astron.nl/telescopes/wsrt-apertif/apertif-dr1-documentation/data-access/data-usage-policy/} or WALLABY \citep{Koribalski2020}.
%TODO is there a references for Apertif???

\section{Conclusion}
\label{sec:sum_conclude}
In this work, we present new optical longslit spectra for 27 low-mass galaxies from the \xgass\ \citep{Catinella2018} and \xcoldgass\ surveys \citep{Saintonge2017}. By combining the new data with data from \xgass\ and \xcoldgass, we investigated the relation between gas-phase oxygen abundance, gas content, and star formation. In particular, we focused on metallicity gradients and the local metallicity at different galactocentric radii and their correlation to global galaxy properties. Our findings can be summarised as follows:
% \LEt{ Bullets in the conclusions are fine.}
\begin{itemize}
    \item While there is a number of global galaxy properties that correlate with the metallicity gradient, various statistical analyses suggest that only the stellar mass surface density $\mu_\star$ drives metallicity gradients. Other correlations come about as \must\ correlates with these global galaxy properties.
    \item The correlation between $\mu_\star$ and metallicity gradient can be interpreted with the help of chemo-dynamical evolution models of \citet{Pezzulli2016} and \citet{Boissier2000}: The observed correlation can be interpreted as a sign of (i) different spin parameters of the host halo or (ii) different accretion and radial flow patterns in galaxies, depending on their stellar mass surface density.
    \item The local metallicity is correlated with the global \hi\ mass fraction. Although it is surprising that a local measurement should be informed about global galaxy properties, this correlation can actually be modelled with a simple gas regulator model, which is described by a local closed-box model plus an effective yield, which accounts for small radial flows.
    \item When comparing to metallicity gradients in the literature, in particular MaNGA \citep{Bundy2015,Belfiore2017a,Sanchez2018a,Sanchez2016,Sanchez2016a} and SAMI
    \citep{Croom2012, Poetrodjojo2018}, we find that our results agree within the errors for high-mass galaxies. In the lower stellar mass regime we observe relatively steep gradients. These discrepancies can not be explained by sample selection but potentially by small sample statistics. We expect further discussions in the literature over trends with metallicity gradients for galaxies at stellar masses, M$_\star \leq 10^{10}$\,M$_\odot$, or with low stellar mass surface densities (small to no bulges). Furthermore, it is  vital that metallicity gradients are measured from metallicities at similar radial regions. Once data points inwards of 0.5\,$\rm r_{eff,r}$ are included in the gradient measurement, which was not done by the MaNGA team, our results start to differ significantly.
\end{itemize}

In particular, the (local) correlation between metallicity and \hi\ has not yet been studied in great detail across galaxy discs. Upcoming and ongoing surveys such as MaNGA, SAMI, WALLABY, and Apertif, as well as future surveys on MeerKAT will provide more information and further details about local ISM enrichment and radial gas flows.

\begin{acknowledgements}
We would like to thank the referee for a constructive report that helped improving this paper.

KL would like to thank Virginia Kilborn and Gabrielle Pezzulli for valuable discussions during the making of this paper.\\
LC is the recipient of an Australian Research Council Future Fellowship (FT180100066) funded by the Australian Government.\\
Parts of this research were supported by the Australian Research Council Centre of Excellence for All Sky Astrophysics in 3 Dimensions (ASTRO 3D), through project number CE170100013.\\

Besides software packages already mentioned in the main body of this paper, this work has also made use of Python\footnote{http://www.python.org} and the Python packages: astropy \citep{AstropyCollaboration2013}, NumPy\footnote{http://www.numpy.org/}, matplotlib\footnote{https://matplotlib.org/} \citep{Hunter2007}, pandas\citep{Reback2020,McKinney2010} and seaborn\footnote{https://seaborn.pydata.org/}. Furthermore TOPCAT has been used \citep{Taylor2005}. \\

This research has made use of the VizieR catalogue access tool, CDS,
Strasbourg, France \citep{Ochsenbein2000}; the SIMBAD database,
operated at CDS, Strasbourg, France \citep{Wenger2000}; TOPCAT
\citep{Taylor2005}; the "Aladin sky atlas" developed at CDS, Strasbourg
Observatory, France \citep{Bonnarel2000,Boch2014}; NASA’s Astrophysics Data
System Bibliographic Services.\\

Funding for the SDSS and SDSS-II has been provided by the Alfred P. Sloan
Foundation, the Participating Institutions, the National Science
Foundation, the U.S. Department of Energy, the National Aeronautics and Space
Administration, the Japanese Monbukagakusho, the Max Planck Society, and
the Higher Education Funding Council for England. The SDSS Web Site is
http://www.sdss.org/.

The SDSS is managed by the Astrophysical Research Consortium for the
Participating Institutions. The Participating Institutions are the American
Museum of Natural History, Astrophysical Institute Potsdam, University of
Basel, University of Cambridge, Case Western Reserve University, University
of Chicago, Drexel University, Fermilab, the Institute for Advanced Study, the
Japan Participation Group, Johns Hopkins University, the Joint Institute
for Nuclear Astrophysics, the Kavli Institute for Particle Astrophysics and
Cosmology, the Korean Scientist Group, the Chinese Academy of Sciences
(LAMOST), Los Alamos National Laboratory, the Max-Planck-Institute for
Astronomy (MPIA), the Max-Planck-Institute for Astrophysics (MPA), New
Mexico State University, Ohio State University, University of Pittsburgh,
University of Portsmouth, Princeton University, the United States Naval Observatory,
and the University of Washington.\\

This project makes use of the MaNGA-Pipe3D dataproducts. We thank the IA-UNAM
MaNGA team for creating this catalogue, and the Conacyt Project CB-285080 for
supporting them.
\end{acknowledgements}
%%%%%%%%%%%%%%%%%%%%%%%%%%%%%%%%%%%%%%%%%%%%%%%%%%



\bibliographystyle{aa}
\bibliography{../../../Bib_Files/paper_xGASS_xCOLDGASS-metallicity_gradients.bib}

%%%%%%%%%%%%%%%%%%%%%%%%%%%%%%%%%%%%%%%%%%%%%%%%%%

%%%%%%%%%%%%%%%%% APPENDICES %%%%%%%%%%%%%%%%%%%%%

\begin{appendix}
\section{Comparison to the MaNGA sample}
\label{app:compare_manga}
\begin{figure}
    \center
    \includegraphics[width=3.15in]{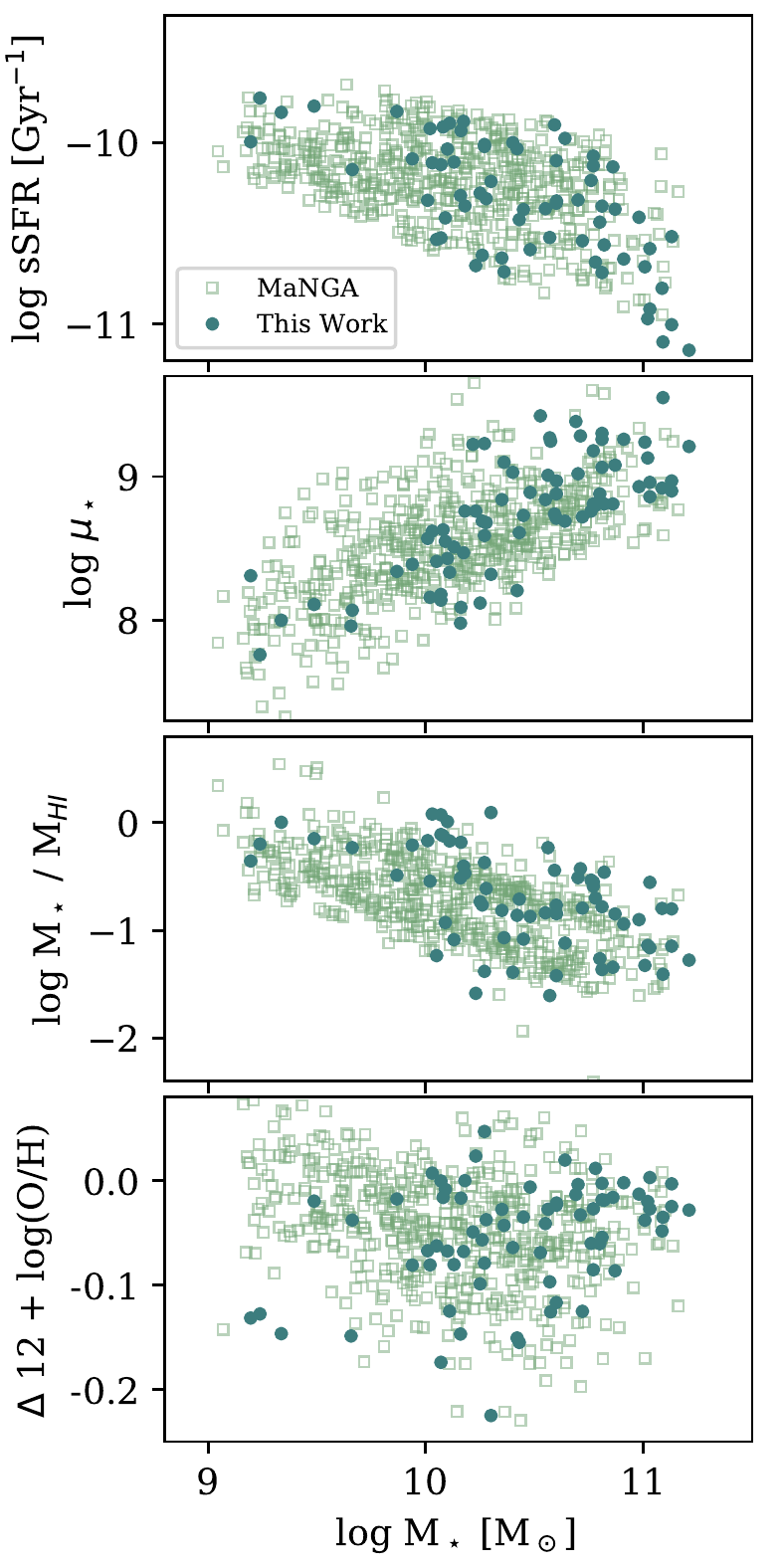}
    \caption{Comparison of the sample in this work with the MaNGA star-forming galaxy sample used in Sect. \ref{sec:compare_manga}. }
    \label{fig:manga_sample}
\end{figure}

In order to test for sample differences, which might induce the different metallicity gradient trends observed in Sect.~\ref{sec:compare_manga}, we investigate the distribution of the two samples on the star formation main sequence, the stellar mass surface brightness versus stellar mass plane, and the \hi\ mass fraction versus stellar mass plane in Fig.~\ref{fig:manga_sample}.

As can be seen, even the four galaxies in our sample, which have relatively steep metallicity gradients at low stellar masses, are on all other plots within the parameter space covered by the MaNGA sample. Thanks to selecting star-forming galaxies only, the sensitivity of the MaNGA \hi\ data, allows detections down to similar \hi\ contents as measured in our galaxies. Furthermore, no significant differences in morphology (as traced by stellar mass surface density) can be observed. Thus, it seems unlikely that differences in the metallicity gradients versus stellar mass plane arise due to sample selection effects.

\section{Individual Gradients}
\begin{figure*}
    \center
        \includegraphics[width=2in]{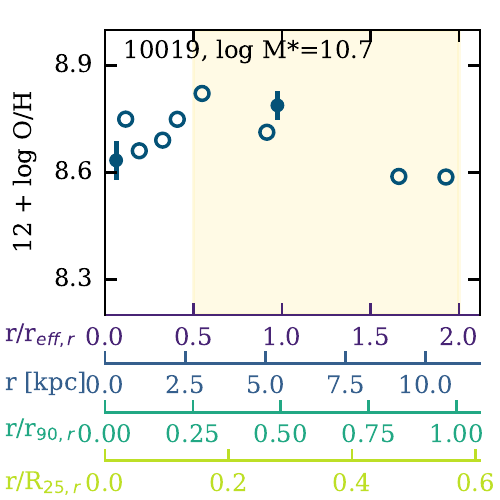} \hfill
        \includegraphics[width=2in]{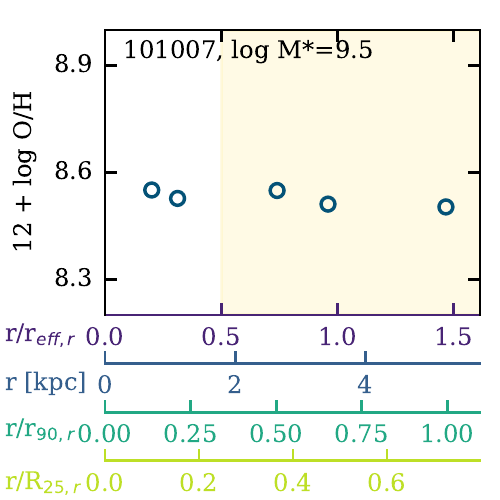} \hfill
        \includegraphics[width=2in]{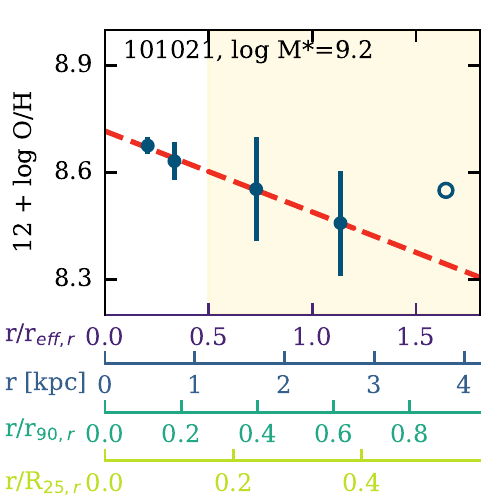} \\
        \includegraphics[width=2in]{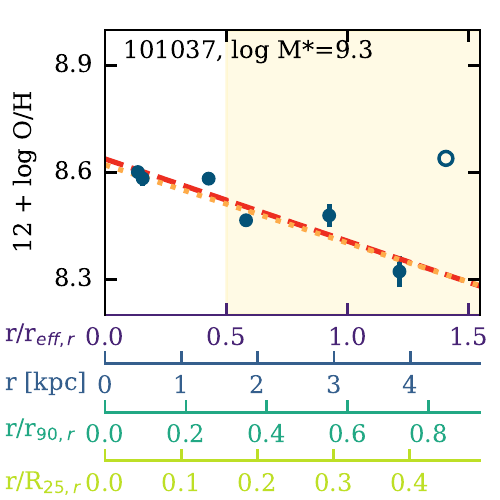}\hfill
        \includegraphics[width=2in]{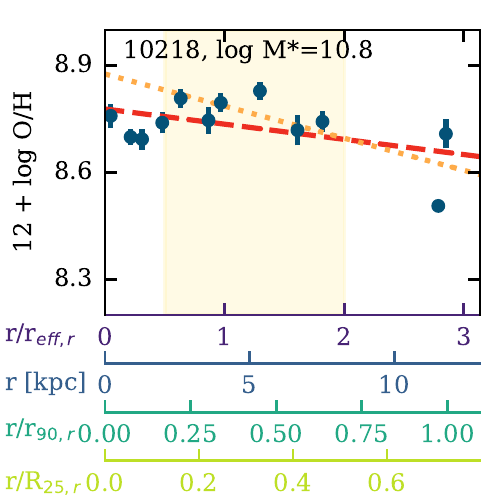}\hfill
        \includegraphics[width=2in]{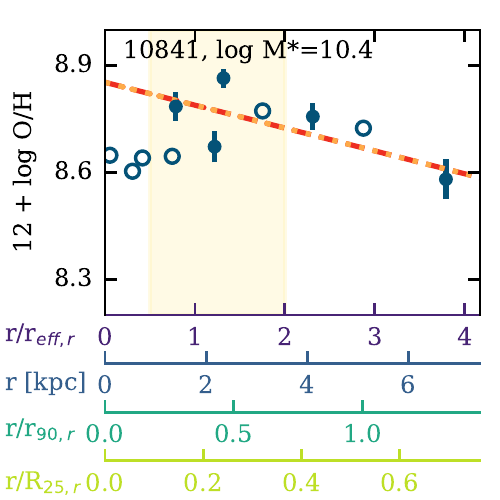} \\
        \includegraphics[width=2in]{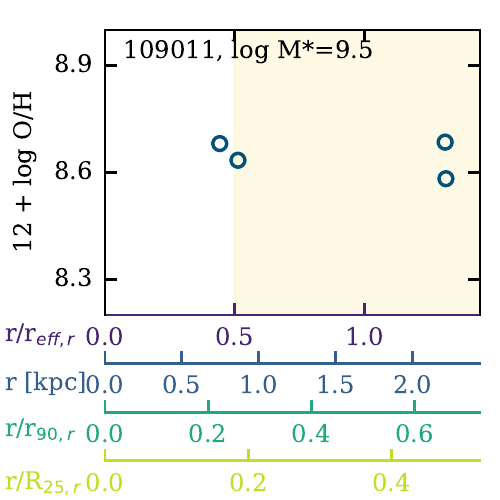}\hfill
        \includegraphics[width=2in]{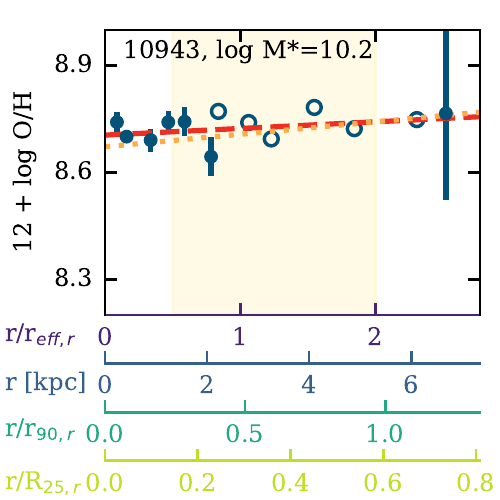}\hfill
        \includegraphics[width=2in]{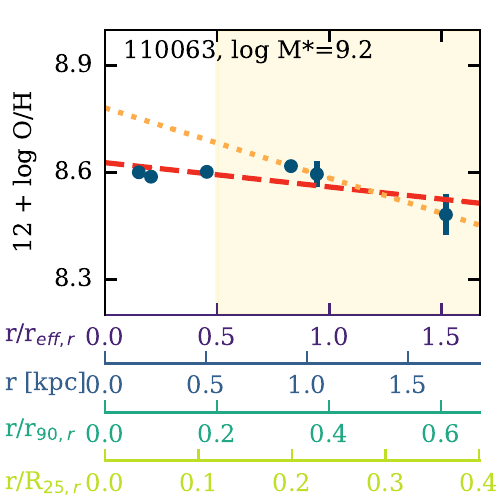}\\
        \includegraphics[width=2in]{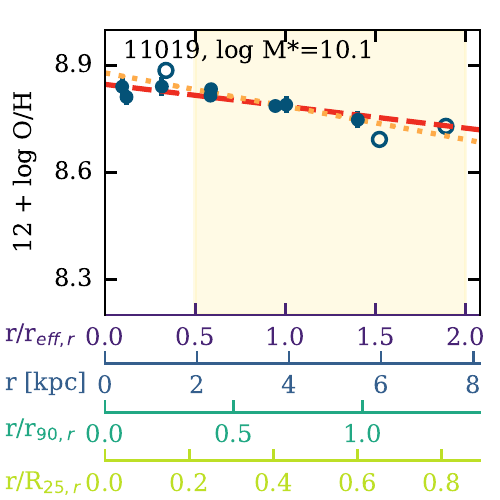}\hfill
        \includegraphics[width=2in]{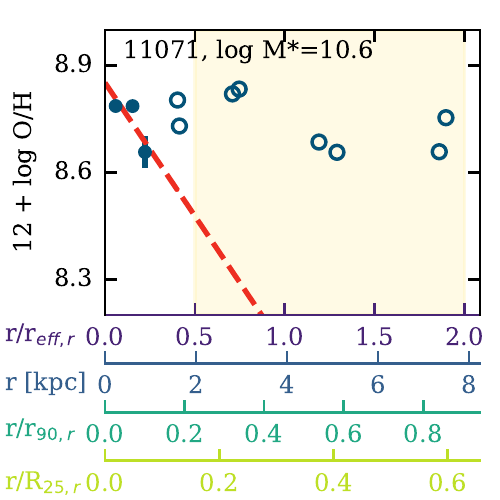}\hfill
        \includegraphics[width=2in]{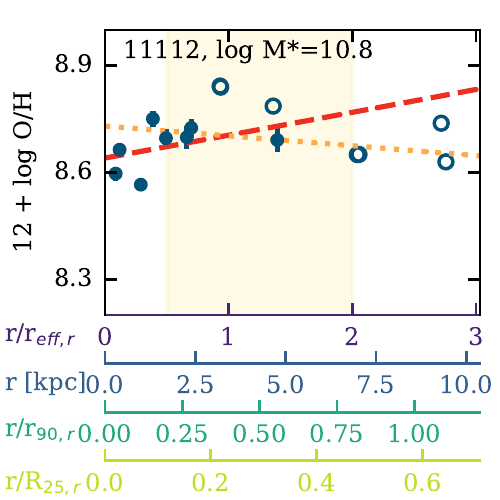}
    \label{fig:app1}
    \caption{Radial metallicity profiles for the different sample galaxies. In each panel, the metallicity (estimated with the \citet{Pettini2004} $O3N2$ method) is shown as a function of radius. On the main x-axis, the radius is normalised by $\rm r_{eff,r}$, additionally we also give the x-axis in units of kpc, normalised by $\rm r_{90,r}$, $\rm r_{50,r}$ and $\rm R_{25,r}$ for orientation. Filled circles mark all metallicity measurements that meet our quality criteria (see Sect.~\ref{sec:grad_fit}) and open circles that measurements that do not meet our criteria. The red dashed line shows the linear fit used to measure the metallicity gradient from the full metallicity profile and the orange dotted line the linear fit used to measure the metallicity gradient without the central 0.5\,$\rm r_{eff,r}$. The yellow shaded area, marks the radial regions between 0.5 and 2.0\,$\rm r_{eff,r}$,
    where e.\,g. publications based on CALIFA and MaNGA data products perform their metallicity gradient measurements. The text in the top right corner gives the GASS ID of the galaxy and the stellar mass. }
\end{figure*}

\setcounter{figure}{0}
\begin{figure*}
    \center
        \includegraphics[width=2in]{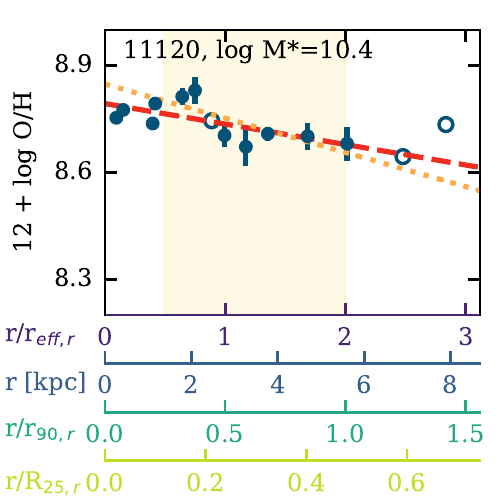}\hfill
        \includegraphics[width=2in]{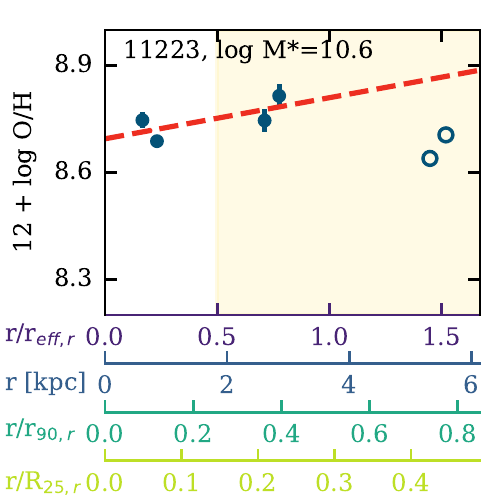}\hfill
        \includegraphics[width=2in]{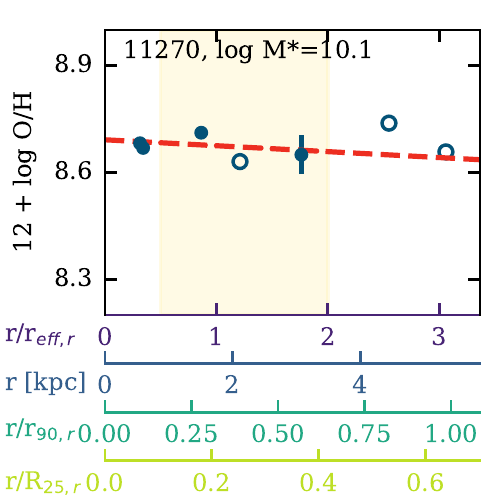}\\
        \includegraphics[width=2in]{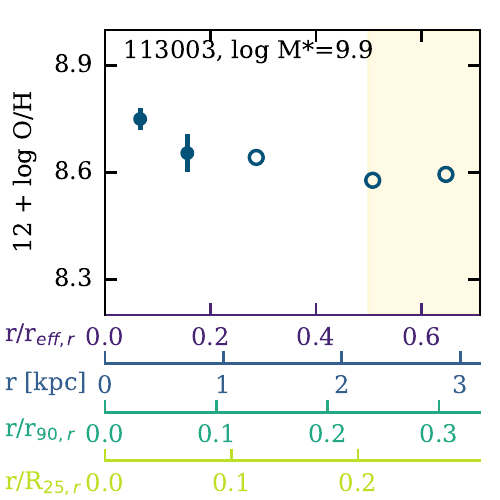}\hfill
        \includegraphics[width=2in]{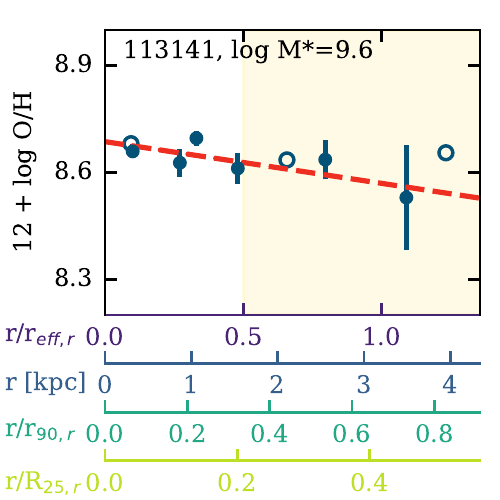}\hfill
        \includegraphics[width=2in]{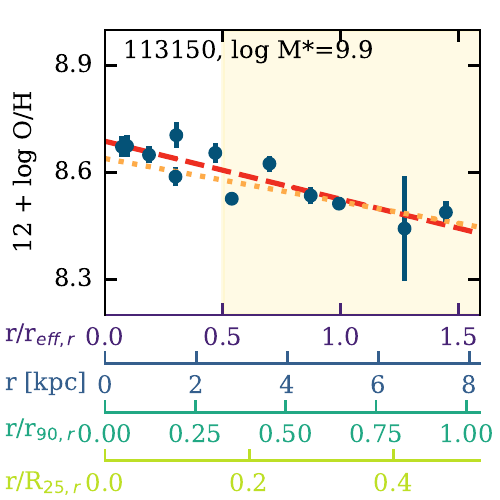}\\
        \includegraphics[width=2in]{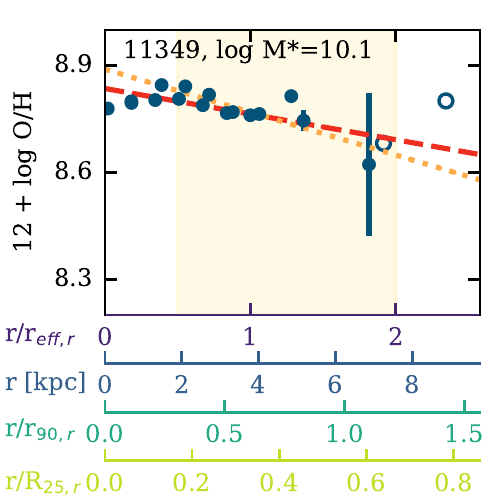}\hfill
        \includegraphics[width=2in]{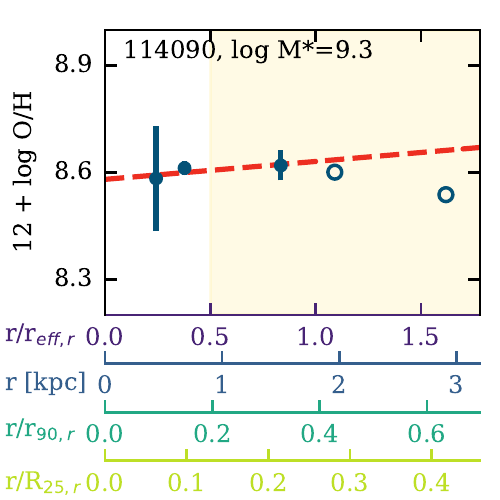}\hfill
        \includegraphics[width=2in]{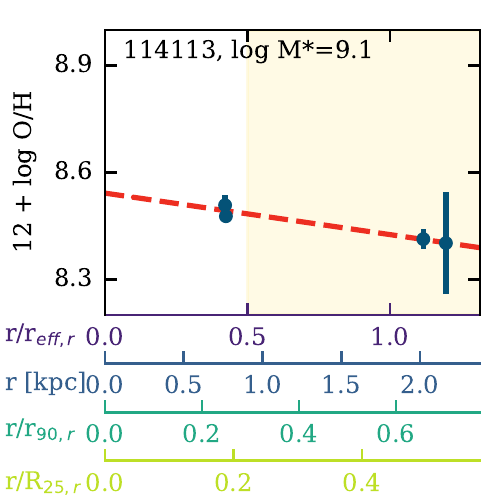}\\
        \includegraphics[width=2in]{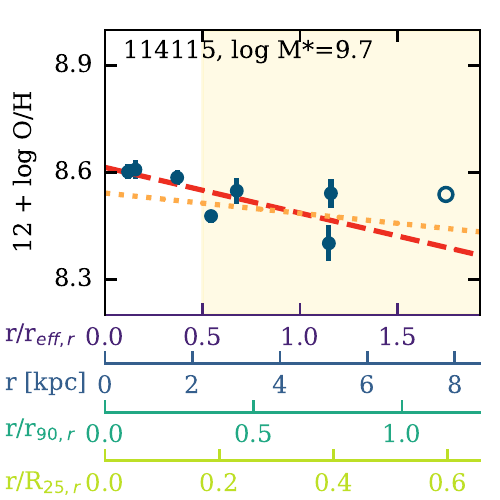}\hfill
        \includegraphics[width=2in]{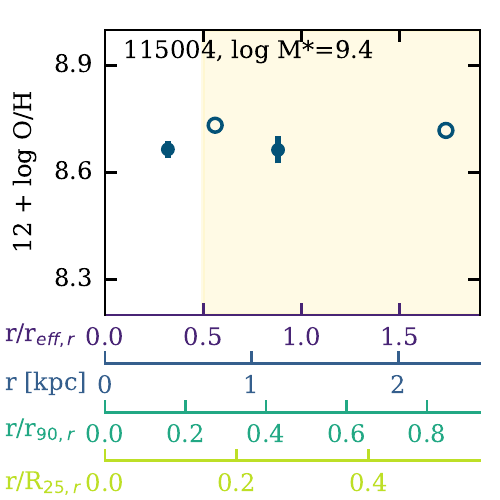}\hfill
        \includegraphics[width=2in]{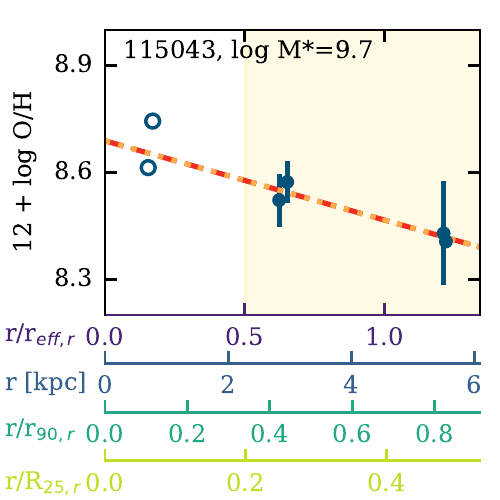}
    \caption{continued. }
\end{figure*}

\setcounter{figure}{0}
\begin{figure*}
    \center
        \includegraphics[width=2in]{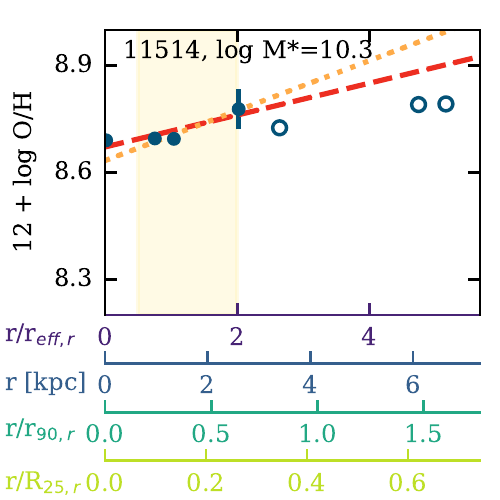}\hfill
        \includegraphics[width=2in]{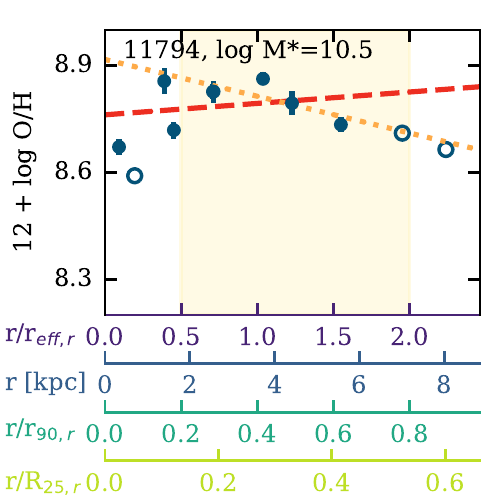}\hfill
        \includegraphics[width=2in]{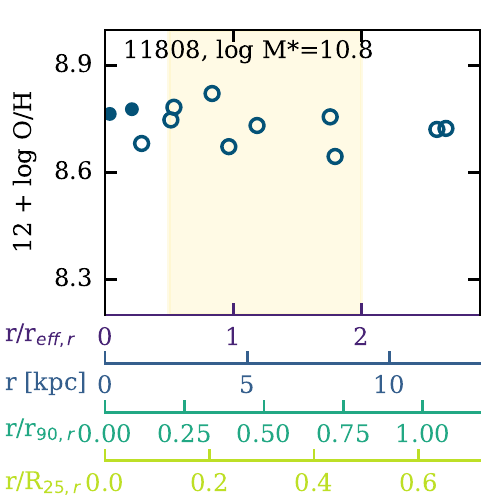}\\
        \includegraphics[width=2in]{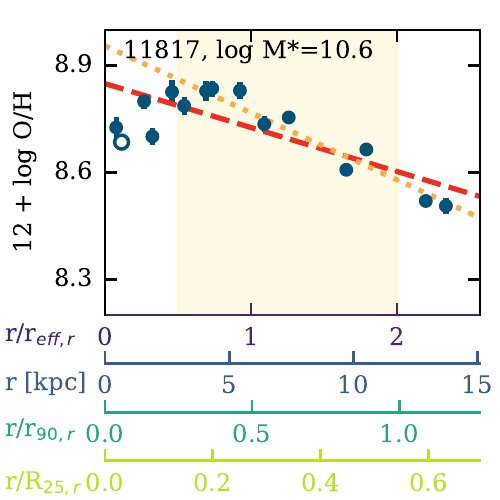}\hfill
        \includegraphics[width=2in]{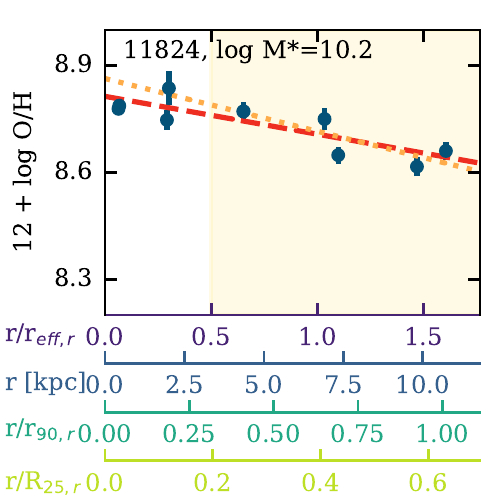}\hfill
        \includegraphics[width=2in]{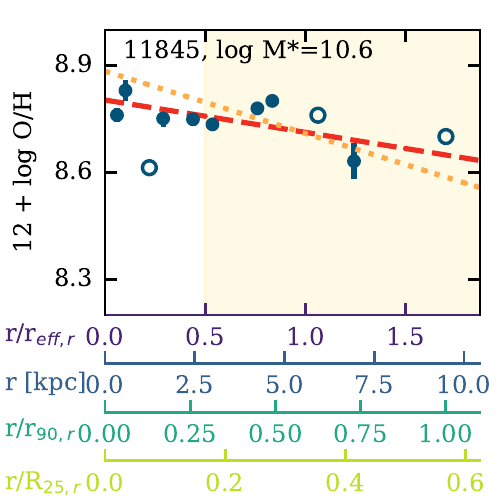}\\
        \includegraphics[width=2in]{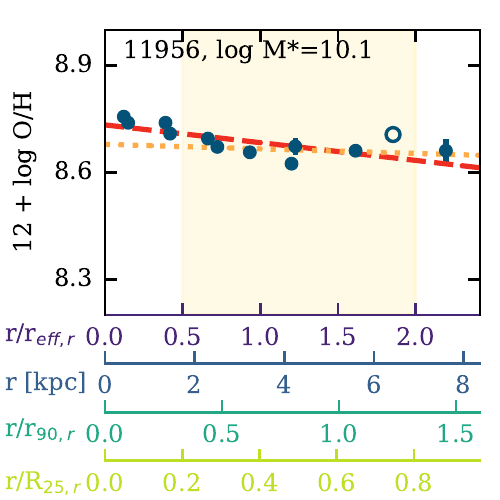}\hfill
        \includegraphics[width=2in]{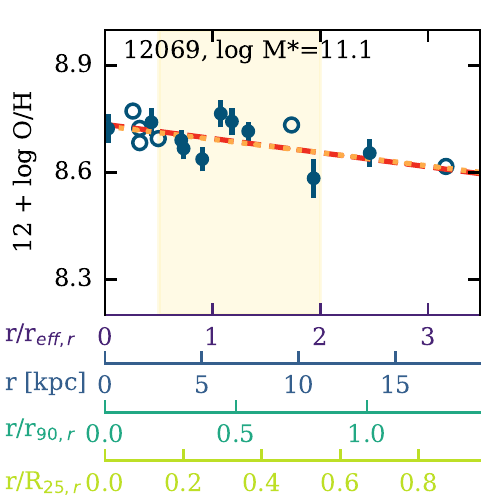}\hfill
        \includegraphics[width=2in]{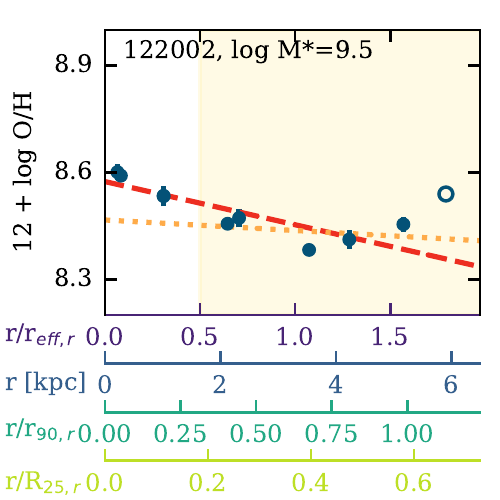}\\
        \includegraphics[width=2in]{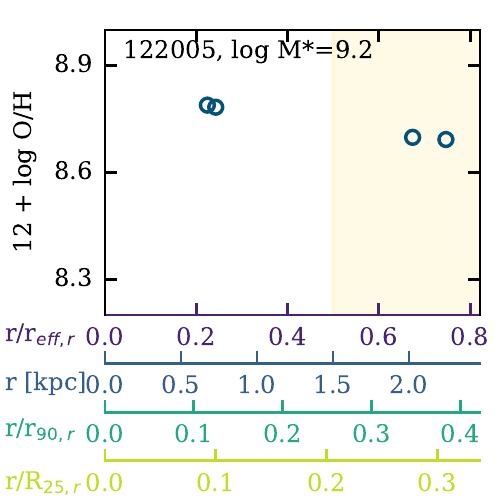}\hfill
        \includegraphics[width=2in]{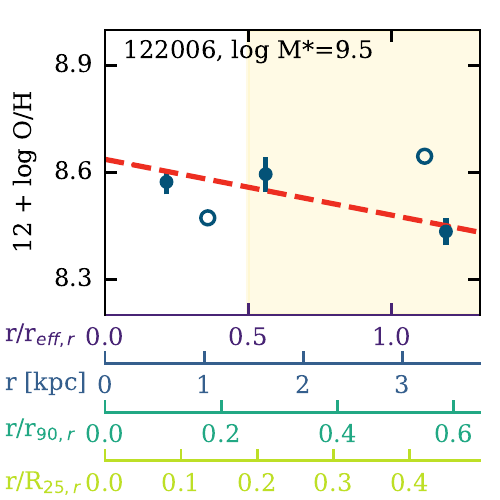}\hfill
        \includegraphics[width=2in]{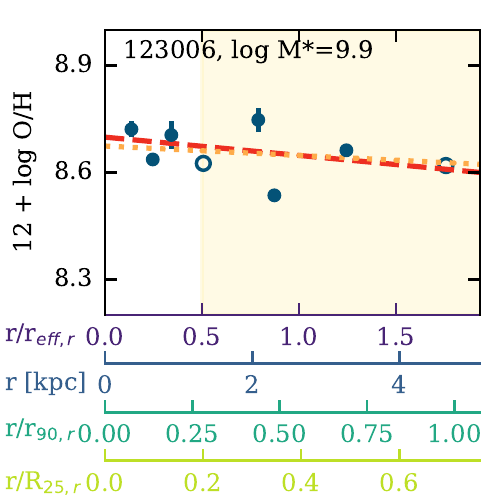}
    \caption{continued. }
\end{figure*}

\setcounter{figure}{0}
\begin{figure*}
    \center
        \includegraphics[width=2in]{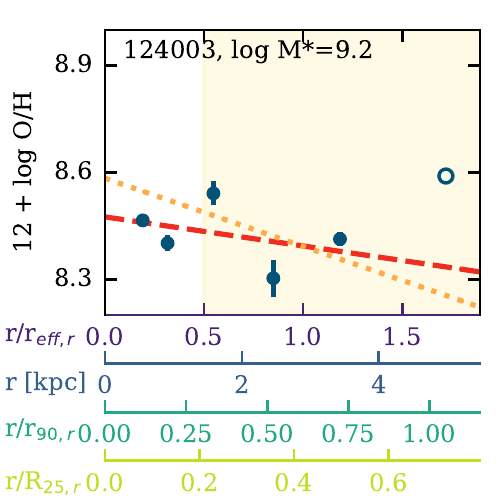}\hfill
        \includegraphics[width=2in]{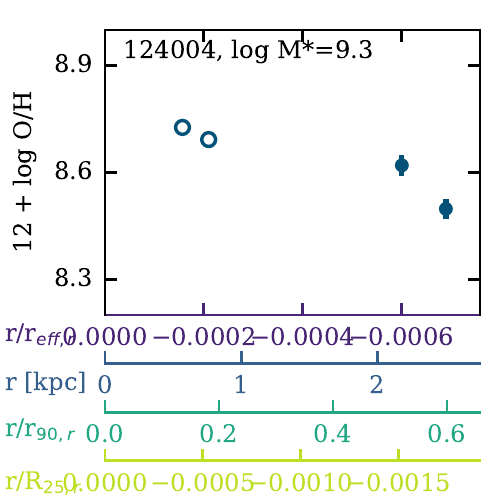}\hfill
        \includegraphics[width=2in]{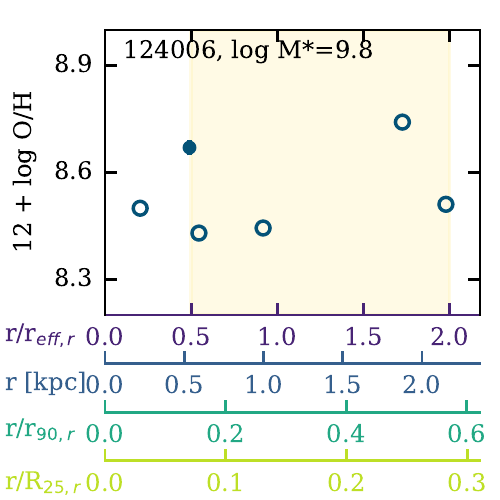}\\
        \includegraphics[width=2in]{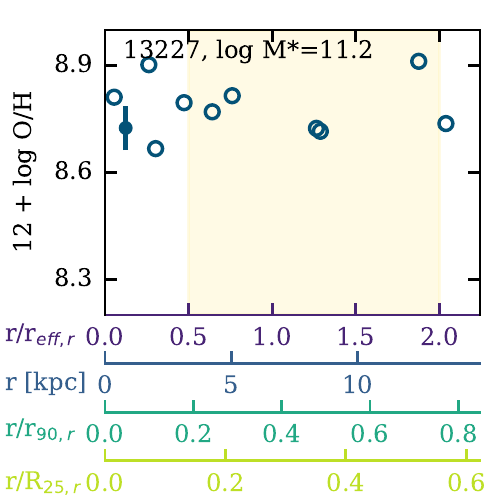}\hfill
        \includegraphics[width=2in]{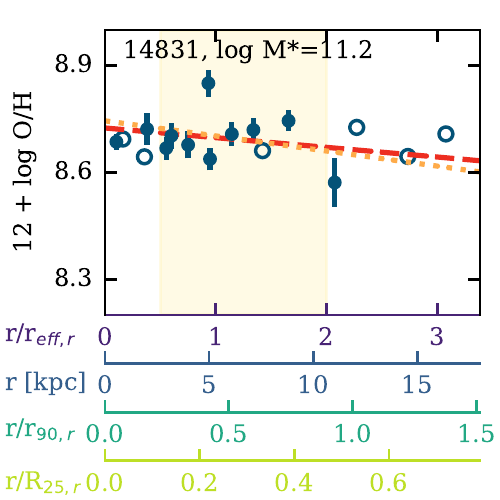}\hfill
        \includegraphics[width=2in]{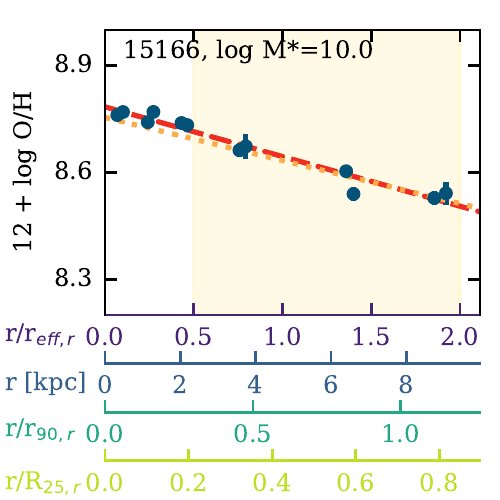}\\
        \includegraphics[width=2in]{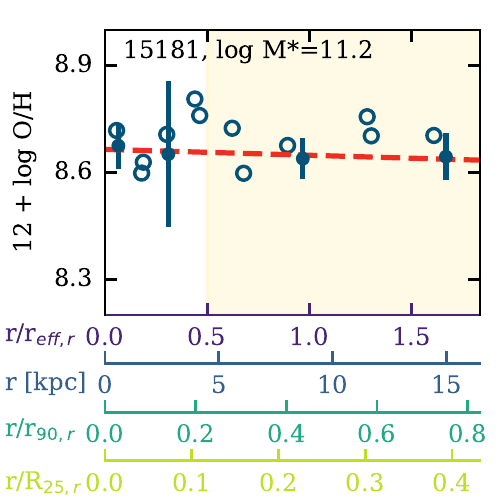}\hfill
        \includegraphics[width=2in]{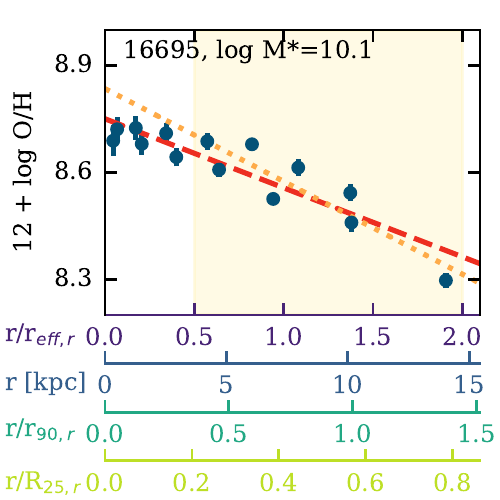}\hfill
        \includegraphics[width=2in]{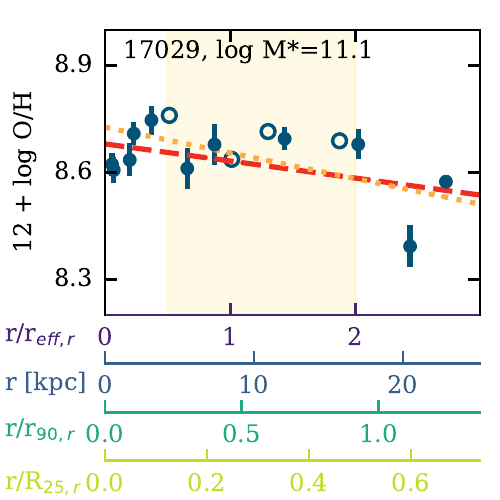}\\
        \includegraphics[width=2in]{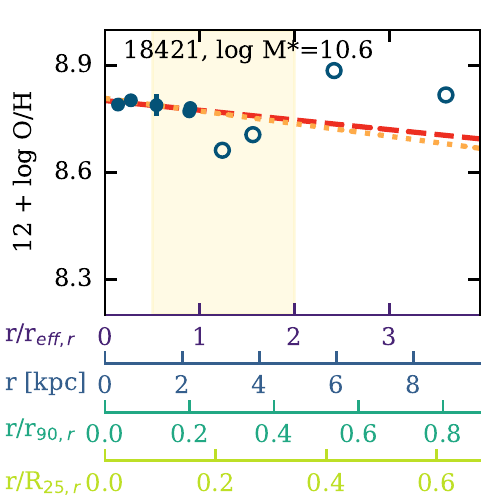}\hfill
        \includegraphics[width=2in]{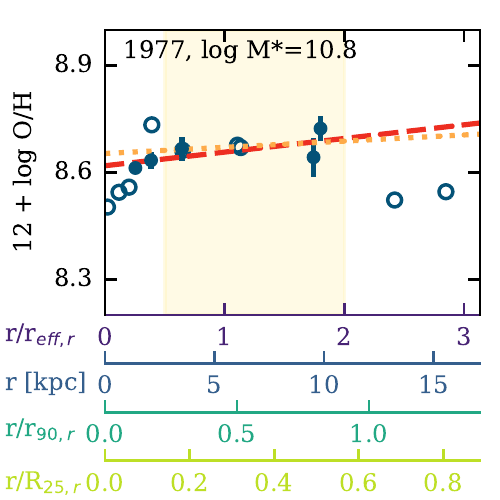}\hfill
        \includegraphics[width=2in]{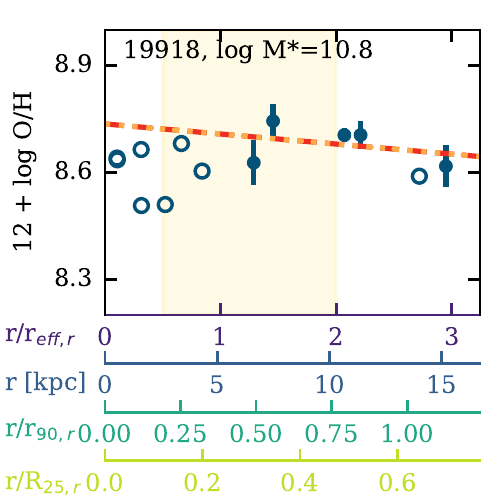}
    \caption{continued. }
\end{figure*}

\setcounter{figure}{0}
\begin{figure*}
    \center
        \includegraphics[width=2in]{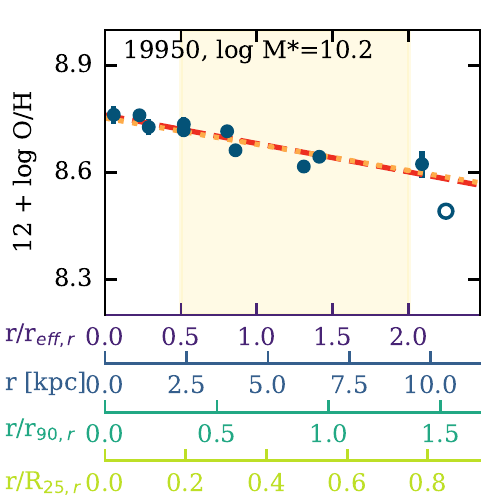}\hfill
        \includegraphics[width=2in]{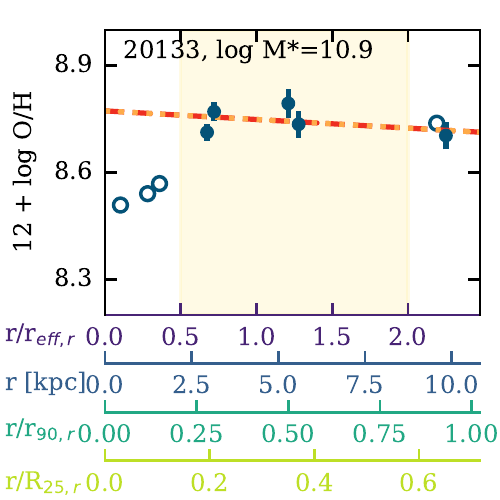}\hfill
        \includegraphics[width=2in]{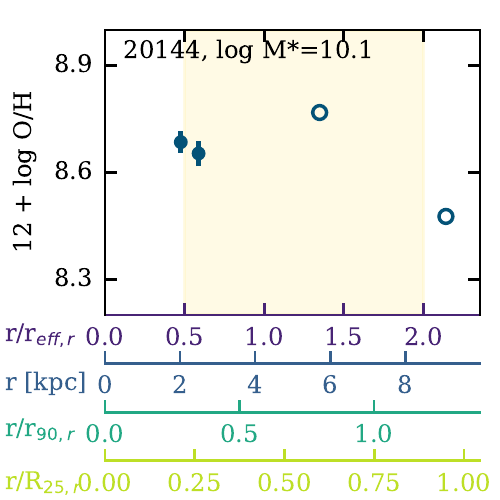}\\
        \includegraphics[width=2in]{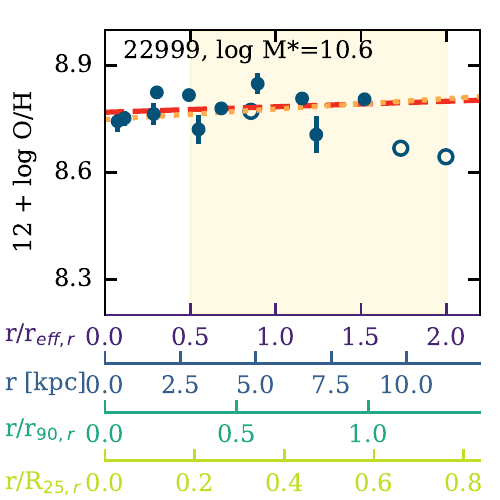}\hfill
        \includegraphics[width=2in]{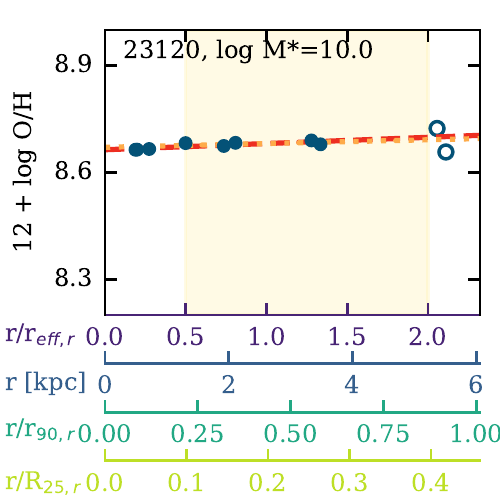}\hfill
        \includegraphics[width=2in]{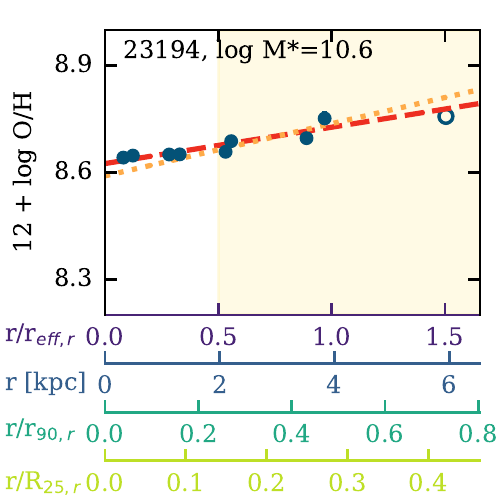}\\
        \includegraphics[width=2in]{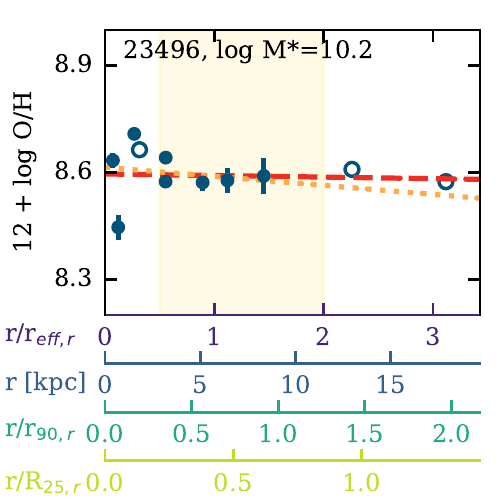}\hfill
        \includegraphics[width=2in]{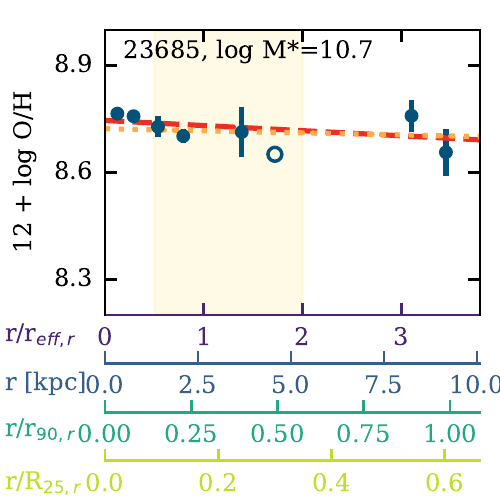}\hfill
        \includegraphics[width=2in]{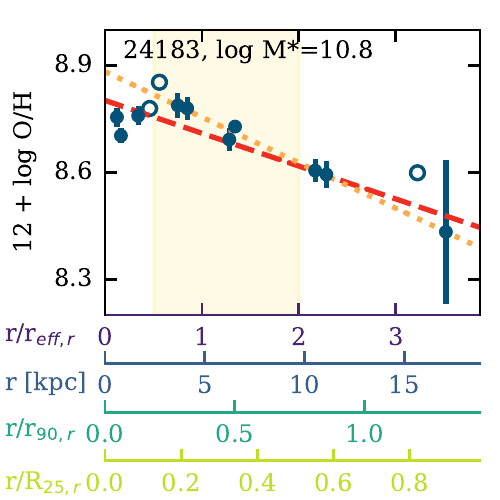}\\
        \includegraphics[width=2in]{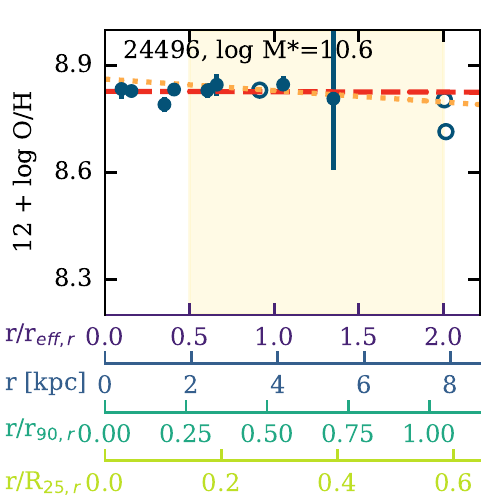}\hfill
        \includegraphics[width=2in]{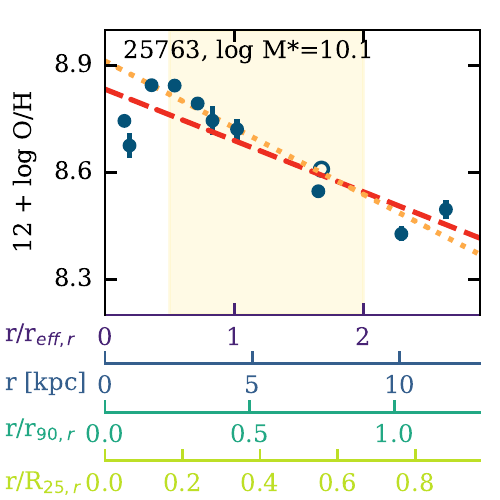}\hfill
        \includegraphics[width=2in]{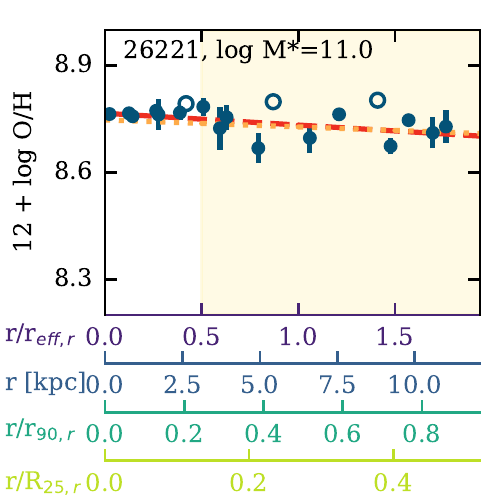}
    \caption{continued. }
\end{figure*}

\setcounter{figure}{0}
\begin{figure*}
    \center
        \includegraphics[width=2in]{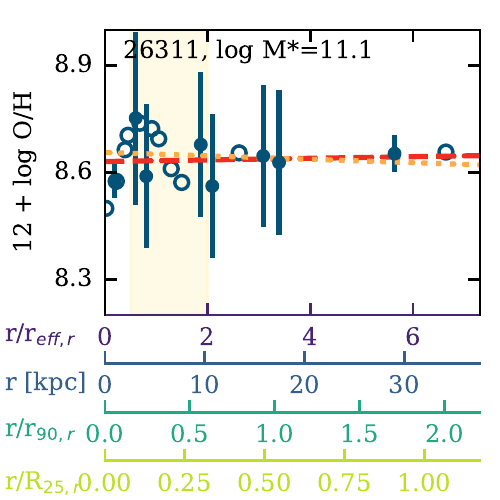}\hfill
        \includegraphics[width=2in]{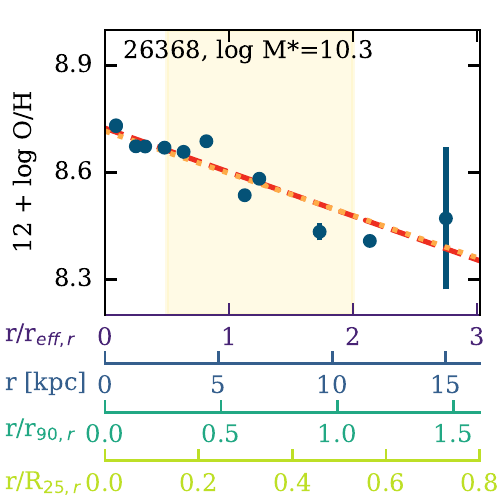}\hfill
        \includegraphics[width=2in]{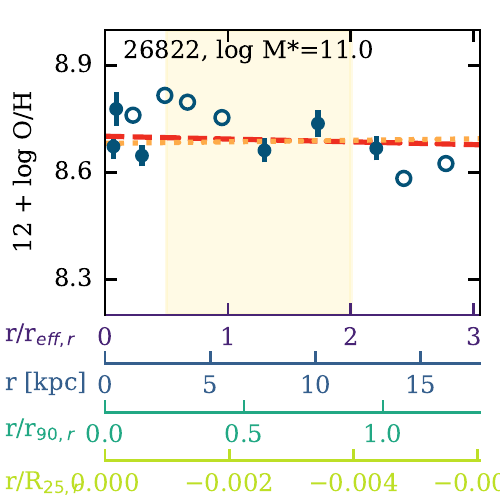}\\
        \includegraphics[width=2in]{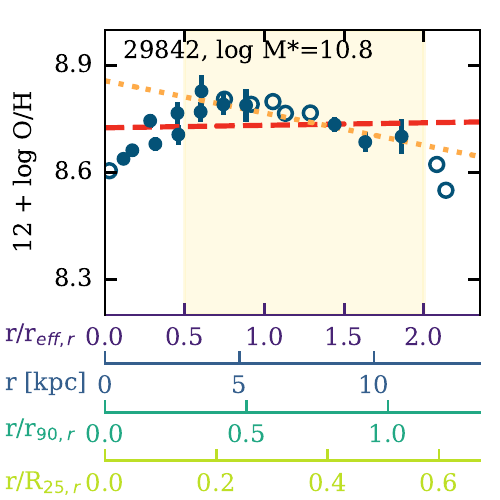}\hfill
        \includegraphics[width=2in]{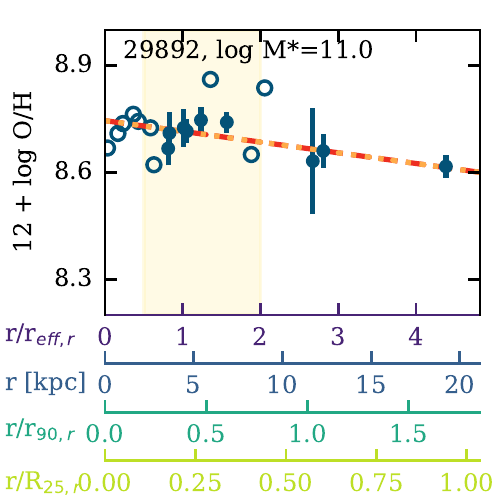}\hfill
        \includegraphics[width=2in]{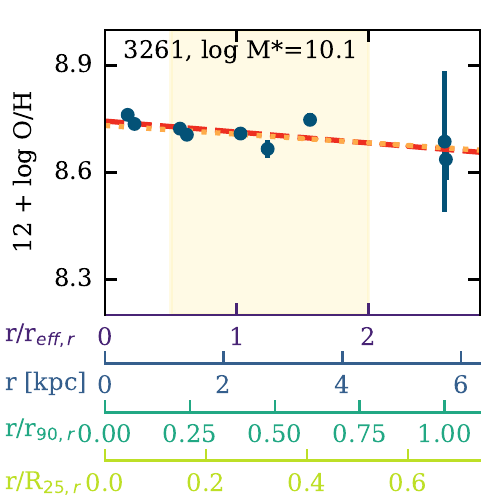}\\
        \includegraphics[width=2in]{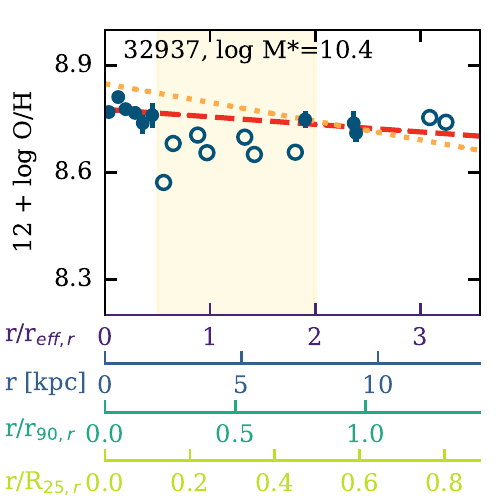}\hfill
        \includegraphics[width=2in]{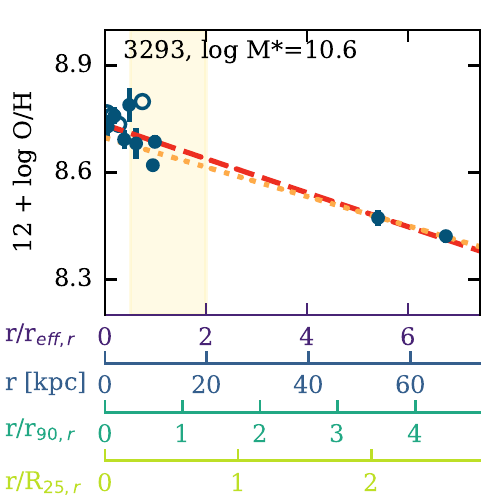}\hfill
        \includegraphics[width=2in]{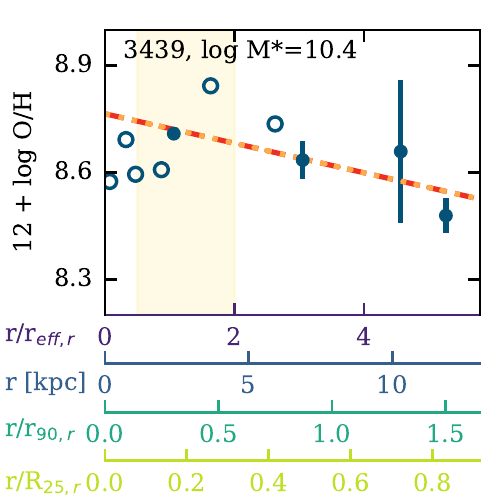}\\
        \includegraphics[width=2in]{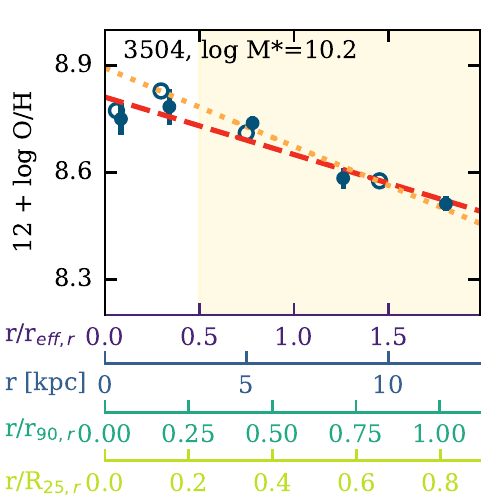}\hfill
        \includegraphics[width=2in]{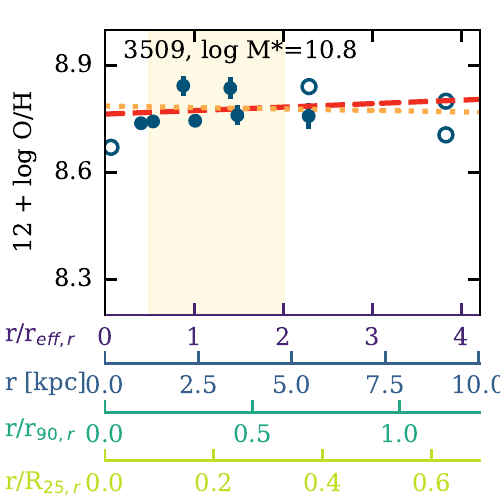}\hfill
        \includegraphics[width=2in]{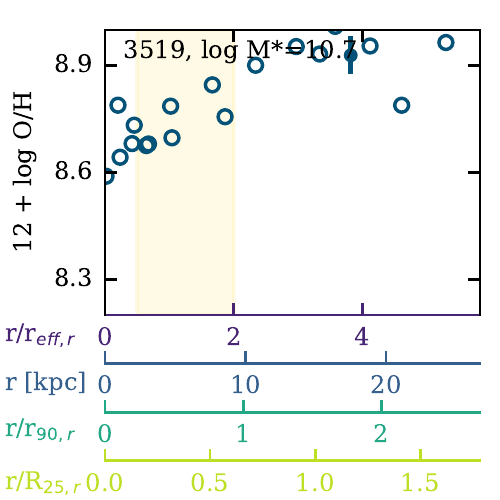}
    \caption{continued. }
\end{figure*}

\setcounter{figure}{0}
\begin{figure*}
    \center
        \includegraphics[width=2in]{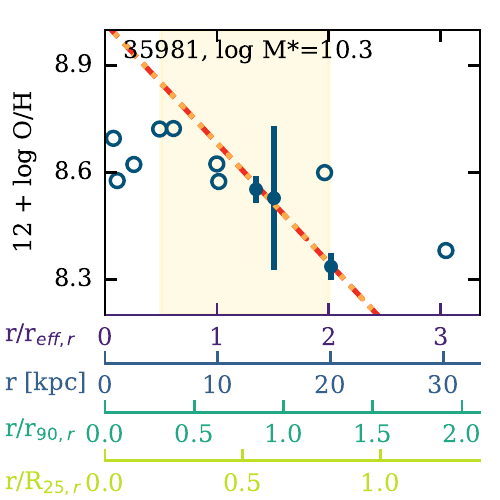}\hfill
        \includegraphics[width=2in]{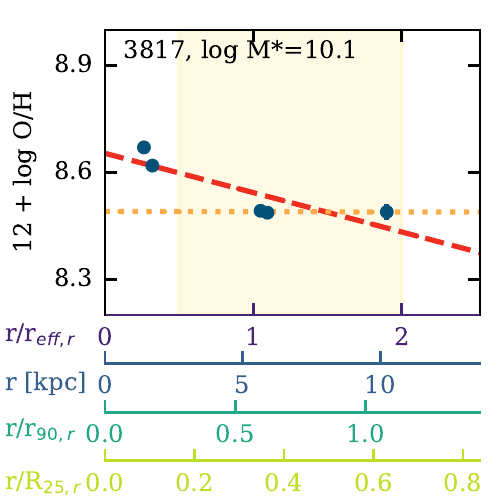}\hfill
        \includegraphics[width=2in]{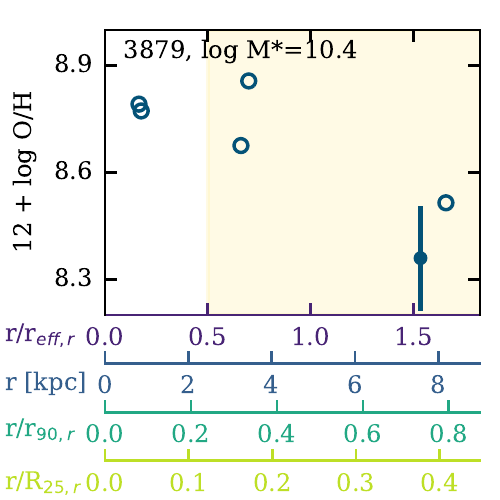}\\
        \includegraphics[width=2in]{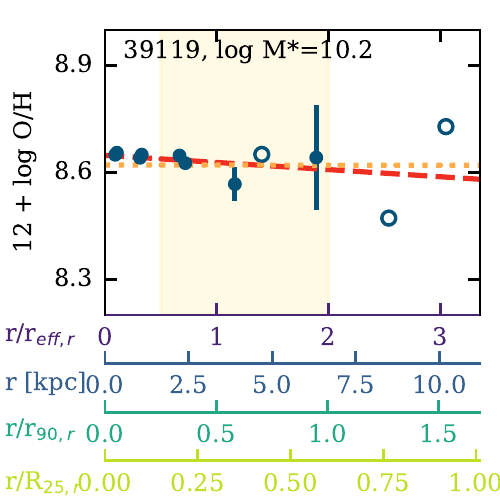}\hfill
        \includegraphics[width=2in]{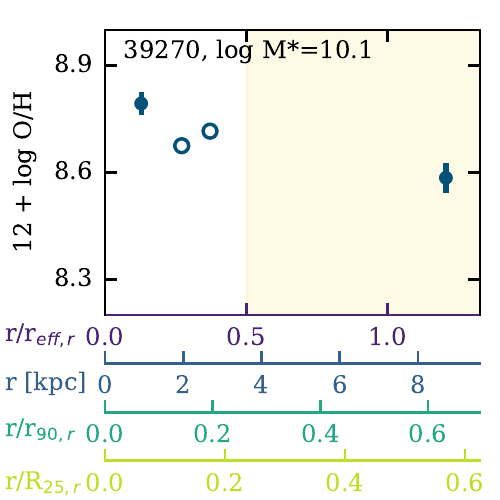}\hfill
        \includegraphics[width=2in]{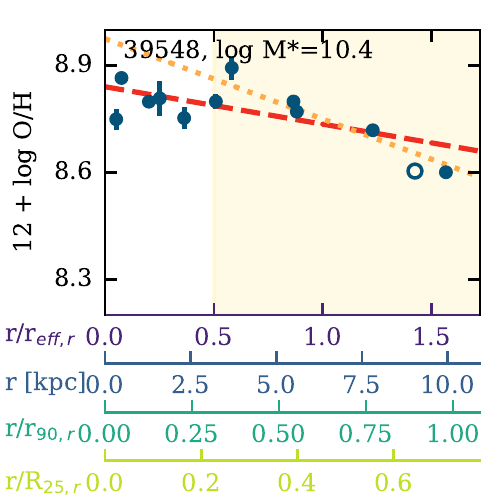}\\
        \includegraphics[width=2in]{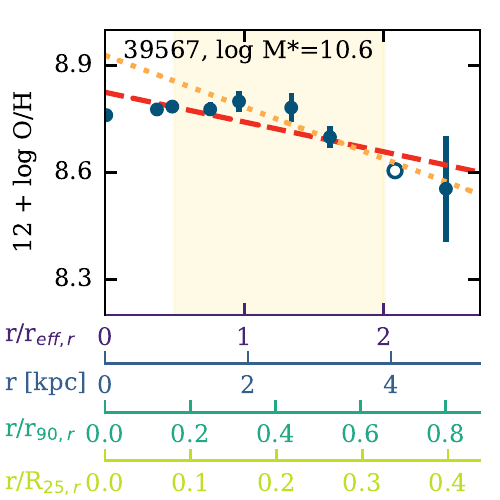}\hfill
        \includegraphics[width=2in]{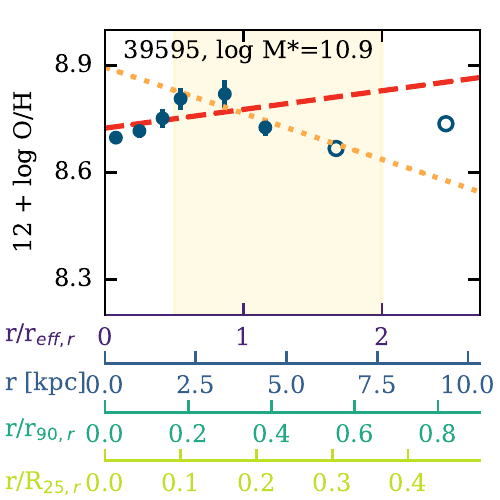}\hfill
        \includegraphics[width=2in]{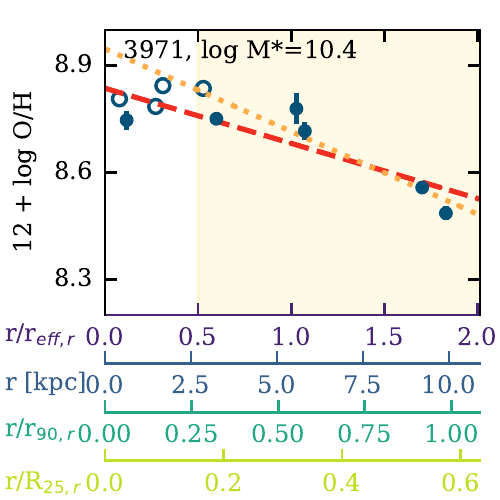}\\
        \includegraphics[width=2in]{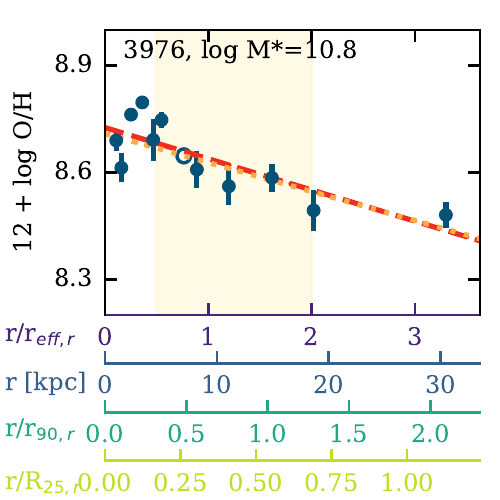}\hfill
        \includegraphics[width=2in]{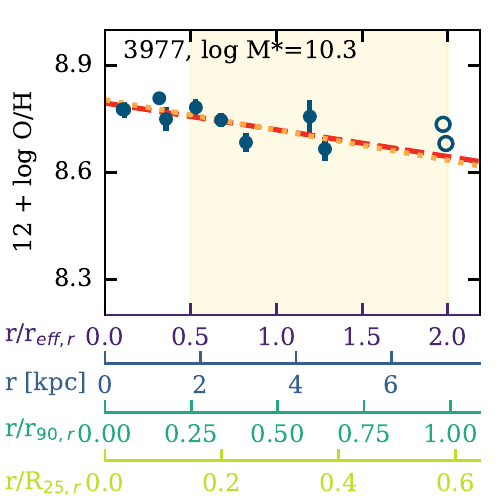}\hfill
        \includegraphics[width=2in]{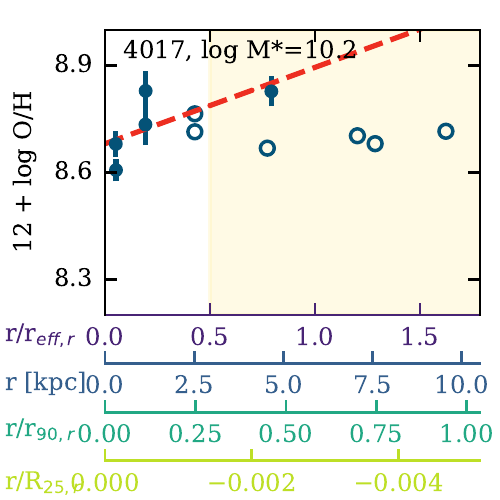}
    \caption{continued. }
\end{figure*}

\setcounter{figure}{0}
\begin{figure*}
    \center
        \includegraphics[width=2in]{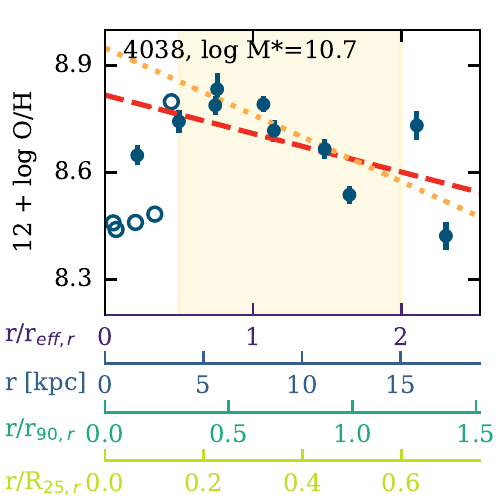}\hfill
        \includegraphics[width=2in]{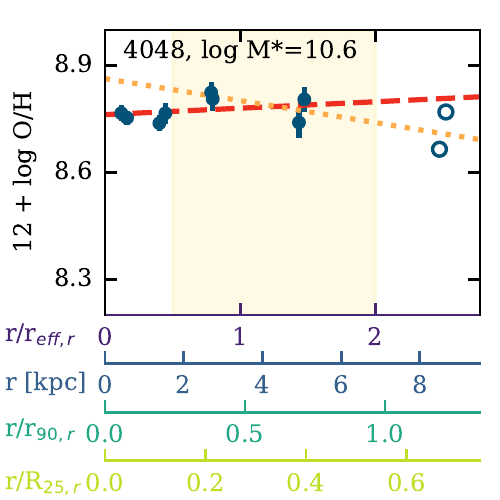}\hfill
        \includegraphics[width=2in]{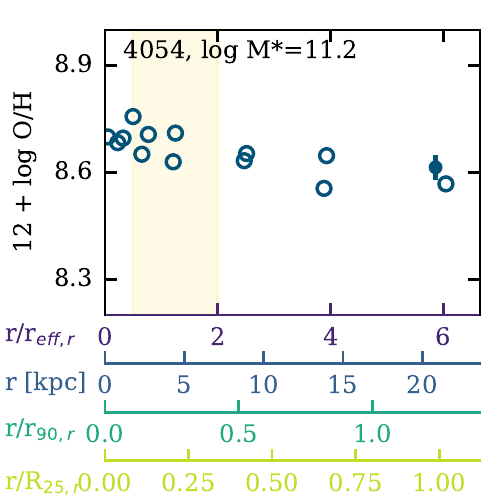}\\
        \includegraphics[width=2in]{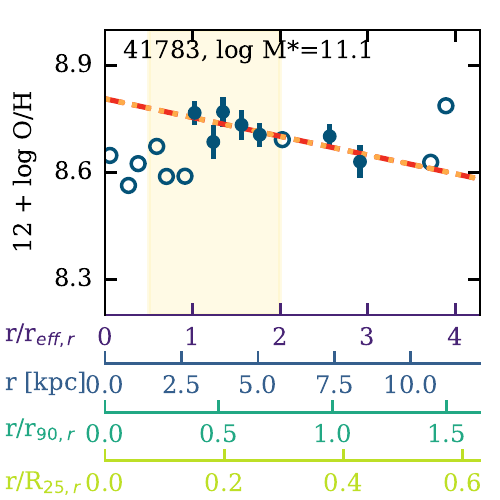}\hfill
        \includegraphics[width=2in]{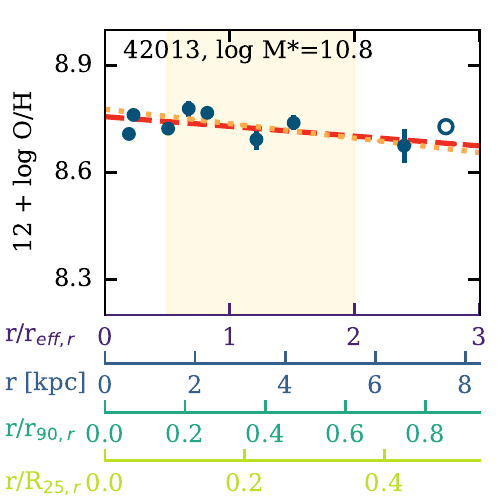}\hfill
        \includegraphics[width=2in]{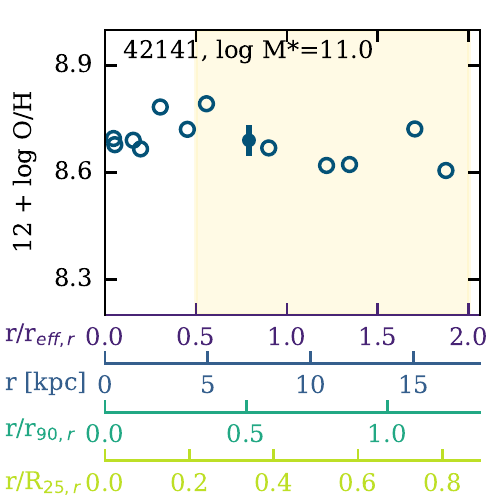}\\
        \includegraphics[width=2in]{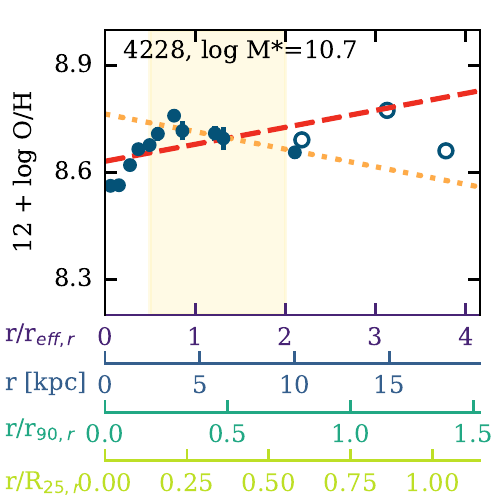}\hfill
        \includegraphics[width=2in]{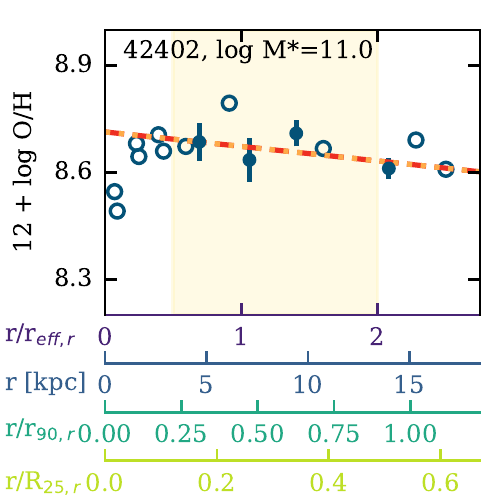}\hfill
        \includegraphics[width=2in]{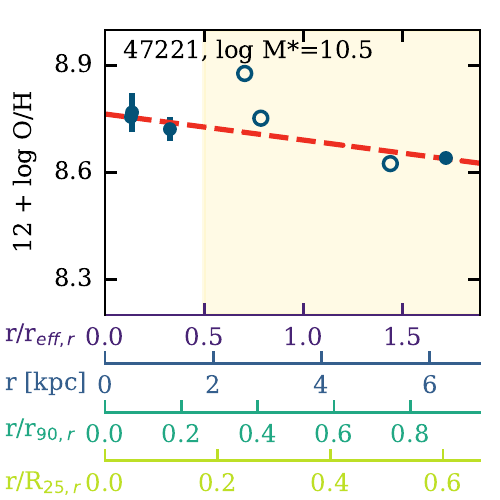}\\
        \includegraphics[width=2in]{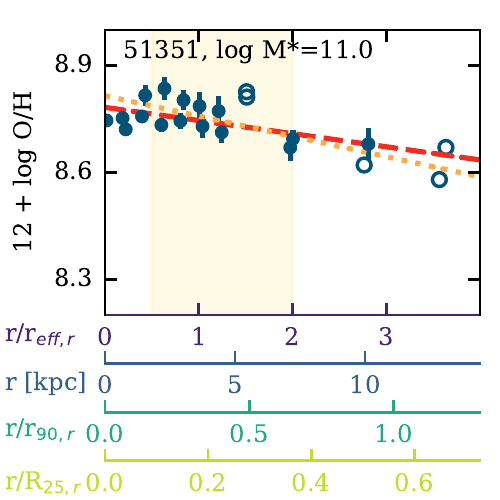}\hfill
        \includegraphics[width=2in]{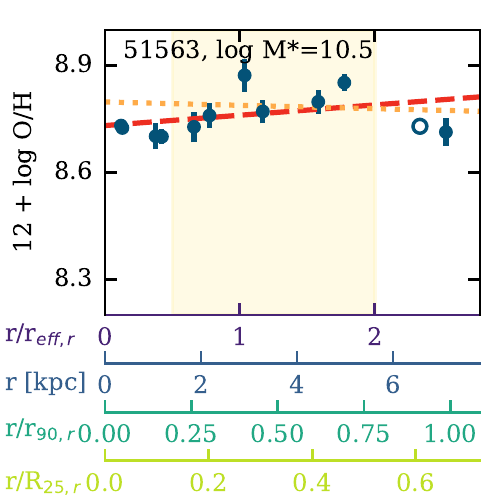}\hfill
        \includegraphics[width=2in]{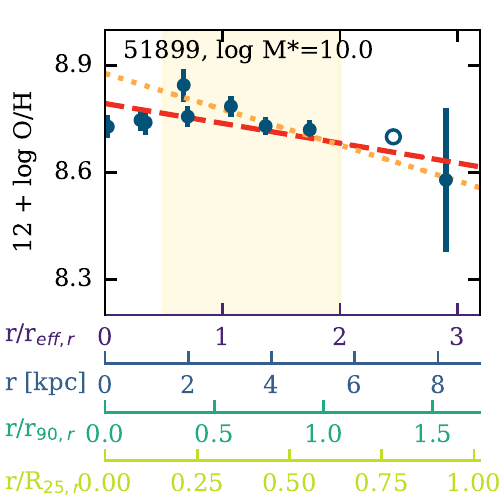}
    \caption{continued. }
\end{figure*}

\setcounter{figure}{0}
\begin{figure*}
    \center
        \includegraphics[width=2in]{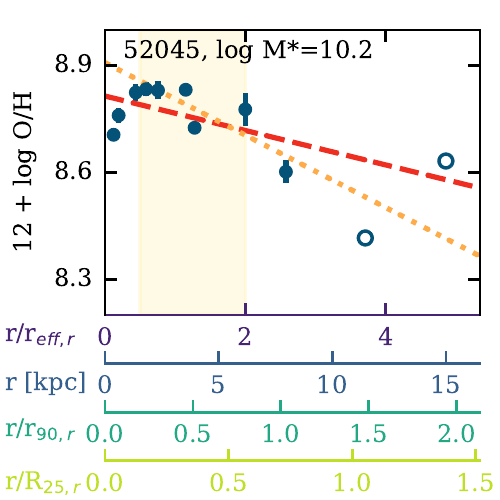}\hfill
        \includegraphics[width=2in]{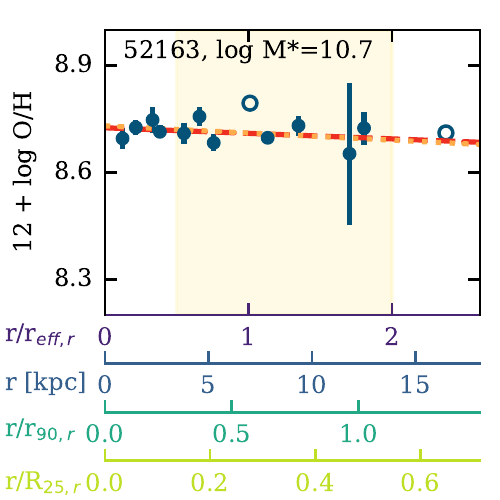}\hfill
        \includegraphics[width=2in]{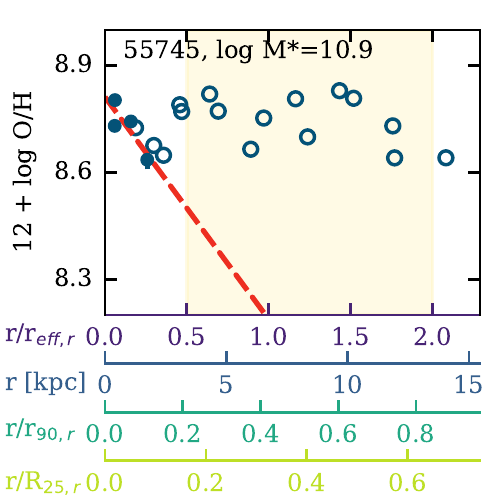}\\
        \includegraphics[width=2in]{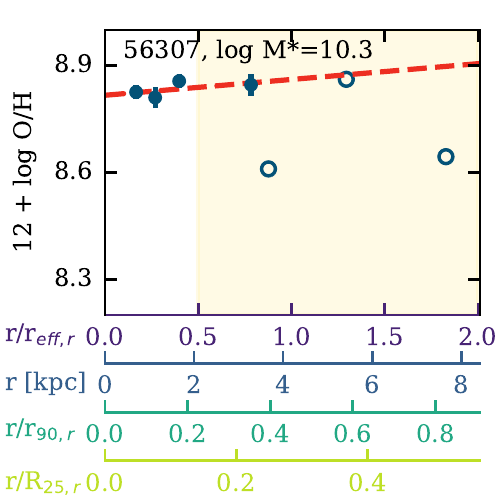}\hfill
        \includegraphics[width=2in]{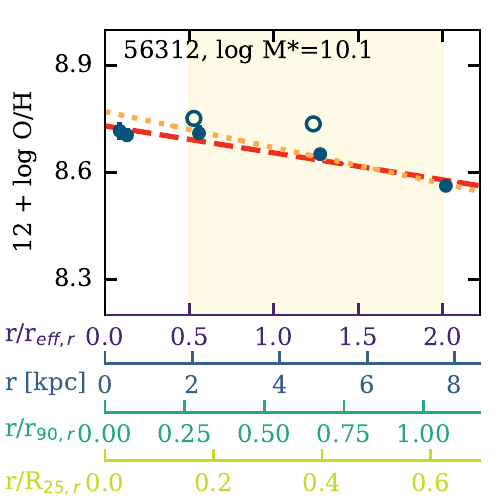}\hfill
        \includegraphics[width=2in]{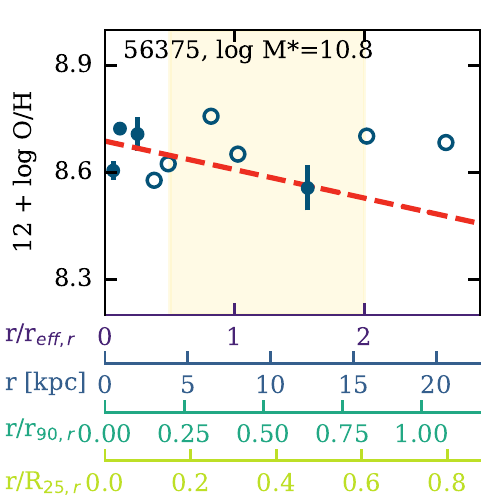}\\
        \includegraphics[width=2in]{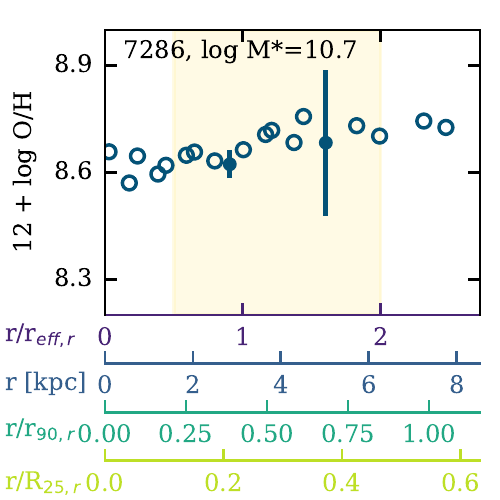}\hfill
        \includegraphics[width=2in]{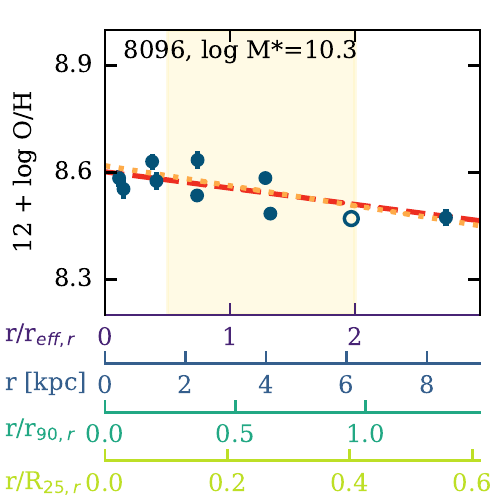}\hfill
        \includegraphics[width=2in]{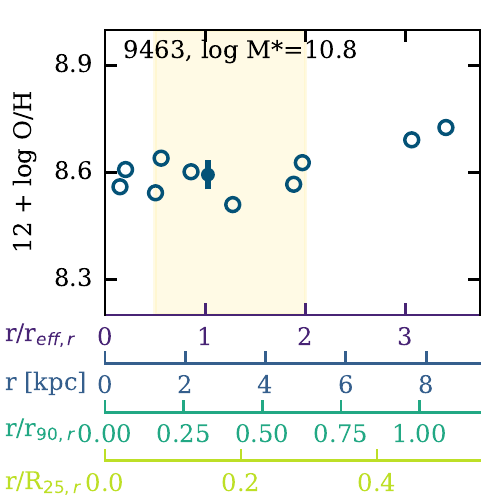}\\
        \includegraphics[width=2in]{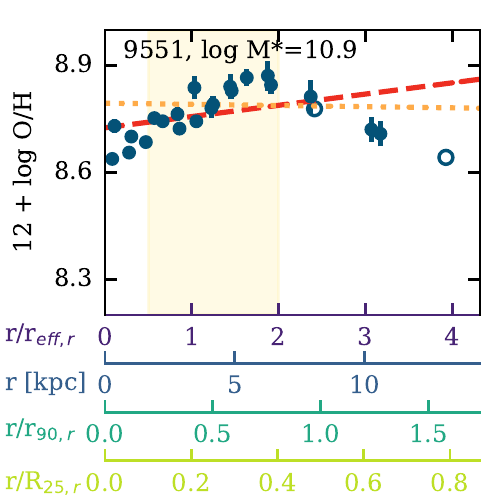}
    \caption{continued. }
\end{figure*}

\end{appendix}
%%%%%%%%%%%%%%%%%%%%%%%%%%%%%%%%%%%%%%%%%%%%%%%%%%
\end{document}